\documentclass[aps,amsfonts,pra,twocolumn,showpacs]{revtex4-1}
\usepackage{epsfig,amsmath,amssymb,bm,epsf,graphicx,psfrag,hyperref, bbm}
\usepackage{color, comment, verbatim}
\usepackage{ragged2e}
\usepackage{subfigure}
\def\bra#1{\langle#1\vert}
\def\ket#1{\vert#1\rangle}
\newcommand{\bc}{\begin{center}}
\newcommand{\ec}{\end{center}}
\newcommand{\be}{\begin{equation}}
\newcommand{\ee}{\end{equation}}
\newcommand{\bea}{\begin{eqnarray}}
\newcommand{\eea}{\end{eqnarray}}

\begin{document}

\title{Symmetry-protected topologically ordered states for universal  quantum computation}

\author{Hendrik Poulsen Nautrup}
\email{h.poulsennautrup@gmail.com}
\affiliation{Department of Physics and Astronomy, Stony Brook University, Stony Brook, NY 11794, United States}
\author{Tzu-Chieh Wei}
\email{tzu-chieh.wei@stonybrook.edu}
\affiliation{C.N. Yang Institute for Theoretical Physics and Department of Physics and Astronomy, Stony Brook University, Stony Brook, NY 11794, United States}
\date{\today}

\begin{abstract}
Measurement-based quantum computation is a model for quantum information processing utilizing local measurements on suitably entangled resource states for the implementation of quantum gates. A complete characterization for universal resource states is still missing. It has been shown that symmetry-protected topological order in one dimension can be exploited for the protection of certain quantum gates in measurement-based quantum computation. In this paper we show that the two-dimensional plaquette states on arbitrary lattices exhibit nontrivial symmetry-protected topological order in terms of symmetry fractionalization and that they are universal resource states for quantum computation. Our results of the nontrivial symmetry-protected topological order on arbitrary 2D lattices are based on an extension of the recent construction by Chen, Gu, Liu and Wen {[Phys. Rev. B \textbf{87}, 155114 (2013)]} on the square lattice. 
\end{abstract}
\maketitle

\section{Introduction}
Quantum computers provide, for certain classes of problems, an exponential speedup over classical computers~\cite{NielsenChuang00}. Among many approaches, quantum computation can be implemented  by local measurement on suitably entangled resource states~\cite{Oneway, Oneway2,MBQCAKLT,ZengKwek}, which is known as measurement-based quantum computation (MBQC). Resource states are said to be universal for MBQC if we can implement a universal set of quantum gates by local measurement so that any quantum circuit can be efficiently simulated. This model of computation can also be understood in terms of teleportation-based quantum computation in the valence-bond solid (VBS) formalism~\cite{Verstraete}, as well as the so-called correlation-space quantum computation~\cite{Gross,correlationspace}. A local measurement in MBQC corresponds to a joint measurement on partons of the VBS (which are to be understood as virtual qubits) within one physical site~\cite{Verstraete} and the resources in the two equivalent models are related by a mathematical projection. Since the discovery of the one-way computer~\cite{Oneway} on the square-lattice cluster state~\cite{Cluster}, there have been quite a few other universal resources states uncovered, including, for example, graph states~\cite{universality}, the tricluster state~\cite{Chen2009}, weighted graph states~\cite{GrossEtAl}, modified toric code states~\cite{GrossEtAl}, and Affleck-Kennedy-Lieb-Tasaki (AKLT) states~\cite{AKLT} on various lattices~\cite{WeiAffleckRaussendorf11,Miyake11,Wei13,WeiEtAl,WeiRaussendorf15} and their deformations~\cite{Gross,DarmawanBrennenBartlett}. However, as of present, there still does not exist a complete characterization of all universal resources.

An intriguing connection between the resourcefulness in MBQC and certain phases of matter was discovered in Ref.~\cite{symmetryMBQC}, where the authors show that there exists a property of many-body states, namely symmetry-protected topological (SPT) order in 1D, that can be utilized for the protection of certain quantum gates in MBQC, even though 1D quantum states do not accommodate universal quantum computation. In Ref. \cite{universalSPT1D,abhi} the utility of SPT phases for quantum computation in 1D has also been demonstrated for other symmetry groups beyond the $Z_2\times Z_2$ symmetry. SPT phases are gapped quantum systems that have certain symmetries which are not spontaneously broken in the nondegenerate ground state~\cite{1DSPT, 1DSPTcomplete, SPTRG, SPTtensor, SPTLU,ChenScience,cohomology}. Systems with SPT order (SPTO) can be robust against small local perturbations that respect the symmetry~\cite{robustness1, robustness2}.  Ground states of SPT phases can be mapped to product states with local unitary (LU) transformations and thus, are short-range entangled (SRE). Hence, they all belong to the same phase as the trivial (product) state if no symmetries are imposed. However, if we restrict ourselves to symmetric LU transformations, SPT ordered systems are distinct from the trivial phase~\cite{SPTtensor, SPTLU, SPTRG}. Due to the potential advantage that SPTO may bring to MBQC, one question that arises naturally is whether we can find a novel universal resource state that exhibits nontrivial SPTO in two spatial dimensions or higher.

In Refs.~\cite{ChenScience,cohomology} the authors first discovered a consistent relation between the third group cohomology $\mathcal{H}^3(G,U(1))$ of a symmetry group $G$ and SPTO in (2+1)D bosonic systems. Particularly, they prove that each nontrivial element of the third group cohomology corresponds to a distinct, nontrivial SPT phase. Specifically in Ref.~\cite{cohomology}, they discuss a fixed point wave function in its canonical form, showing that it exhibits nontrivial SPTO with respect to symmetry representations constructed from nontrivial elements in  $\mathcal{H}^3(G,U(1))$. In this approach different SPT phases are not encoded in the wave function but in the different symmetry fractionalizations of the symmetry transformation. The canonical form is defined on a square lattice constituting a plaquette-like entanglement structure among partons on neighboring sites. In an in-depth discussion, the authors of Ref.~\cite{cohomology} explain that the canonical fixed point wavefunction can be related to any other, more generic SRE state in the same phase through symmetric LU evolution.

In this paper, we show that the canonical plaquette-like entanglement structure that displays nontrivial SPTO enables universal MBQC. Moreover, we shall show that such entanglement structure defined on general 2D lattices is a universal resource. But does it also display nontrivial SPTO as in the square lattice? To answer this, we explicitly extend the original symmetry construction to arbitrary lattices in two spatial dimensions. With the symmetry representation generalized, we can then show that these plaquette states and their generalizations on all 2D lattices indeed exhibit nontrivial SPTO, depending on the symmetry.  We show that these states are universal resources for MBQC as long as their underlying graphs reside in the supercritical phase of percolation. This means that all (generalized) plaquette states on 2D regular or quasicrystallene lattices  are both nontrivial SPT states and universal resources for MBQC.

It is interesting that whether the plaquette states and their generalizations are universal resouce states depends only on whether the underlying graphs are in the supercritical phase of percolation. This is exactly the same property that governs the universality of graph states, which was shown in, e.g., regular lattices~\cite{universality}, faulty lattices~\cite{Browne} and random planar graphs~\cite{WeiAffleckRaussendorf12}. As a contrast, the AKLT states, which also exhibit nontrivial SPTO only if translation invariance is preserved~\cite{cohomology,Zengbook}, are known to be universal only for some regular lattices~\cite{WeiAffleckRaussendorf11,Miyake11,Wei13,WeiEtAl,WeiRaussendorf15}. 

The rest of the paper is organized as follows. In Sec.~\ref{review}, we familiarize the reader with the mathematical framework of group cohomology and its graphic representation. Since the results have already been proven in e.g. Ref. \cite{cohomology}, we will only outline the most relevant properties. In the last part of this section, we will review the aforementioned canonical form. In Sec.~\ref{construction}, we illustrate the issue of defining the symmetry representation on aribitrary 2D lattices and provide a solution. We also construct SRE plaquette states on arbitrary lattices and verify that they exhibit nontrivial SPT order.
  In Sec.~\ref{universality}, we prove that this class of symmetry-protected states is universal for MBQC. We summarize and discuss extensions to our results in Sec.~\ref{summary}.

%%%%%%%%%%%%%%%%%%%%%%%%%%%%%%%%%%%%%%%%%%%%%%%%%%%%%%%%%%%%%%%%%%%%%%%%%%%%%%%%%%%%%%%%%%%%%%%%%%%%%%%%%%%%%%%%%%%%%%%%%%%%%%%%%%%%%%%%%%%%%%%%%%%%%%%%%%%%%%%%%%%%%%%%%%%%%%%%%%%%%%%%%%%%%%%%%%%%%%%%%%%%%%%%%%%%%%%%%%%%%%%%%%%%%%%%%%%%%%%%%%%%%%%%%%%%%%%%%%%%%%%%%%%%%%%%%%%%%%%%%%%%%%%%%%%%%%%%%%%%%%%%%%%%%%

\section{Review of relevant definitions and results}\label{review}

Symmetry-protected topoplogical phases possess nontrivial topological order only if a certain symmetry of the system is perserved, as opposed to the intrinsic topological order for which no symmetry is needed~\cite{1DSPT,1DSPTcomplete,SPTRG,SPTLU,SPTtensor,ChenScience}. Such a SPTO cannot be characterized by a local order parameter, either. Ground states of  nontrivial SPT phases cannot be continuously connected to trivial product states without closing the gap or breaking the protected
symmetry. From the entanglement viewpoint, these states possess the so-called short-ranged entanglement  as opposed to the long-ranged entanglement in the intrinsic topologically ordered states~\cite{SPTLU}. It turns out that group cohomology provides a complete classification for 1D SPT phases~\cite{1DSPTcomplete} and a comprehensive classification for two and higher dimensions~\cite{cohomology}. Here we review several subjects that will be essential for this work, including  group cohomology and the canonical form for short-ranged entangled states. We follow the seminal paper by Chen, Gu, Liu and Wen~\cite{cohomology}.
We remark that 
the classification can actually be understood via other approaches: (1) anomalous symmetry action at the boundary~\cite{ElseNayak}, (2) the nonlinear sigma model with a topological theta term~\cite{BiXu01,BiXu02}, and (3) gauge fields for symmetry twists~\cite{WangGuWen}, and (4) cobordism~\cite{Kapustin}, all leading to SPT phases beyond cohomology~\cite{VishwanathSenthil}.
   Readers who are familiar with the canonical form of short-ranged entangled states in 2D and group cohomology can skip this section and go directly to Sec.~\ref{construction} for our extension of the original symmetry representation for SPT phases on arbitrary lattices in two spatial dimensions.

\subsection{SRE many-body states and their canonical form}\label{canonical}
\begin{figure}
	\subfigure[]{
		\includegraphics[width=0.2\textwidth]{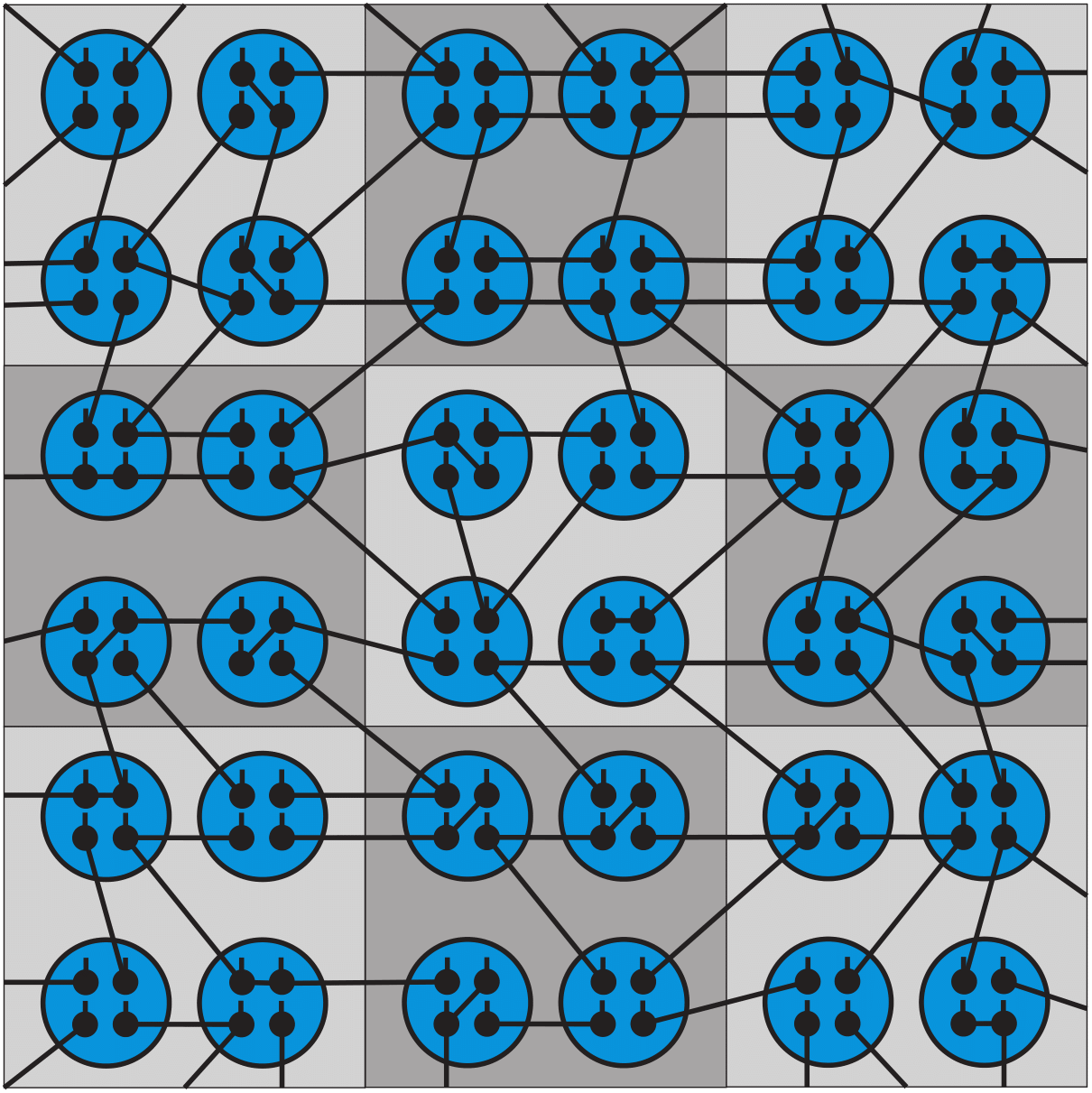}
		\label{tensorSRE}}
	\subfigure[]{
		\includegraphics[width=0.2\textwidth]{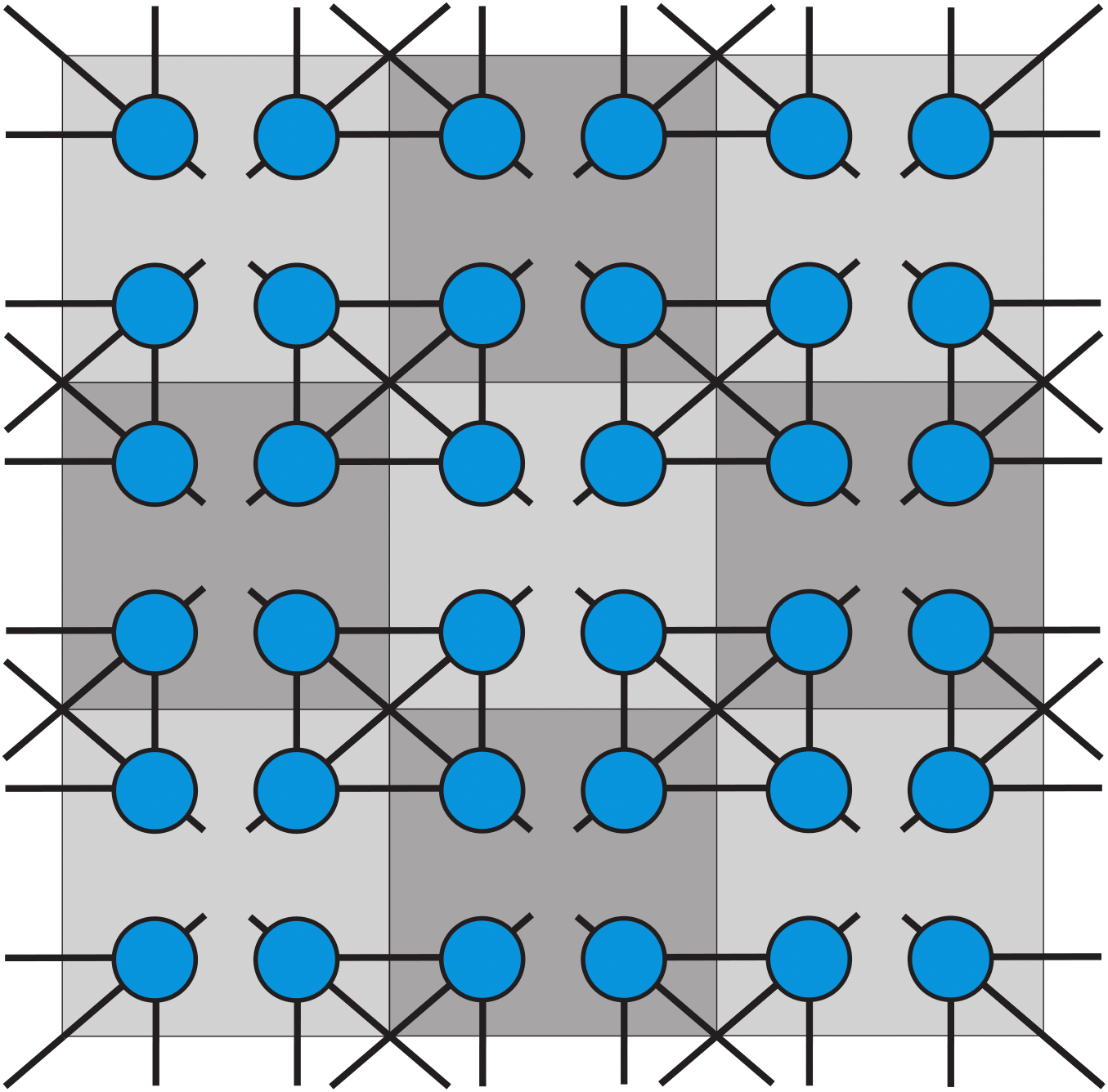}
		\label{tensorcanonical}}
		\caption{(Color online) (a) Left panel: An exemplary tensor network representation of a SRE state. Black dots represent degrees of freedom within each site (blue circles). Entanglement exists between connected partons. Together but not necessarily individually, black dots form a linear representation of $G$ on each site. (b) Right panel: Canonical form of a SRE tensor network state after rearranging degrees of freedom within shaded squares that represent sites in the canonical form.}\label{fig:canonicalSRE}
\end{figure}

In Ref. \cite{cohomology} it was argued intuitively that a SRE state in a distinct SPT phase can be transformed to a symmetric tensor network state (see Fig.~\ref{tensorcanonical}) by symmetric LU transformations. It was also argued that this local unitary transformation can be accomplished by applying a constant-depth quantum circuit.
 A wave function of a generic many-body Hamiltonian is usually very complicated. Fortunately, the above statement allows us to formulate a simple SRE tensor network state that can be regarded as a fixed-point wave function for more complicated systems in the same phase. Though a generic proof is missing, Ref.~\cite{cohomology} covers an in-depth analysis and provides convincing argument, but it will not be repeated here in detail. Essentially the argument goes as follows.

Given a tensor network representation of any SRE state (see Fig.~\ref{tensorSRE}), we can choose squares of size $L$ such that entanglement only exists between neighboring squares and within. Within each square, we then can change the basis and rearrange the degrees of freedom using symmetric unitary transformations. As a result, we obtain a tensor network state in its canonical form (see Fig.~\ref{tensorcanonical}).

Given a SRE state $\ket{\psi}$ in its canonical form, we ideally want to identify all pairs $(U_i, \ket{\psi})$ that exhibit nontrivial SPT order, with $U_i$ being an on-site, global symmetry representation acting on site $i$. Therefore, we have to clarify how two distinct pairs $(U_i, \ket{\psi})$ and $(U'_i, \ket{\psi'})$ describing the same phase can be related. There are basically two equivalence relations: (1) Two phases have the same SPT order if their on-site symmetry representations only differ by a certain class of unitary transformation $W^i$ on each effective site. This is a map $(U_i, \ket{\psi})\mapsto (U'_i, \ket{\psi'})$ where $\ket{\psi'}$ has still the same tensor network form. (2)~Adding local degrees of freedom that form a 1D representation of the symmetry does not change the phase. The second equivalence relation can be utilized to reduce the canonical form in Fig.~\ref{tensorcanonical} to a plaquette state as depicted in Fig.~\ref{fig:square}. Note that it was also argued that we can choose $U_i$ to be a symmetry on a local region of the canonical form instead of a global symmetry.

One approach to classify SPT order in Ref.~\cite{cohomology} is to define a simple SRE state $\ket{\psi_{gs}}$ in its canonical form on plaquettes and find a nontrivial on-site representation of the symmetry. Then, more complicated wave functions can presumably be found by the aforementioned equivalence relations. {\it In this work, we shall mainly focus on plaquette states on \textit{arbitrary} lattices and construct an on-site symmetry representation that is defined on any lattices}. In Sec.~\ref{construction} we will illustrate how the original construction of the symmetry operation~\cite{cohomology} needs to be modified in order to have a symmetry representation defined on all lattices and show that this simple modification yields the same classification of phases in terms of group cohomology.

%%%%%%%%%%%%%%%%%%%%%%%%%%%%%%%%%%%%%%%%%%%%%%%%%%%%%%%%%%%%%%%%%%%%%%%%%%%%%%%%%%%%%%%%%%%%%%%%%%%%%%%%%%%%%%%%%%%%%%%%%%%%%%%%%%%%%%%%%%%%%%%%%%%%%%%%%%%%%%%%%%%%%%%%%%%%%%%%%%%%%%%%%%%%%%%%%%%%%%%%%%%%%%%%%%%%%%%%%%%%%%%%%%%%%%%%%%%%%%%%%%%%%%%%%%%%%%%%%%%%%%%%%%%%%%%%%%%%%%%%%%%%%%%%%%%%%%%%%%%%%%%%%%%%%%

\subsection{Group Cohomology $\mathcal{H}^{n}(G,U(1))$}
As group cohomology shall be used to define the symmetry representation, we first give its algebraic definition before discussing its graphic representation, with the latter giving a very useful tool employed in Sec.~\ref{construction}. A more rigorous introduction to group cohomology can be found in Ref.~\cite{introcohomology}.

Let $\omega_n:G^n\rightarrow M$ be a function of $n$ group elements $g_i\in G$ with value in the $G$ module $M$. Let $\mathcal{C}^n(G,M)$ be the space of all such functions which form an Abelian group under the function multiplication in $M$. A $G$ module $M$ is an Abelian group with a multiplication operation compatible with the action of the group $G$, i.e.
\begin{align}
g\cdot (ab)=(g\cdot a)(g\cdot b),\:\:\forall\: g\in G,\:\:\: a,b\in M.
\end{align}
 For our purposes we restrict ourselves to $M=U(1)$ which is just the group consisting of complex phases with a trivial group action $g\cdot a=a$ for $g\in G$ and $a\in U(1)$.

Furthermore, let us define the map $d_n:\mathcal{C}^n(G,U(1))\rightarrow \mathcal{C}^{n+1}(G,U(1))$:
\begin{align}
&(d_n\omega_n)(g_1,g_2,...,g_{n+1})\nonumber\\&=[g_1\cdot\omega_n(g_2,...,g_{n+1})]\omega_n^{(-1)^{n+1}}(g_1,...,g_{n})\nonumber\\&\times\prod_{i=1}^{n}\omega_n^{(-1)^{i}}(g_1,...,g_{i-1},g_ig_{i+1},g_{i+2},...,g_{n+1}).
\end{align}
We utilize this map to define the group of all $n$-coboundaries to be
\begin{align}
&\mathcal{B}^n(G,M)\nonumber\\=&\{\omega_n\in\mathcal{C}^{n}(G,M)\vert\omega_n=d_{n-1}\omega_{n-1}|\nonumber\\&\omega_{n-1}\in\mathcal{C}^{n-1}(G,M)\},
\end{align}
and the group of all $n$-cocycles to be
\begin{align}
\mathcal{Z}^n(G,M)=\{\omega_n\in\mathcal{C}^{n}(G,M)\vert d_{n}\omega_{n}=1\}.
\end{align}
$\mathcal{B}^n(G,M)$ and $\mathcal{Z}^n(G,M)$ are Abelian groups satisfying $\mathcal{B}^n(G,M)\subset\mathcal{Z}^n(G,M)$. The $n$th group cohomology is the quotient group of $n$-cocycles and $n$-coboundaries:
\begin{align}
\mathcal{H}^n(G,M)=\mathcal{Z}^n(G,M)/\mathcal{B}^n(G,M).
\end{align}
Since we are solely considering two spatial dimensions and $M=U(1)$, let us discuss $n=3$ as an example. From
\begin{align}
&(d_3\omega_3)(g_1,g_2,g_3,g_4)\nonumber\\&=\frac{\omega_3(g_2,g_3,g_4)\omega_3(g_1,g_2g_3,g_4)\omega_3(g_1,g_2,g_3)}{\omega_3(g_1g_2,g_3,g_4)\omega_3(g_1,g_2,g_3g_4)}
\end{align}
we obtain the 3-cocyles
\begin{align}
\label{eqn:3coclye}
&\mathcal{Z}^3(G,U(1))\nonumber\\&=\left\{\omega_3\Big|\frac{\omega_3(g_2,g_3,g_4)\omega_3(g_1,g_2g_3,g_4)\omega_3(g_1,g_2,g_3)}{\omega_3(g_1g_2,g_3,g_4)\omega_3(g_1,g_2,g_3g_4)}=1\right\}
\end{align}
and the 3-coboundaries
\begin{align}
&\mathcal{B}^3(G,U(1))\nonumber\\&=\left\{\omega_3\Big|\omega_3(g_1,g_2,g_3)=\frac{\omega_2(g_2,g_3)\omega_2(g_1,g_2g_3)}{\omega_2(g_1g_2,g_3)\omega_2(g_1,g_2g_3)}\right\}
\end{align}
which in turn defines the 3rd group cohomology $\mathcal{H}^3(G,U(1))=\mathcal{Z}^3(G,U(1))/\mathcal{B}^3(G,U(1))$.

One exemplary representation of a $3$-cocycle for a group ${Z}_d$ would be (see Ref. \cite{cocyclerepr})
\begin{align}
\omega_3(g_1,g_2,g_3)=\exp\left(\frac{2\pi ic[g_1]}{d^2}([g_2]+[g_3]-[g_2+g_3])\right)\label{cocyclerepr}
\end{align}
for an appropriate $c\in\mathbb{Z}$ and $[g]\equiv g\mod{d}$ for $g\in{Z}_d$ in the regular representation.
\subsection{Graphic representation of group cohomology}\label{graphic}
\begin{figure}
	\centering
	\includegraphics[width=0.4\textwidth]{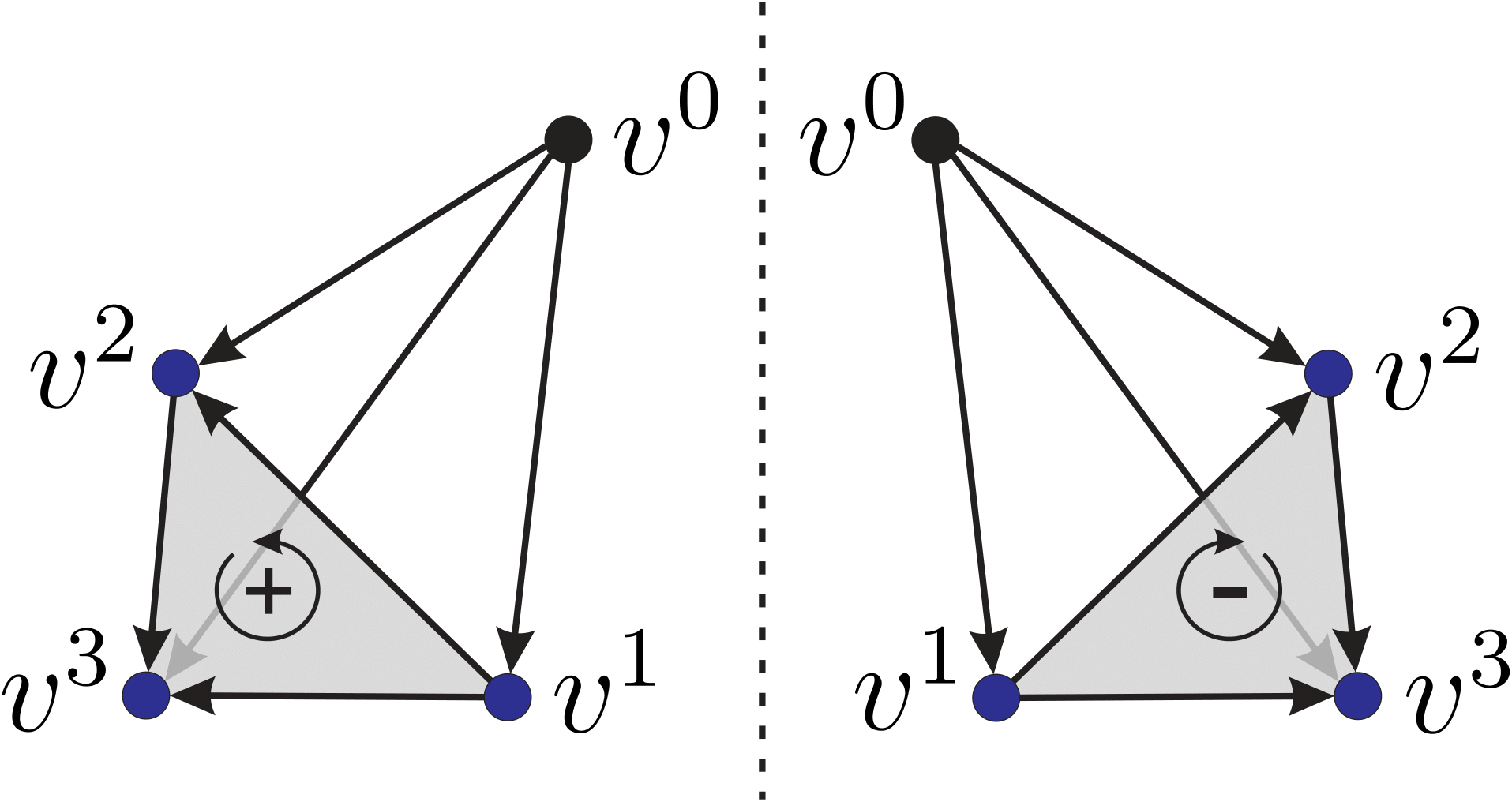}
	\caption{(Color online) Graphic representation of a $3$-cochain $\nu_3^{\pm 1}(v^0,v^1,v^2,v^3)$. The representation is unique up to a reflection along the dotted line. The two mirrored tetrahedrons are identified with opposite chiralities $\pm1$.}\label{cochain}
\end{figure}

We now review the graphical representation of the group cohomology. It turns out that $n$-cocycles $\omega_n$ can be represented graphically if we consider so-called $n$-cochains $\nu_n$ instead.
There is a one-to-one relation between the formerly discussed functions $\omega_n:G^n\rightarrow M$ and cochains. A $n$-cochain $\nu_n$ is a map $G^{n+1}\rightarrow M$ that obeys
\begin{align}
g\cdot \nu_n(g_0,g_1,...,g_n)=\nu_n(gg_0,gg_1,...,gg_n),
\end{align}
which is related to $\omega_n$ by
\begin{align}
\omega_n(g_1,g_2...,g_n)=\nu_n(1,g_1,g_1g_2,...,g_1\cdots g_n)
\end{align}
It was shown  that any $n$-cochain can be represented uniquely by a branched $n$-simplex~\cite{cohomology}. An arbitrary $n$-simplex can be viewed as a complete graph of its $n+1$ vertices $v^i$. For example, it is a triangle for $n=2$, a tetrahedron for $n=3$ and a pentachoron for $n=4$. Let us also impose a naturally ordered labeling of vertices. A natural order would imply $v^i<v^j$ if $i<j$. Then, the labeling of vertices together with its natural order gives rise to a branching structure. A branching structure on a simplex is an orientation of edges such that there are no oriented loops (see Fig.~\ref{cochain}).

In fact, a branching structure on a simplex uniquely determines the canonical ordering of its vertices. On the other hand, the branching structure of two mirrored simplices is the same. By imposing a direction in space to the branching structure, we can uniquely determine each simplex by assigning opposite chirality $\pm 1$ to mirrored simplices (see Fig.~\ref{cochain}). Given the above two properties, we can represent any $n$-cochain
\begin{align}
\nu_n^{s_{ij...k}}(g_i,g_j,...,g_k)
\end{align}
by a $n$-simplex with chirality $s_{ij...k}=\pm 1$ and canonically ordered vertices $g_i,g_j,...,g_k$.

Even though we can use a $n$-simplex to represent any $n$-cochain, we have to take into account some redundancy implied by the cocycle condition. Since we are predominantly interested in two spatial dimensions, let us only consider $n=3$ from now on. The $3$-cocycle condition~\ref{eqn:3coclye} for cochains takes the form
\begin{align}
\frac{\nu_3(g_1,g_2,g_3,g_4)\nu_3(g_0,g_1,g_3,g_4)\nu_3(g_0,g_1,g_2,g_3)}{\nu_3(g_0,g_2,g_3,g_4)\nu_3(g_0,g_1,g_2,g_4)}=1\label{cocyclecond}
\end{align}
In particular, it was shown that the cocycle condition implies that two cochains represented by simplices with the same surface generate the same phase in $M=U(1)$~\cite{cohomology}. This statement is true even for a complex with oriented loops as long as the simplices within the complex define appropriate $3$-cochains. Due to the one-to-one correspondence between cocycles and cochains, the $3$-cochains satisfying this condition equivalently define $\mathcal{Z}^3(G,U(1))$. 
On the other hand, $3$-coboundaries denoted by $\lambda_3$ are $3$-cochains that obey
\begin{align}
\lambda_3(g_0,g_1,g_2,g_3)=\frac{\mu_2(g_1,g_2,g_3)\mu_2(g_0,g_1,g_3)}{\mu_2(g_0,g_2,g_3)\mu_2(g_0,g_1,g_2)}
\end{align}
with $\mu_2$ being a $2$-cochain but not necessarily a cocycle. Then, the equivalence classes $[\nu_3]:=\{\nu_3,\nu_3'\in\mathcal{Z}^3(G,U(1))|\nu_3=\nu_3'\lambda_3\}$ give rise to the third group cohomology $\mathcal{H}^3(G,U(1))$.

%%%%%%%%%%%%%%%%%%%%%%%%%%%%%%%%%%%%%%%%%%%%%%%%%%%%%%%%%%%%%%%%%%%%%%%%%%%%%%%%%%%%%%%%%%%%%%%%%%%%%%%%%%%%%%%%%%%%%%%%%%%%%%%%%%%%%%%%%%%%%%%%%%%%%%%%%%%%%%%%%%%%%%%%%%%%%%%%%%%%%%%%%%%%%%%%%%%%%%%%%%%%%%%%%%%%%%%%%%%%%%%%%%%%%%%%%%%%%%%%%%%%%%%%%%%%%%%%%%%%%%%%%%%%%%%%%%%%%%%%%%%%%%%%%%%%%%%%%%%%%%%%%%%%%%

\section{Constructing SPT phases on arbitrary lattices in two spatial dimensions}\label{construction}
In this section we first illustrate the issue of defining a consistent symmetry representation when generalizing the construction in Ref.~\cite{cohomology}. It turns out that a simple modification can solve this issue and the symmetry can be defined on arbitrary lattices, where each site does not necessarily have the same number of partons. Moreover, the symmetry operation on the boundary encodes useful information about the SPTO of the bulk.

\subsection{The issue of defining a consistent symmetry operation}\label{crystal}

In accordance with the discussion in Ref.~\cite{cohomology}, we will employ and extend the maximally entangled SRE state $\ket{\psi_{gs}}$ used there and examine the  global, on-site symmetry $U_i(g)$ of a $d$-dimensional finite group $G$. In fact, one key point will be how to properly define such a symmetry with nontrivial action on the system both without and with {\it boundaries}. (In fact, examining the boundary will help to understand that the bulk possesses nontrivial SPTO.) The corresponding exactly solvable, local Hamiltonian will just be the projection onto that particular nondegenerate ground state, in a way similar to the CZX model in Ref.~\cite{edgemodes}. Note that the restriction to finite groups is useful for the discussion of universality for MBQC in Sec.~\ref{universality}.

\begin{figure}
	\centering
	\subfigure[]{
		\includegraphics[width=0.22\textwidth]{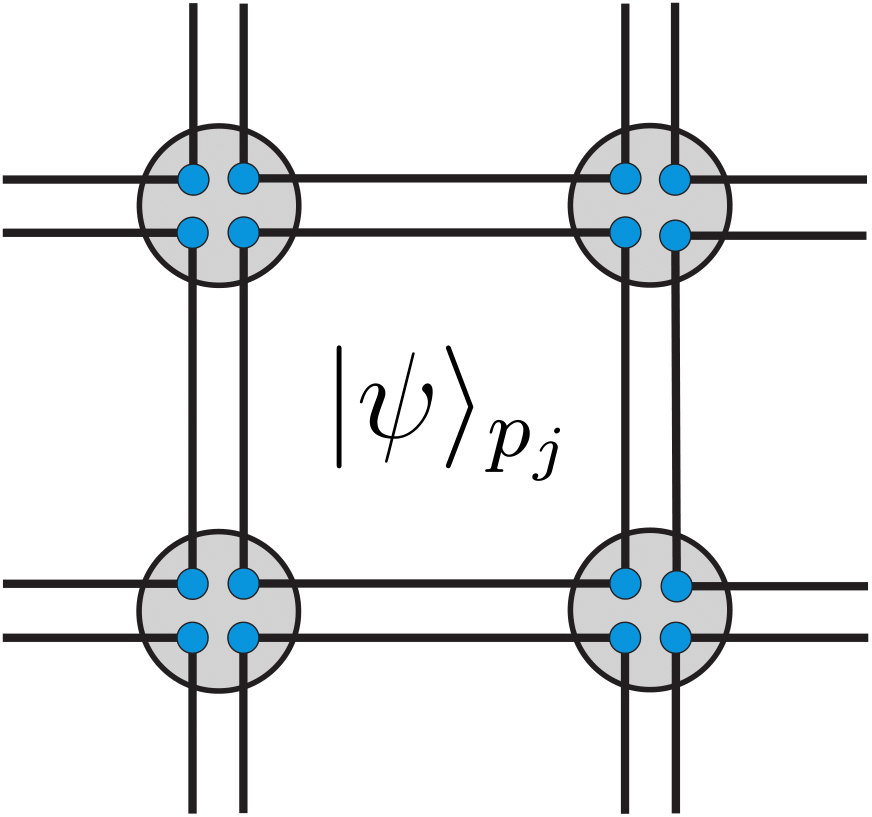}
		\label{fig:square}
	}
	\subfigure[]{
		\includegraphics[width=0.18\textwidth]{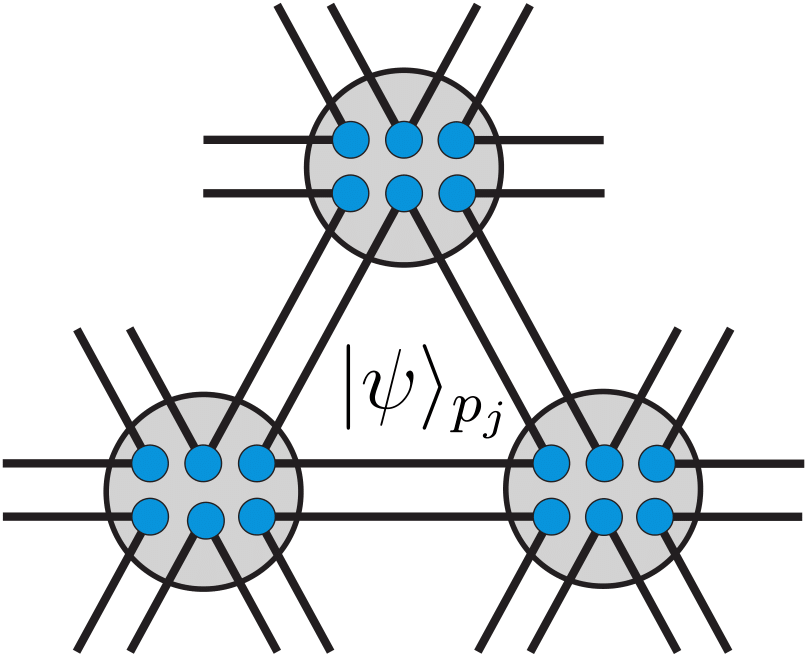}
		\label{fig:triangle}}
	\subfigure[]{
		\includegraphics[width=0.18\textwidth]{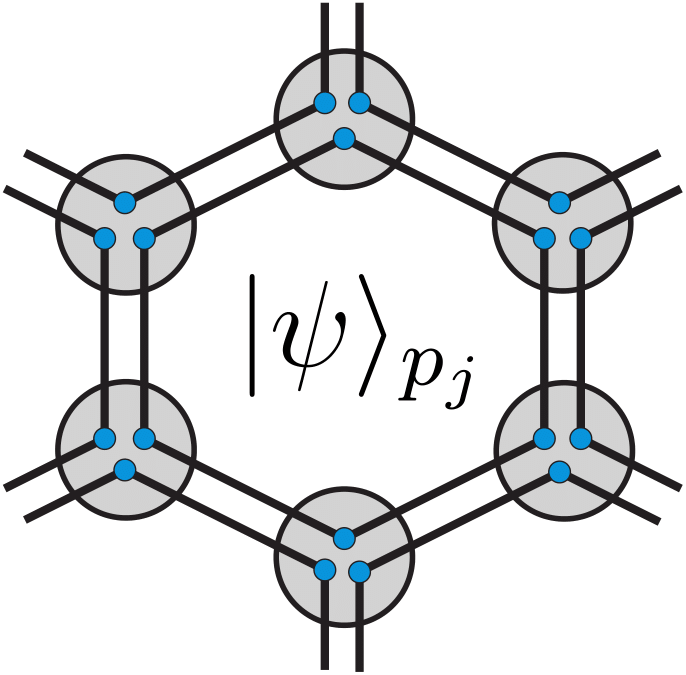}
		\label{fig:honeycomb}}
	\subfigure[]{
		\includegraphics[width=0.22\textwidth]{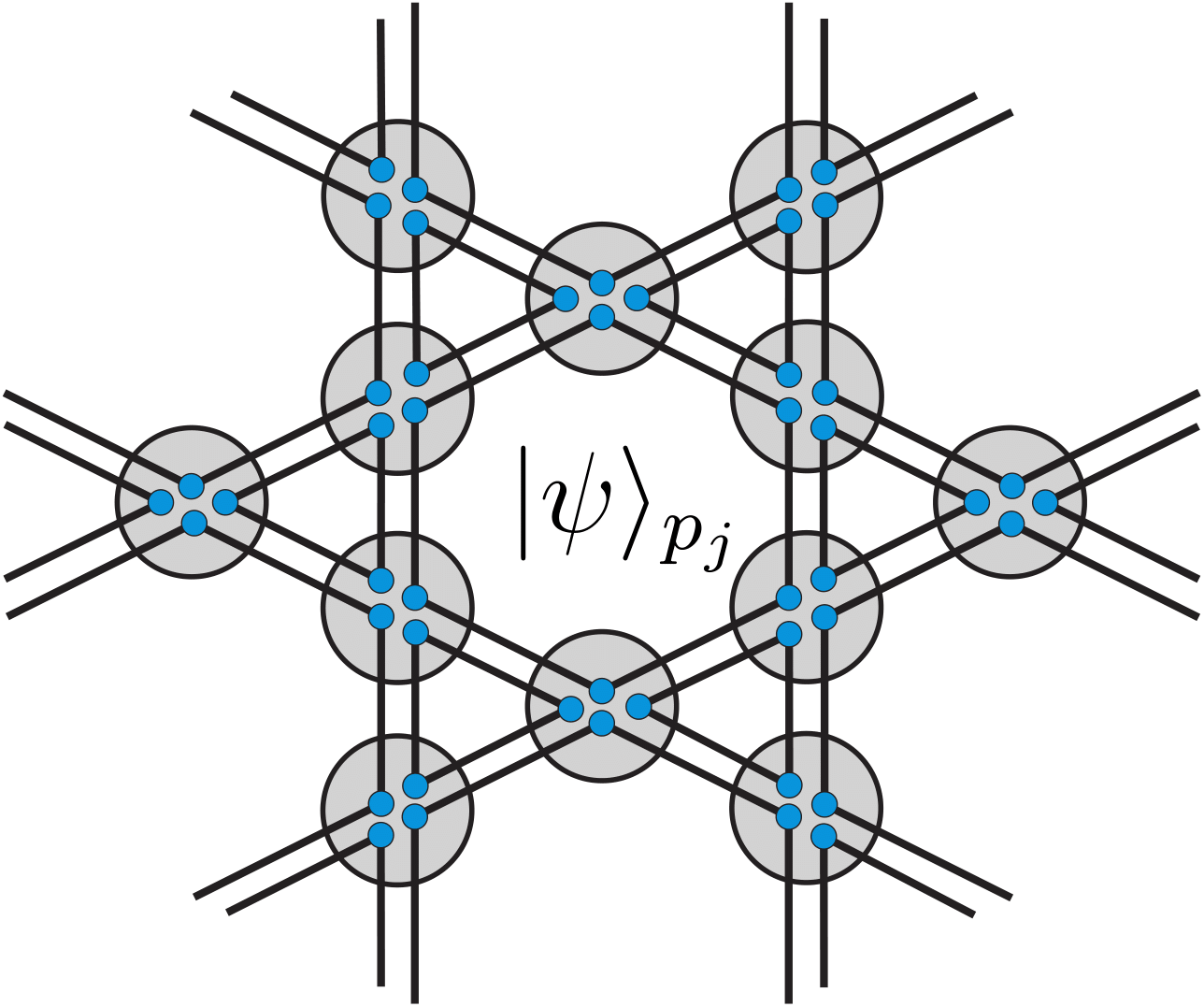}
		\label{fig:kagome}}
	\caption{(Color online) Plaquette states $\ket{\psi}_{p_j}$ defined on (a) square, (b) triangular, (c) honeycomb and (d) kagome lattice. Blue dots are identified with partons and shaded circles with physical sites. Connections indicate entanglement.}\label{fig:plaqs}
\end{figure}

In order to illustrate the issue, let us consider square, triangular and honeycomb lattices for now. We define the maximally entangled SRE state to be
\begin{align}
\ket{\psi_{gs}}&=\bigotimes_{j}\ket{\psi}_{p_j}\nonumber\\&=\frac{1}{d^{n_p/2}}\bigotimes_{j}\sum_{g\in G}\ket{\alpha_1=g, \beta_2=g,...,\zeta_k=g}_{p_j}\label{SREstate}
\end{align}
with parton states $\ket{\alpha_1},\ket{\beta_2},...,\ket{\zeta_k}$ defined on $d$-dimensional Hilbert spaces where group elements label physical states according to their regular representation, $d=|G|$ is the order of the group $G$, $n_p$ is the number of plaquettes and subscript $p_j$ labels the $j$th plaquette, and $k\in \{3,4,6\}$ is the number of partons entangled within triangular, square and hexagonal plaquettes respectively (see Fig.~\ref{fig:plaqs}). This is a maximally entangled state, i.e., the so-called GHZ state, where partons within a plaquette are in the same state in their canonical basis.

\begin{figure}
	\centering
	\subfigure[]{
		\includegraphics[width=0.16\textwidth]{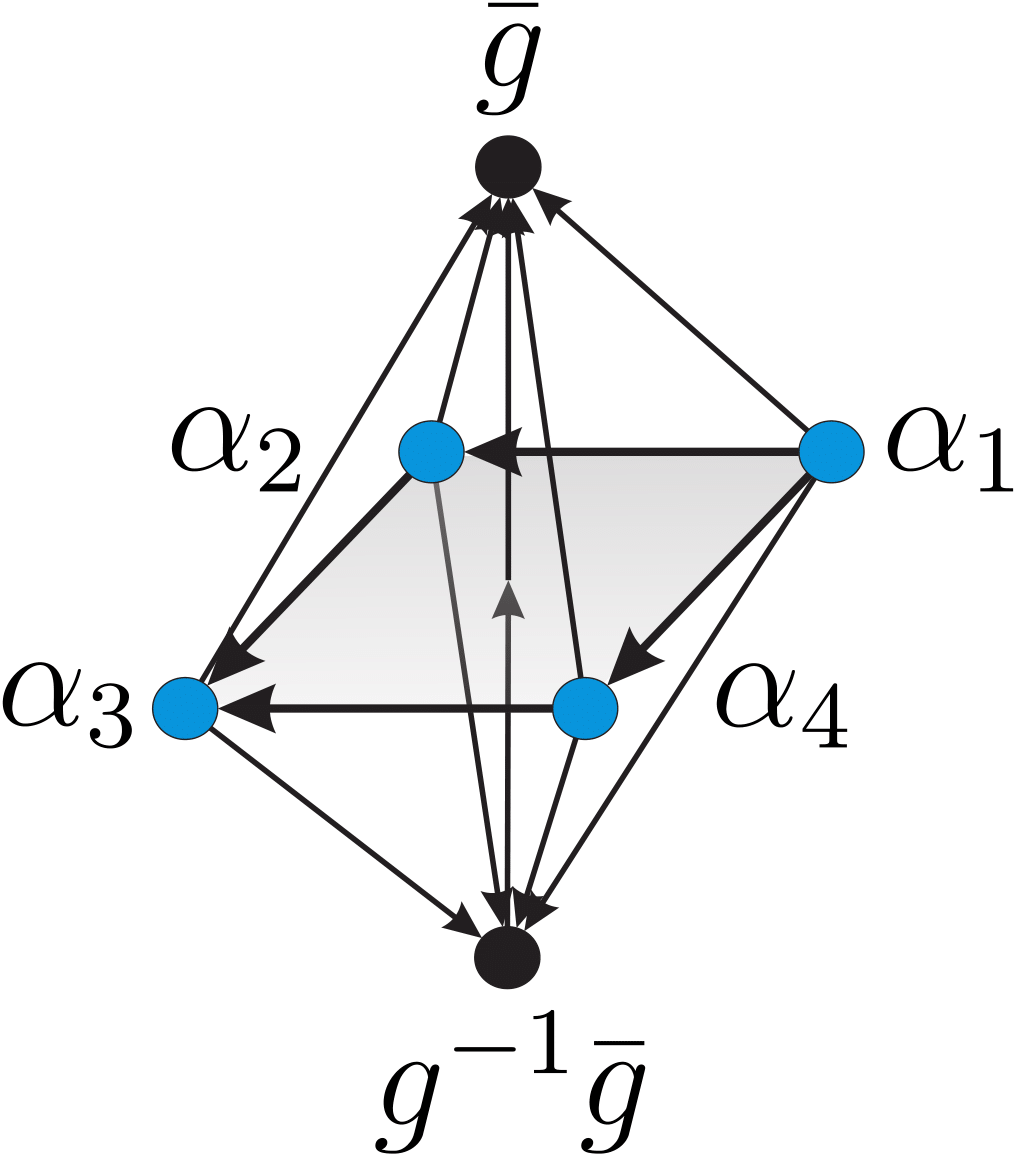}\label{graphicsquare}
	}
	\subfigure[]{
		\includegraphics[width=0.23\textwidth]{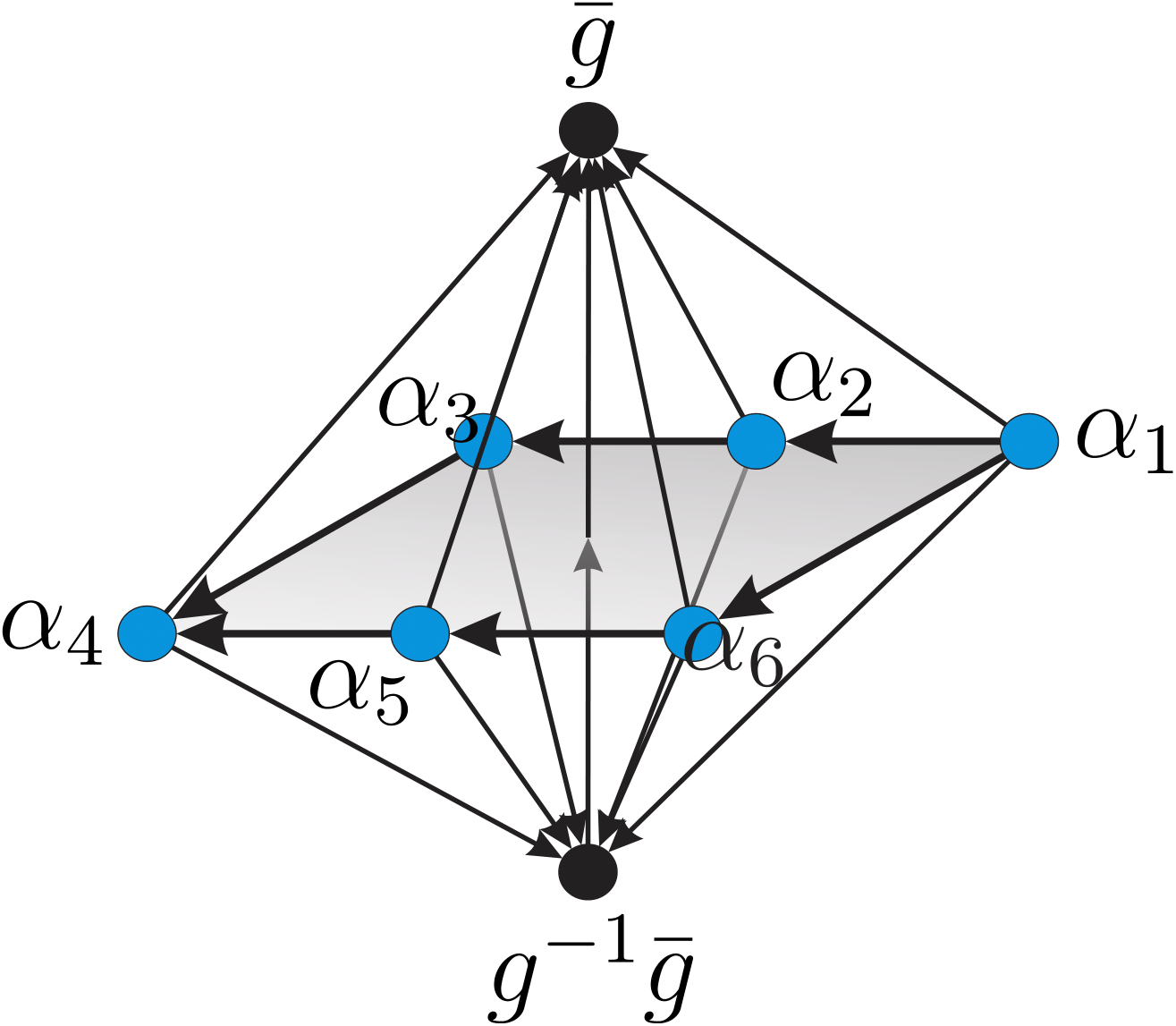}\label{graphictriangle}
	}
	\caption{(Color online) (a) Graphic representation of Eq.~\ref{phasesquare}. (b) Graphic representation of Eq.~(\ref{phasetriangle}). Blue dots represent partons within one physical site.}\label{graphicsquaretri}
\end{figure}

The on-site symmetry on the square-lattice plaquette state is defined in Ref.~\cite{cohomology} via the $3$-cochain $\nu_3$ corresponding to $\omega_3\in \mathcal{H}^{3}(G, U(1))$ as follows,
\begin{align}
& U_i(g)\ket{\alpha_1,\alpha_2,\alpha_3,\alpha_{4}}\nonumber\\& =f_3(\alpha_1,\alpha_2,\alpha_3,\alpha_{4}, g, \bar{g})\ket{g\alpha_1,g\alpha_2,g\alpha_3,g\alpha_{4}},\label{originalsym}
\end{align}
where $\ket{\alpha_1,\alpha_2,\alpha_3,\alpha_{4}}$ is an arbitrary state of partons, $\bar{g}$ is a fixed element in the group $G$, (the labeling is illustrated in Fig.~\ref{graphicsquare}) and
\begin{align}
& f_3(\alpha_1,\alpha_2,\alpha_3,\alpha_{4}, g, \bar{g})\nonumber\\
& =\frac{\nu_3(\alpha_1,\alpha_2,g^{-1}\bar{g},\bar{g})\nu_3(\alpha_2,\alpha_3,g^{-1}\bar{g},\bar{g})}{\nu_3(\alpha_1,\alpha_4,g^{-1}\bar{g},\bar{g})\nu_3(\alpha_4,\alpha_3,g^{-1}\bar{g},\bar{g})}.\label{phasesquare}
\end{align}
It is shown that such a definition gives rise to a linear representation of the group $G$ and the plaquette state is indeed invariant under the action of any $g$ on all sites~\cite{cohomology}. Surprisingly, this construction is easily extended to the triangular lattice with six partons per site. In this special case, the on-site representation $U_i(g)$ of $g\in G$ is
\begin{align}
	& U_i(g)\ket{\alpha_1,\alpha_2,...,\alpha_{6}}\nonumber\\& =f_3(\alpha_1,\alpha_2,...,\alpha_{6}, g, \bar{g})\ket{g\alpha_1,g\alpha_2,...,g\alpha_{6}}\label{originalconstr}
\end{align}
where $\ket{\alpha_1,\alpha_2,...,\alpha_{6}}$ is an arbitrary state of partons, $\bar{g}$ is a fixed element in $G$ and
\begin{align}
	& f_3(\alpha_1,\alpha_2,...,\alpha_{6}, g, \bar{g})\nonumber\\
	& =\frac{\nu_3(\alpha_1,\alpha_2,g^{-1}\bar{g},\bar{g})\nu_3(\alpha_2,\alpha_3,g^{-1}\bar{g},\bar{g})}{\nu_3(\alpha_5,\alpha_4,g^{-1}\bar{g},\bar{g})\nu_3(\alpha_6,\alpha_5,g^{-1}\bar{g},\bar{g})}\nonumber\\
	&\times \frac{\nu_3(\alpha_3,\alpha_4,g^{-1}\bar{g},\bar{g})}{\nu_3(\alpha_1,\alpha_6,g^{-1}\bar{g},\bar{g})}\label{phasetriangle}
\end{align}
defines the phase factor with respect to triangular plaquettes (cf. Fig.~\ref{graphictriangle}). This gives rise to a linear representation of $G$ and a global symmetry of $\ket{\psi_{gs}}$ analogous to the original construction. But how do we generalize this to, say, the honeycomb lattice, where each site contains three partons? 

\begin{figure}
	\centering
	\subfigure[]{
		\centering
		\includegraphics[width=0.22\textwidth]{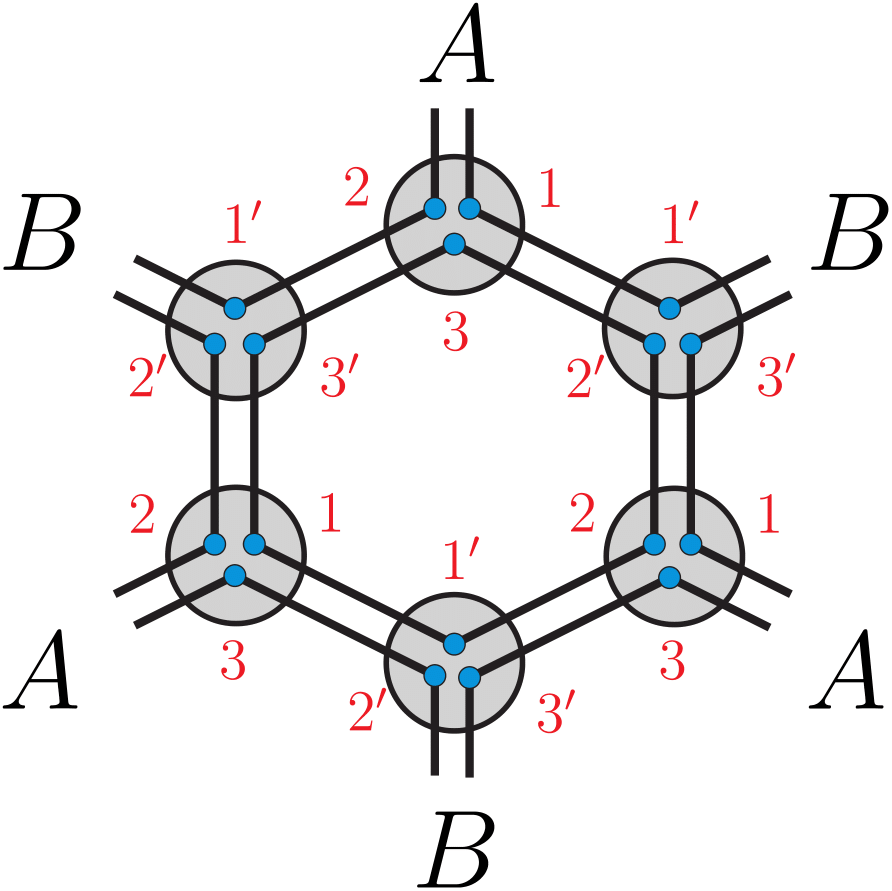}
		\label{extensionhoneycomb}}
	\subfigure[]{
		\includegraphics[width=0.22\textwidth]{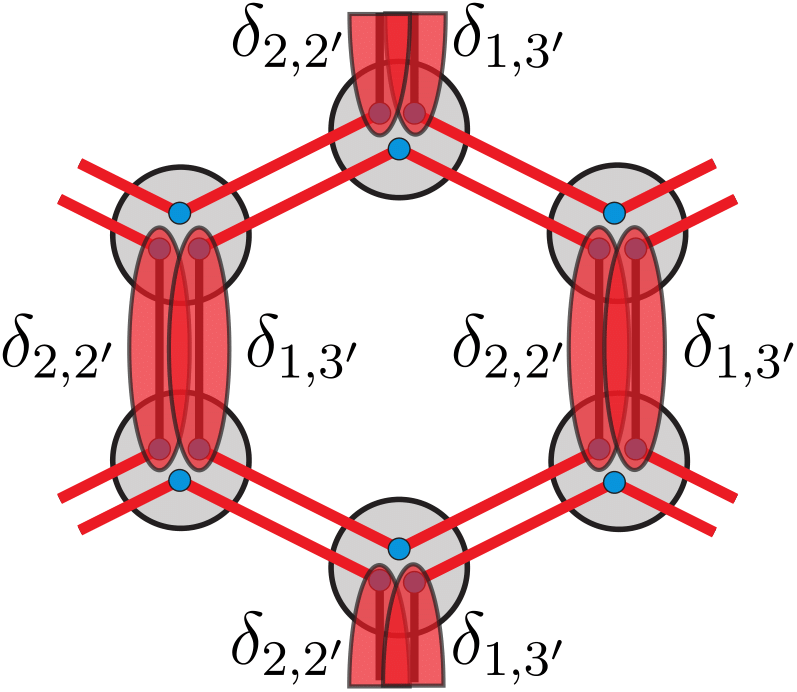}
		\label{merginghoneycomb}}
	\caption{(Color online) (a) In order to extend the original symmetry representation to the honeycomb lattice, we have to define two symmetry representations $U^A_i(g)$ and $U^B_{i}(g)$ on different sublattices $A,B$. Otherwise, global symmetry is not ensured. Red numbers indicate the labeling of partons for $\alpha_1,...,\alpha_3$. (b) Merging as a procedure to obtain plaquette states on a square lattice from the honeycomb lattice. The merging is mathematically a $\delta$-function and is indicated in red with respect to the labeling in (a).}\label{honeycomb}
\end{figure}

On any site of one particular sublattice (say, $A$ in Fig.~\ref{extensionhoneycomb}), a choice of the phase factor $f^A_3$ is 
\begin{align}
	& f^A_3(\alpha_1,\alpha_2,\alpha_3,g,\bar{g})\nonumber\\&=\frac{\nu_3(\alpha_1,\alpha_2,g^{-1}\bar{g},\bar{g})\nu_3(\alpha_2,\alpha_3,g^{-1}\bar{g},\bar{g})}{\nu_3(\alpha_1,\alpha_3,g^{-1}\bar{g},\bar{g})},\label{phasefail}
\end{align}
and it will satisfy the group associativity. But on any other site of sublattice $B$ if we define the symmetry action to be of similar form, then the plaquette state on the honeycomb lattice will {\it not} be invariant under the global on-site symmetry.
 
It turns out by careful examination that the choice of the factor $f^B_3$ on $B$ sublattice sites as below,
\begin{align}
& f^B_3(\alpha_1,\alpha_2,\alpha_3,g,\bar{g})\nonumber\\&=\frac{\nu_3(\alpha_1,\alpha_2,g^{-1}\bar{g},\bar{g})}{\nu_3(\alpha_1,\alpha_3,g^{-1}\bar{g},\bar{g})\nu_3(\alpha_3,\alpha_2,g^{-1}\bar{g},\bar{g})},\label{phaseextension}
\end{align}
will, together with Eq.~(\ref{phasefail}), indeed give rise to a global symmetry for the plaquette state. This is because opposite chirality for the graphic representation of a cochain is defined by the mirror image and two partons connected by a line in Fig.~\ref{extensionhoneycomb} are in a maximally entangled state. That is, phase factors in Fig.~\ref{bondsym} arising between parallel edges are chosen such that they cancel. We can see that whether or not the symmetry representation is a global symmetry depends on how we define the on-site branching structure of $f^I_3$ on sublattice $I$. In the next section, we will generalize this argument to arbitrary 2D lattices.

Let us point out that just as we can relate the honeycomb lattice to the square lattice by merging vertices, we can relate the symmetry representations defined by Eq.~(\ref{phasefail}) and~(\ref{phaseextension}) on the honeycomb lattice to the representation given by Eq.~(\ref{phasesquare}) on the square lattice by merging sites along edges (see Fig.~\ref{merginghoneycomb}), i.e. by forcing pairs of merged partons into the same state.
% % % % % % % % % % % % % % % % % % % % % % % % % % % % % % % % % % % % % % % % % % % % % % % % % % % % % % % % % % % % % % % % % % % % % % % % % % % % % % % % % % % % % % % % % % % % % % % % % % %

%%%%%%%%%%%%%%%%%%%%%%%%%%%%%%%%%%%%%%%%%%%%%%%%%%%%%%%%%%%%%%%%%%%%%%%%%%%%%%%%%%%%%%%%%%%%%%%%%%%%%%%%%%%%%%%%%%%%%%%%%%%%%%%%%%%%%%%%%%%%%%%%%%%%%%%%%%%%%%%%%%%%%%%%%%%%%%%%%%%%%%%%%%%%%%%%%%%%%%%%%%%%%%%%%%%%%%%%%%%%%%%%%%%%%%%%%%%%%%%%%%%%%%%%%%%%%%%%%%%%%%%%%%%%%%%%%%%%%%%%%%%%%%%%%%%%%%%%%%%%%%%%%%%%%%

\subsection{SPT phases from plaquette states on arbitrary lattices}\label{quasicrystal}
\begin{figure}
	\centering
	\subfigure[]{
		\includegraphics[width=0.22\textwidth]{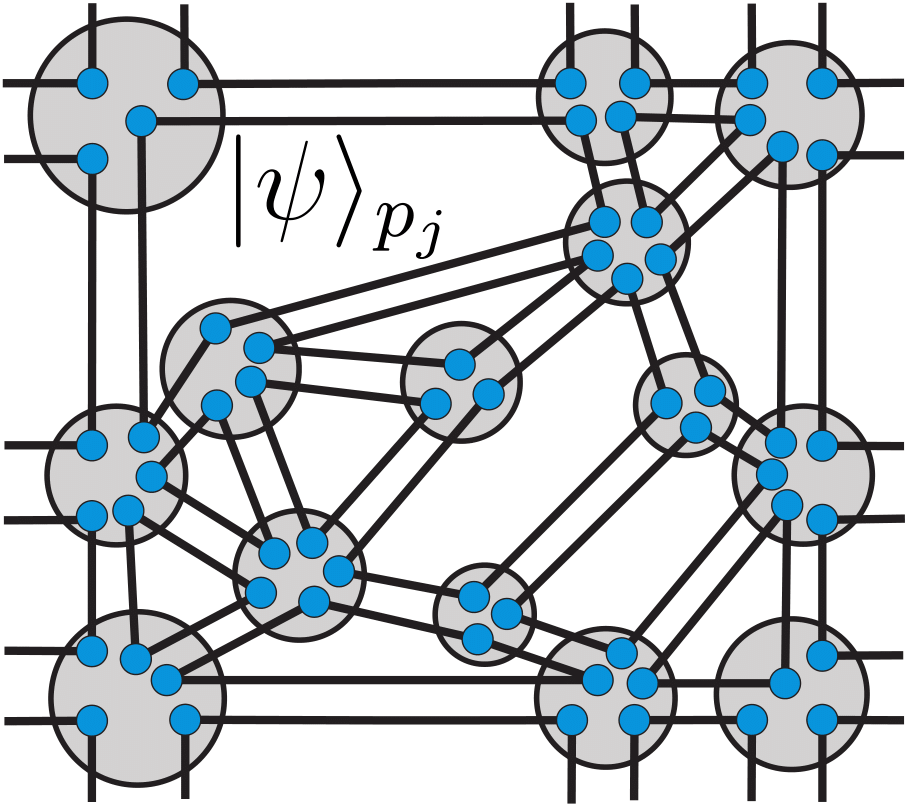}
		\label{fig:plaq}}
	\subfigure[]{
		\includegraphics[width=0.22\textwidth]{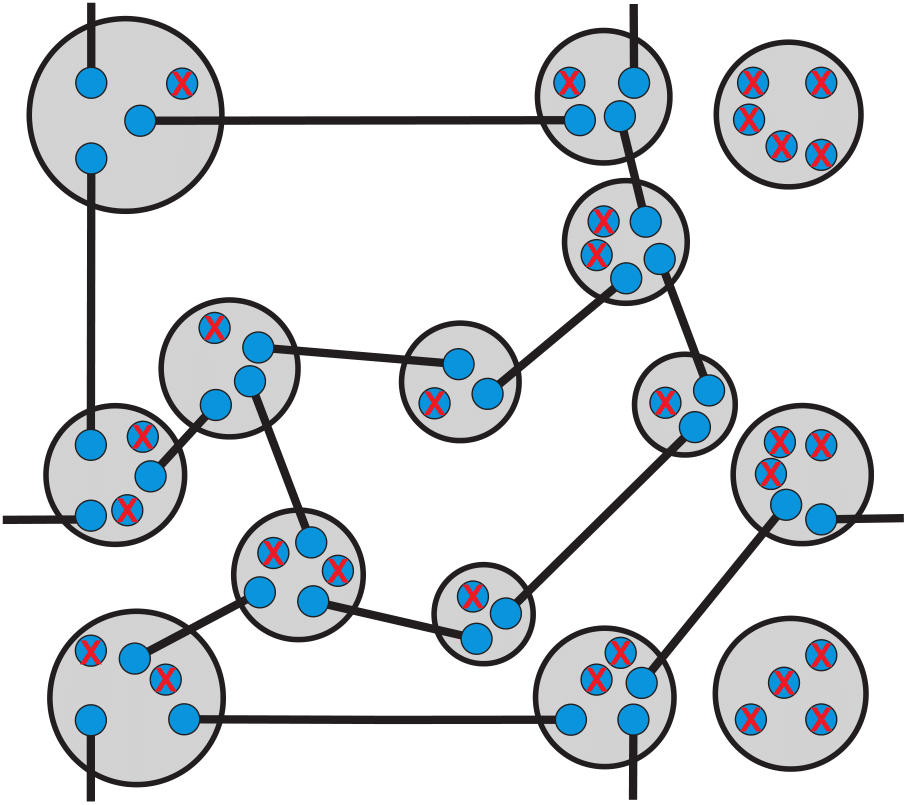}
		\label{fig:plaqbond}}
	\caption{(Color online) (a) (Generalized) plaquette states $\ket{\psi}_{p_j}$ on an arbitrary lattice that obeys conditions (1) and (2), i.e. that adjacent plaquettes share an edge and that we can impose periodic boundary conditions. (b) Performing measurements in $\ket{\pm}$- or $\tilde{Z}^m\ket{\tilde{+}}$-basis (cf. Eq. \ref{patterngeneral}) on virtual qubits that are marked by red crosses, gives rise to a valence-bond state.}\label{plaqtobondstate}
\end{figure}

The argument in the previous section can be extended to arbitrary lattices. In Fig.~\ref{fig:kagome} the kagome lattice serves as an example for an Archimedean lattice but the prescription can be generalized to regular lattices beyond this case. In fact, the construction is consistent on an arbitrary combination of plaquettes (see Fig.~\ref{fig:plaq}) so long as (1) neighboring plaquettes share an edge and (2) the graph can be embedded on a two-dimensional torus, i.e. we impose periodic boundary conditions.

We start again by defining a maximally entangled state $\ket{\psi_{gs}}$ on plaquettes around cells on the lattice equivalent to Eq.~(\ref{SREstate}) with $k$ being the number of partons within a plaquette. Note that $k$ is not necessarily constant across arbitrary lattices.

\begin{figure}
	\centering
	\includegraphics[width=0.3\textwidth]{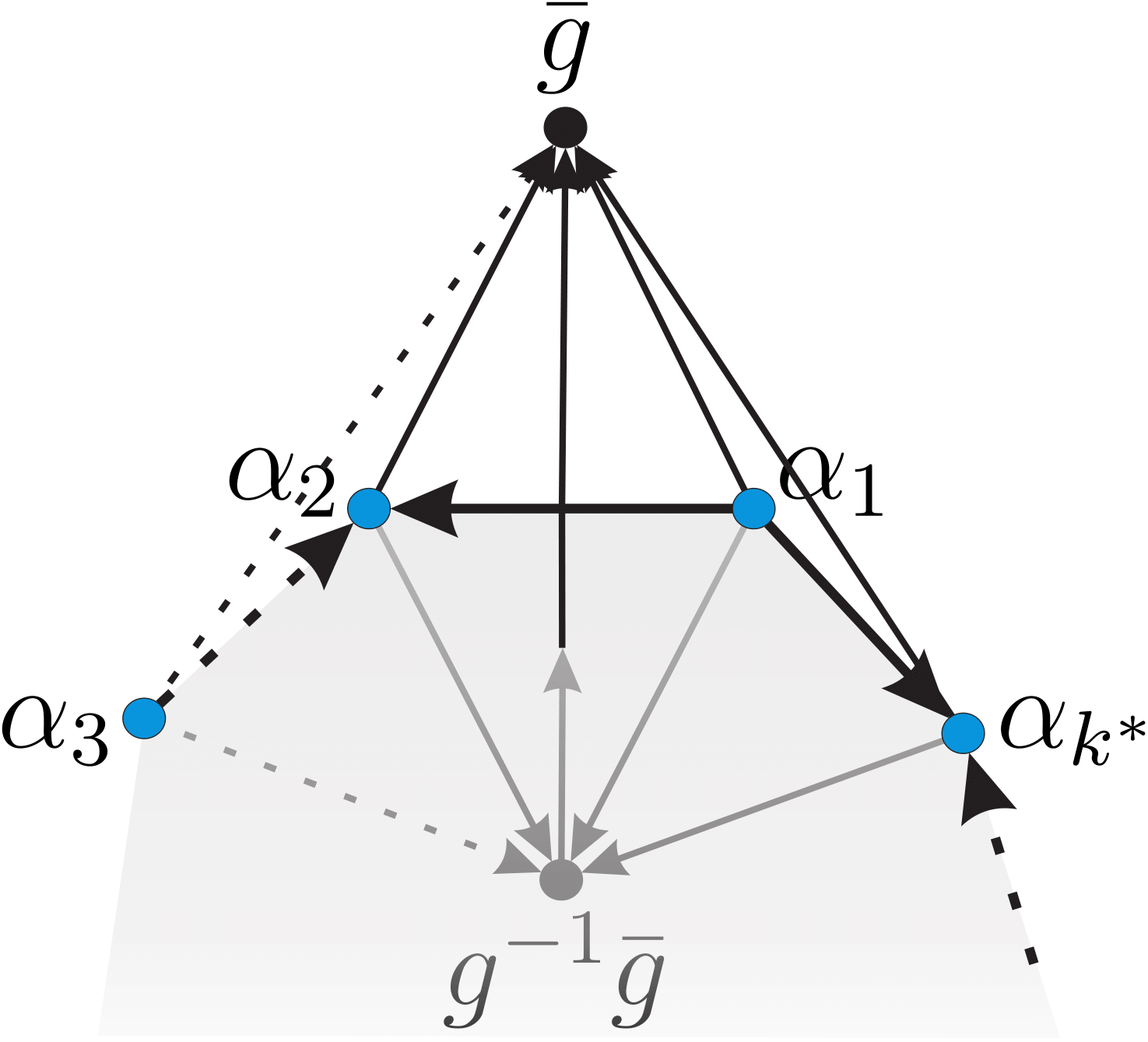}
	\caption{(Color online) Graphic representation of an exemplary phase factor $f^I_3(\alpha_1,...,\alpha_{k^*},g,\bar{g})$ as in Eq.~\ref{f}. For the illustrated cutout we find $\{i_a\}=\{1,...,k^*-1\}$ and $\{i_b\}=\{2,...,k^*\}$. The $\{\alpha_i\}$-plane is illustrated by the shaded area.}\label{complex}
\end{figure}

The on-site representation $U^I_{i}(g)$ of $g\in G$ acting on site $i$ within the sublattice $I$ is then given by
\begin{align}
& U_i^I(g)\ket{\alpha_1,\alpha_2,...,\alpha_{k^*}}\nonumber\\
&=f^I_3(\alpha_1,\alpha_2,...,\alpha_{k^*}, g, \bar{g})\ket{g\alpha_1,g\alpha_2,...,g\alpha_{k^*}}\label{U}
\end{align}
with $k^*$ being the number of partons within one physical site, $\bar{g}$ being a fixed element in $G$ and 
\begin{align}
& f^I_3(\alpha_1,\alpha_2,...,\alpha_{k^*}, g, \bar{g})\nonumber\\
&=\prod_{\{i_a\},\{i_b\}}\frac{\nu_3(\alpha_{i_a},\alpha_{i_a+1},g^{-1}\bar{g},\bar{g})}{\nu_3(\alpha_{i_b+1},\alpha_{i_b},g^{-1}\bar{g},\bar{g})}\label{f}
\end{align}
where we assume that partons are labeled counterclockwise, i.e. $i\rightarrow i+1$ counting counterclockwise, and $\alpha_{k^*+1}\equiv \alpha_1$ and $\alpha_{0}\equiv \alpha_{k^*}$. The sets of indices $\{i_a\},\{i_b\}\subset\{1,...,k^*\}$ define the branching structure of $f^I_3$ and are chosen such that $\{i_a\}\cap\{i_b\}=\emptyset$ and $\{i_a\}\cup\{i_b\}=\{1,...,k^*\}$. On the one hand, $\{i_a\}$ gives counterclockwise branched tetrahedrons in the $\{\alpha_i\}$-plane of the graphic representation of $f_3^I$ in Fig.~\ref{complex}. On the other hand, $\{i_b\}$ defines clockwise orientated tetrahedrons. Naturally, each phase factor (and consequently, each symmetry representation) is uniquely specified by the tuple $(I,\{i_a\},\{i_b\})$.

In order to illustrate the notation, let us consider the kagome lattice as an example. It turns out that we can ensure that $U^I_i(g)$ defines a global symmetry if we divide the lattice into three sublattices $A,B,C$ as illustrated in Fig.~\ref{extensionkagome}. With respect to the depicted labeling of partons, the corresponding phase factors are
\begin{align}
& f^A_3(\alpha_1,...,\alpha_4,g,\bar{g})\nonumber\\
&=\frac{\nu_3(\alpha_1,\alpha_2,g^{-1}\bar{g},\bar{g})\nu_3(\alpha_2,\alpha_3,g^{-1}\bar{g},\bar{g})}{\nu_3(\alpha_1,\alpha_4,g^{-1}\bar{g},\bar{g})\nu_3(\alpha_4,\alpha_3,g^{-1}\bar{g},\bar{g})}\label{fkagome1}
\end{align}
and 
\begin{align}
& f^{B,C}_3(\alpha_1,...,\alpha_4,g,\bar{g})\nonumber\\
&=\frac{\nu_3(\alpha_2,\alpha_3,g^{-1}\bar{g},\bar{g})\nu_3(\alpha_3,\alpha_4,g^{-1}\bar{g},\bar{g})}{\nu_3(\alpha_2,\alpha_1,g^{-1}\bar{g},\bar{g})\nu_3(\alpha_1,\alpha_4,g^{-1}\bar{g},\bar{g})}.\label{fkagome2}
\end{align}
In agreement with Eq.~(\ref{f}) the phase factors can be identified with tuples $(A,\{1,2\},\{3,4\})$, $(B,\{2,3\},\{1,4\})$ and $(C,\{2,3\},\{1,4\})$ respectively.
In the remaining part of this section, we will show that $U^I_i(g)$ is a linear representation of $G$ for any $\{i_a\},\{i_b\}$ and that there is indeed a choice of $I$ together with a choice of $\{i_a\},\{i_b\}$ such that $U^I_i(g)$ is a global symmetry of plaquette states defined on any lattice. Furthermore, we will see that the symmetry action on the system with boundary is the same for this model as it is for the original construction. Thus, it gives rise to the {\it same\/} classification of SPT phases.

% % % % % % % % % % % % % % % % % % % % % % % % % % % % % % % % % % % % % % % % % % % % % % % % % % % % % % % % % % % % % % % % % % % % % % % % % % % % % % % % % % % % % % % % % % % % % % % % % % %

\subsection{$U^I_i(g)$ is a linear representation}\label{Ulinear}
\begin{figure}
	\centering
	\subfigure[]{
		\includegraphics[width=0.22\textwidth]{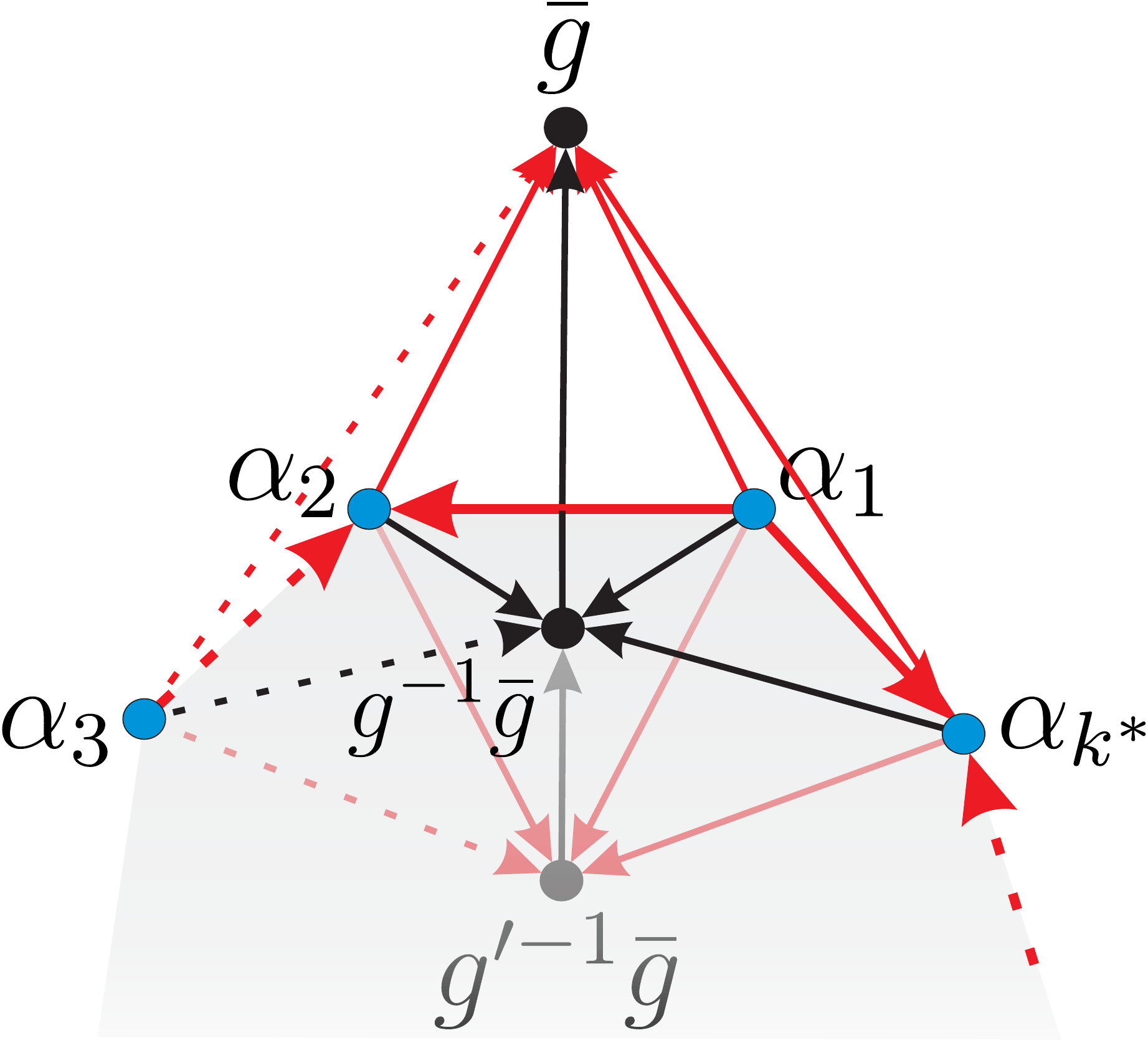}
}
	\subfigure[]{
		\includegraphics[width=0.22\textwidth]{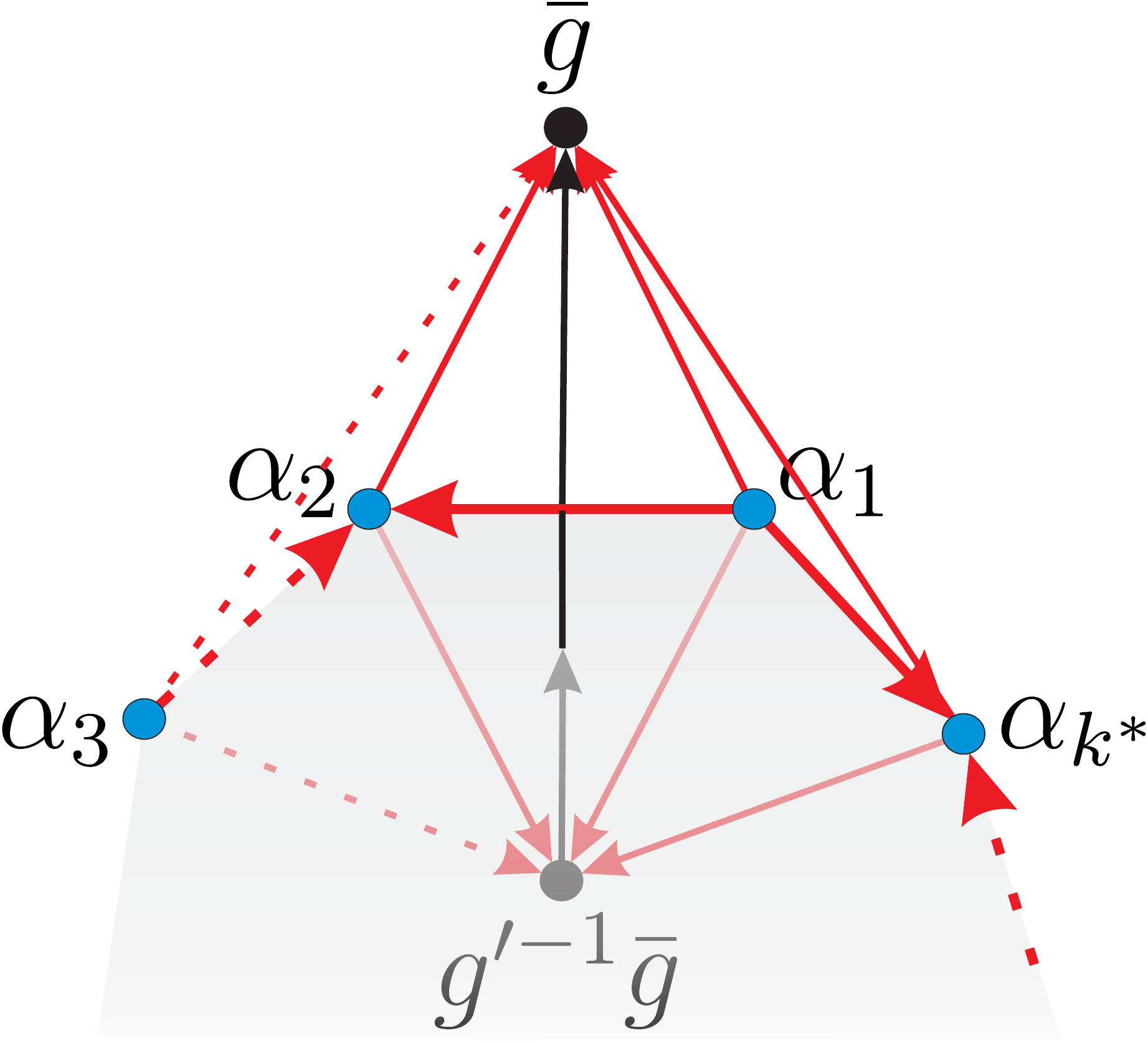}
}
	\caption{(Color online) (a) Graphic representation of Eq.~\ref{phasecompare1} for one choice of $\{i_a\},\{i_b\}$. (b) Graphic representation of Eq.~\ref{phasecompare2} for the same choice of $\{i_a\},\{i_b\}$. Since the complexes have the same surface, they represent the same phase factor (cf. Sec. \ref{graphic}). In fact, for any choice of branching $\{i_a\},\{i_b\}$ for both complexes, the graphic representations will have the same surface. The $\{\alpha_i\}$-plane is illustrated by the shaded area.}\label{graphiccompare}
\end{figure}

In order to show that $U^I_i(g)$ is a linear representation of $G$, we verify that
\begin{align}
U_i^I(g'g^{-1})U_i^I(g)\ket{\alpha_1,\alpha_2,...,\alpha_{k*}}=U_i^I(g')\ket{\alpha_1,\alpha_2,...,\alpha_{k*}}.\label{linear}
\end{align}
Obviously, $\ket{\alpha_j}\mapsto \ket{g'\alpha_j}$ $\forall j$ for both sides of the equation. Therefore, let us compare the phase factors. The left side of Eq. \ref{linear} yields
\begin{align}
&\prod_{\{i_a\},\{i_b\}}\frac{\nu_3(\alpha_{i_a},\alpha_{i_a+1},g^{-1}\bar{g},\bar{g})}{\nu_3(\alpha_{i_b+1},\alpha_{i_b},g^{-1}\bar{g},\bar{g})}\frac{\nu_3(g\alpha_{i_a},g\alpha_{i_a+1},gg'^{-1}\bar{g},\bar{g})}{\nu_3(g\alpha_{i_b+1},g\alpha_{i_b},gg'^{-1}\bar{g},\bar{g})}
\end{align}
which can be rewritten as
\begin{align}
&\prod_{\{i_a\},\{i_b\}}\frac{\nu_3(\alpha_{i_a},\alpha_{i_a+1},g^{-1}\bar{g},\bar{g})\nu_3(\alpha_{i_a},\alpha_{i_a+1},g'^{-1}\bar{g},g^{-1}\bar{g})}{\nu_3(\alpha_{i_b+1},\alpha_{i_b},g^{-1}\bar{g},\bar{g})\nu_3(\alpha_{i_b+1},\alpha_{i_b},g'^{-1}\bar{g},g^{-1}\bar{g})}\label{phasecompare1}
\end{align}
since $g\cdot \nu_3(g_0,g_1,g_3)=\nu_3(gg_0,gg_1,gg_3)$ by definition. From the action of $U_i^I(g')$ we obtain a phase factor
\begin{align}
\prod_{\{i_a\},\{i_b\}}\frac{\nu_3(\alpha_{i_a},\alpha_{i_a+1},g'^{-1}\bar{g},\bar{g})}{\nu_3(\alpha_{i_b+1},\alpha_{i_b},g'^{-1}\bar{g},\bar{g})}.\label{phasecompare2}
\end{align}
Comparing the graphic representation of the two phase factors reveals that they are the same (see Fig.~\ref{graphiccompare}).
Hence, we have proven that $U^{I}_i(g)$ is a linear representation of $G$. This property is completely independent of our choice of $\{i_a\},\{i_b\}$ and thus, of the underlying lattice.
\subsection{$U^I_i(g)$ is a global symmetry}\label{Uglobal}
To prove that there exists a choice of sublattices $I$ and $\{i_a\},\{i_b\}$ such that $U_i^I(g)$ is a global symmetry of $\ket{\psi_{gs}}$ on any lattice, we have to show that there exists such a choice so that the resulting phase factor $F_3$ in
\begin{align}
\underset{i,I}{\bigotimes}U_i^I(g)\ket{\psi_{gs}}=F_3\ket{\psi_{gs}}
\end{align} 
equals unity.

\begin{figure}
	\centering
	\includegraphics[width=0.48\textwidth]{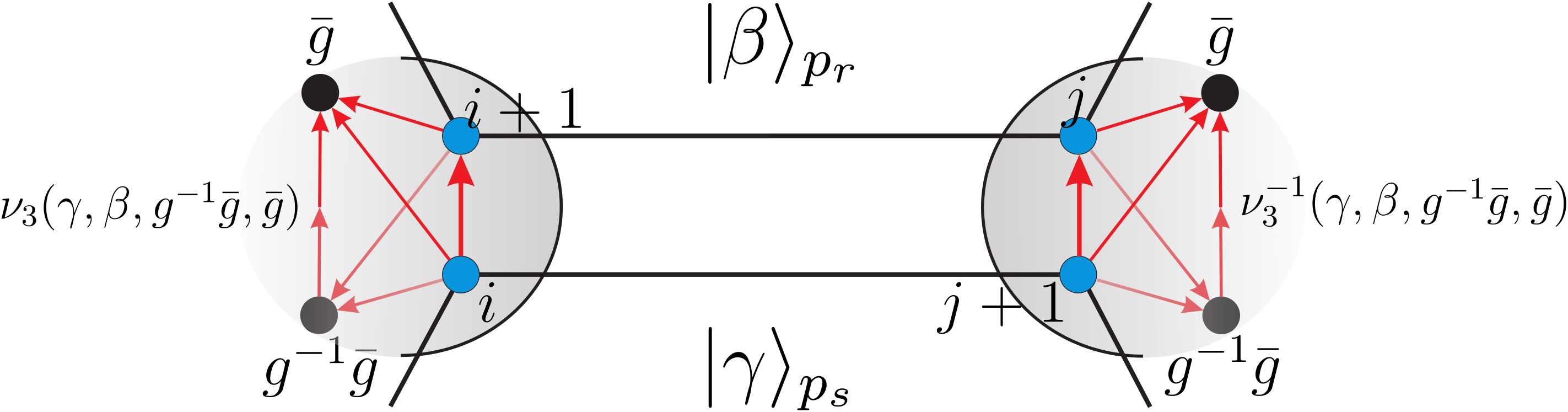}
	\caption{(Color online) Partons $i+1$ and $j$ are in a maximally entangled state $\ket{\beta}$ and partons $i$ and $j+1$ are entangled in $\ket{\gamma}$. The graphic representations of exemplary phase factors arising between partons is depicted in red. The two opposite phase factors cancel if one can be transformed into the other by reflection along the edges. This is guaranteed if the orientation of the edges of branched tetrahedrons between partons is parallel.}\label{bondsym}
\end{figure}

 Let us understand why phase factors cancel in the first place. Consider two pairs of maximally entangled partons as depicted in Fig.~\ref{bondsym}. We want to choose the symmetry representation such that phase factors arising between parallel edges cancel. The tetrahedrons representing the phase factors have opposite chirality if one can be transformed into the other by reflection. That is, the phase factors in Fig.~\ref{bondsym} cancel if the branched edges between partons are orientated parallelly.
\begin{figure}
	\centering
	\subfigure[]{
		\includegraphics[width=0.23\textwidth]{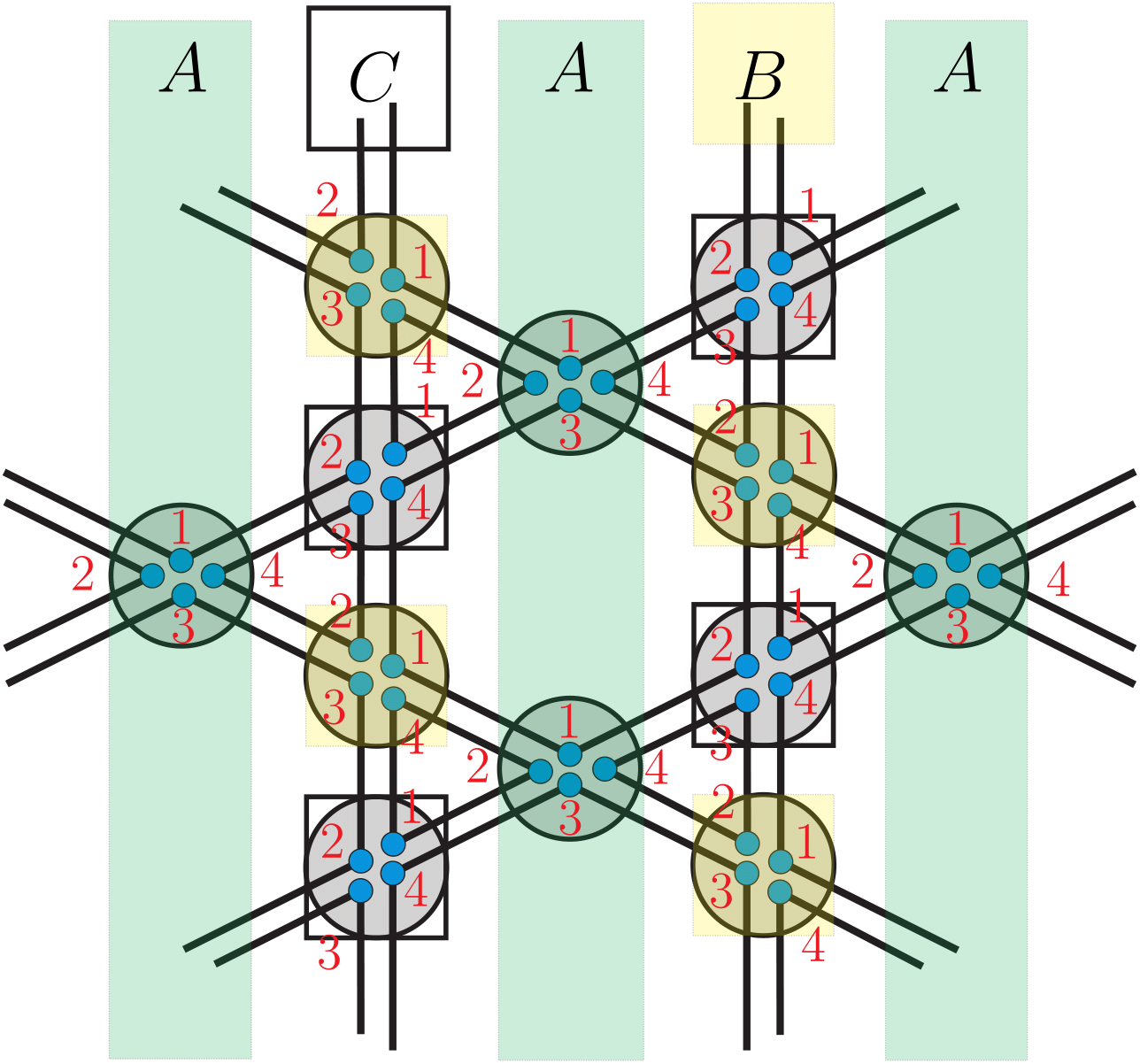}
		\label{extensionkagome}}
	\subfigure[]{
		\includegraphics[width=0.22\textwidth]{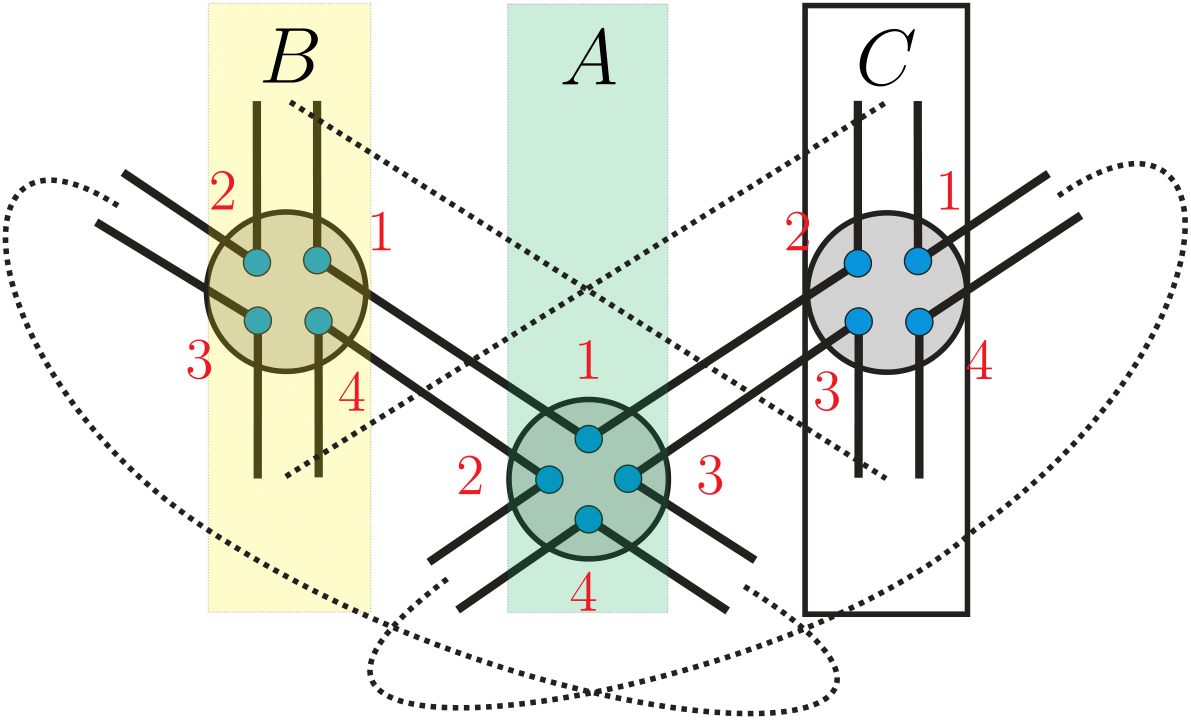}
		\label{unitcell}}
	\caption{(Color online) (a) Dividing the kagome lattice into sublattices $A,B,C$ as depicted by colored columns and boxes together with a labeling of partons as indicated in red, ensures that $U^I_i(g)$ defines a global symmetry if we impose periodic boundary conditions between edges along straight lines. (b) A unit cell of the kagome lattice. Assuming periodic boundary conditions as depicted by dashed lines, this unit cell generates the extended lattice by translation.}\label{symkagome}
\end{figure}
As long as we assume that all plaquettes share an edge with each neighboring plaquette while imposing periodic boundary conditions, we can choose a branching between partons such that the two graphic representations of phase factors involving all four partons can be transformed into each other by reflection along such edges, i.e. the phase factors cancel (see Fig.~\ref{bondsym}). In other words, there exists a choice of sublattices $I$ and $\{i_a\},\{i_b\}$ such that $U_i^I(g)$ is a global symmetry for any lattice that obeys these conditions. This is because we can choose w.l.o.g. an arbitrary branching on each site. This statement is very general and leaves room for a lot of ambiguity, i.e., we may end up choosing different branching structures without respecting the periodicity of the lattices. Fortunately, in the case of regular lattices, we have to consider only a finite region. Therefore, consider the minimal region of some regular lattice with plaquettes such that embedding the region on a $2$-torus gives rise to the infinite extension of this lattice, i.e., the unit cell of a regular lattice. Let us further assume that edges of such a region never cut through vertices. Such a unit cell contains all structural and symmetry information of the macroscopic lattice. In particular, all edges between vertices of the extended lattice can be found within one unit cell. This is the key point since the argument for global symmetry as discussed before is based on reflections along edges. Given the previous considerations, we can find a choice of $\{(I,\{i_a\},\{i_b\})\}$ on the unit cell such that it defines a global symmetry with respect to periodic boundary conditions. By definition each vertex in the unit cell defines one sublattice. Then, $\{(I,\{i_a\},\{i_b\})\}$ also defines a global symmetry on any extended lattice with periodic boundary conditions that preserve the lattice configuration. This is because any edge configuration of such a macroscopic lattice, can be found in the unit cell. That is, an edge in the unit cell connecting two vertices of sublattices $(I,\{i_a\},\{i_b\})$ and $(I',\{i'_a\},\{i'_b\})$ respectively, connects the same sublattices in the extended lattice in the same way. This is why the choice of $\{(I,\{i_a\},\{i_b\})\}$ in Eq.~(\ref{fkagome1}) and~(\ref{fkagome2}) gives rise to a global symmetry on the unit cell depicted in Fig.~\ref{unitcell} and any extension to it as long as it is still embedded on a $2$-torus that preserves the lattice configuration (e.g. Fig.~\ref{extensionkagome}). 

\begin{figure}
	\centering
	\subfigure[]{
		\includegraphics[width=0.22\textwidth]{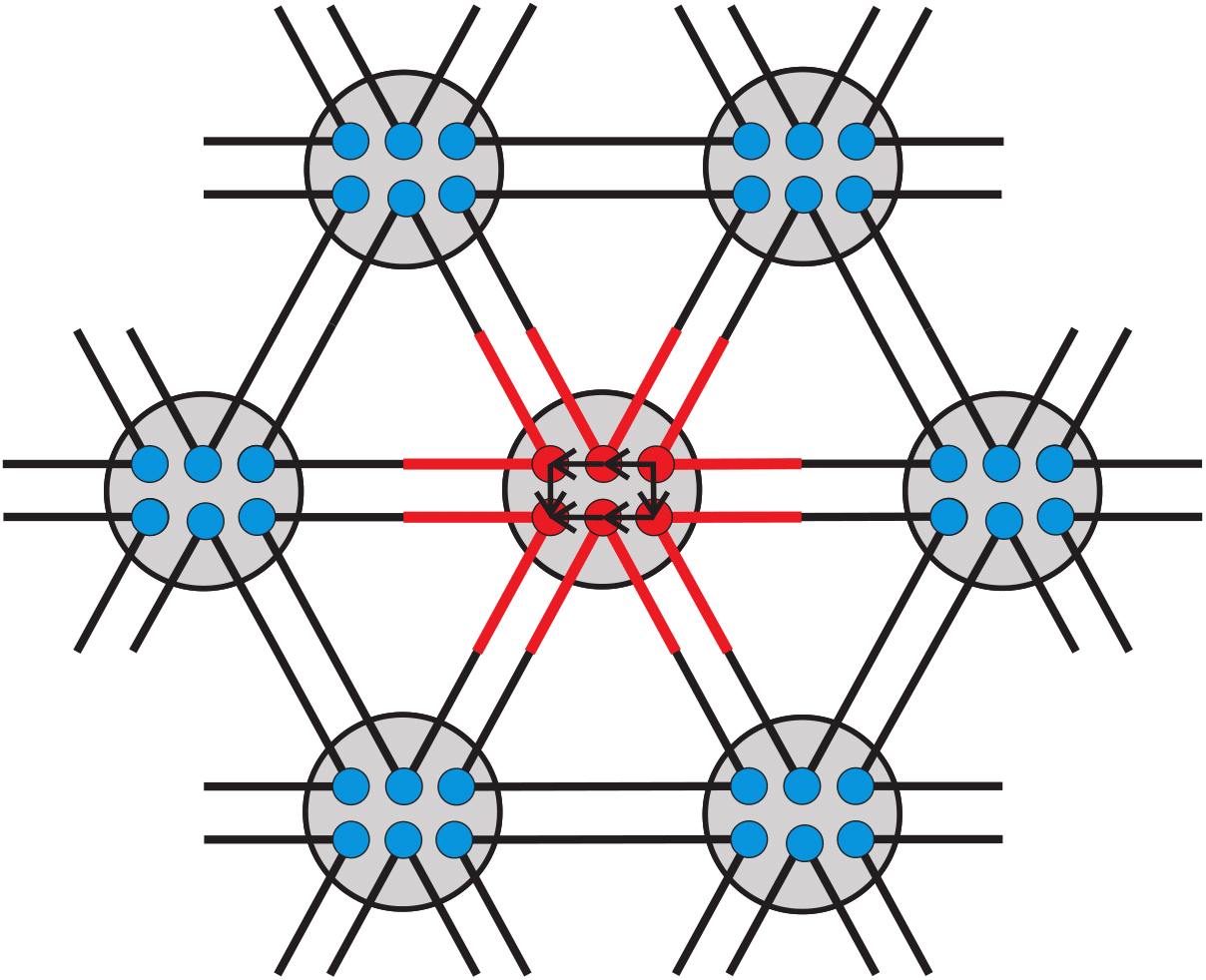}
		
	}
	\subfigure[]{
		\includegraphics[width=0.22\textwidth]{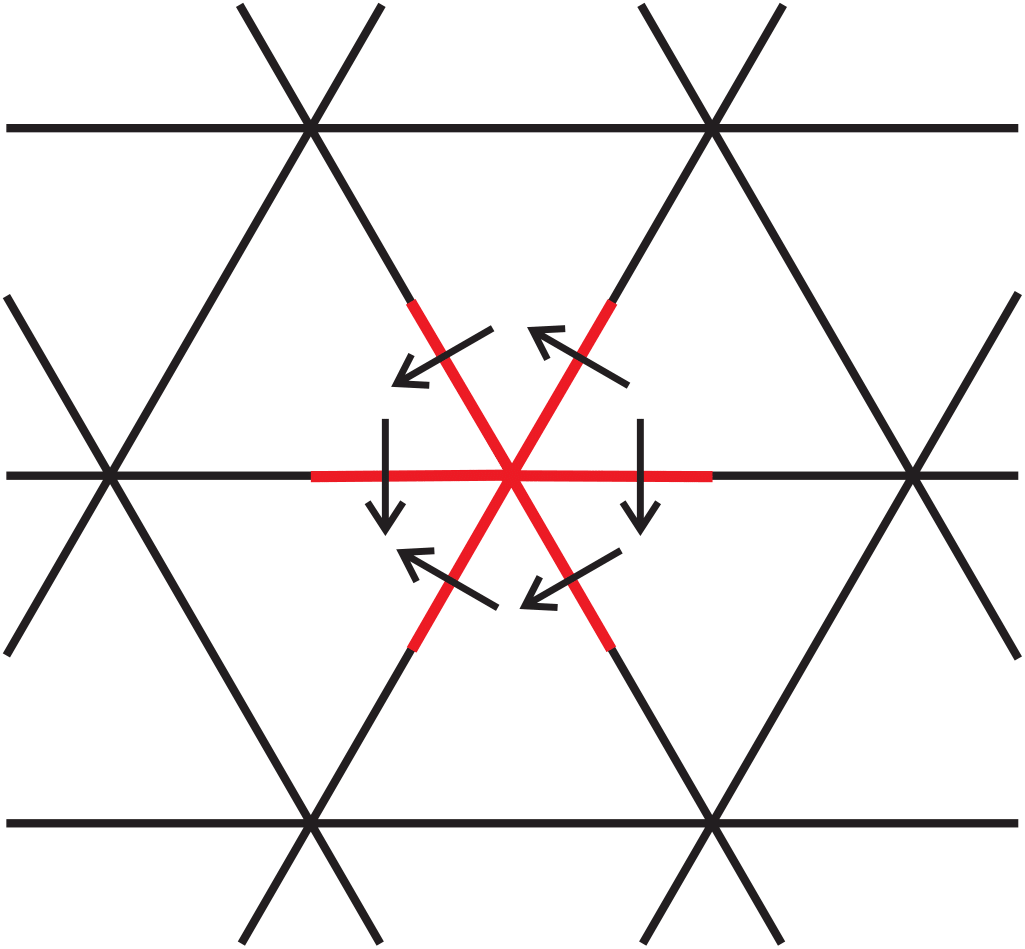}
		%\label{fig:plaqtrianglebond}
	}
	
	\subfigure[]{
		\includegraphics[width=0.22\textwidth]{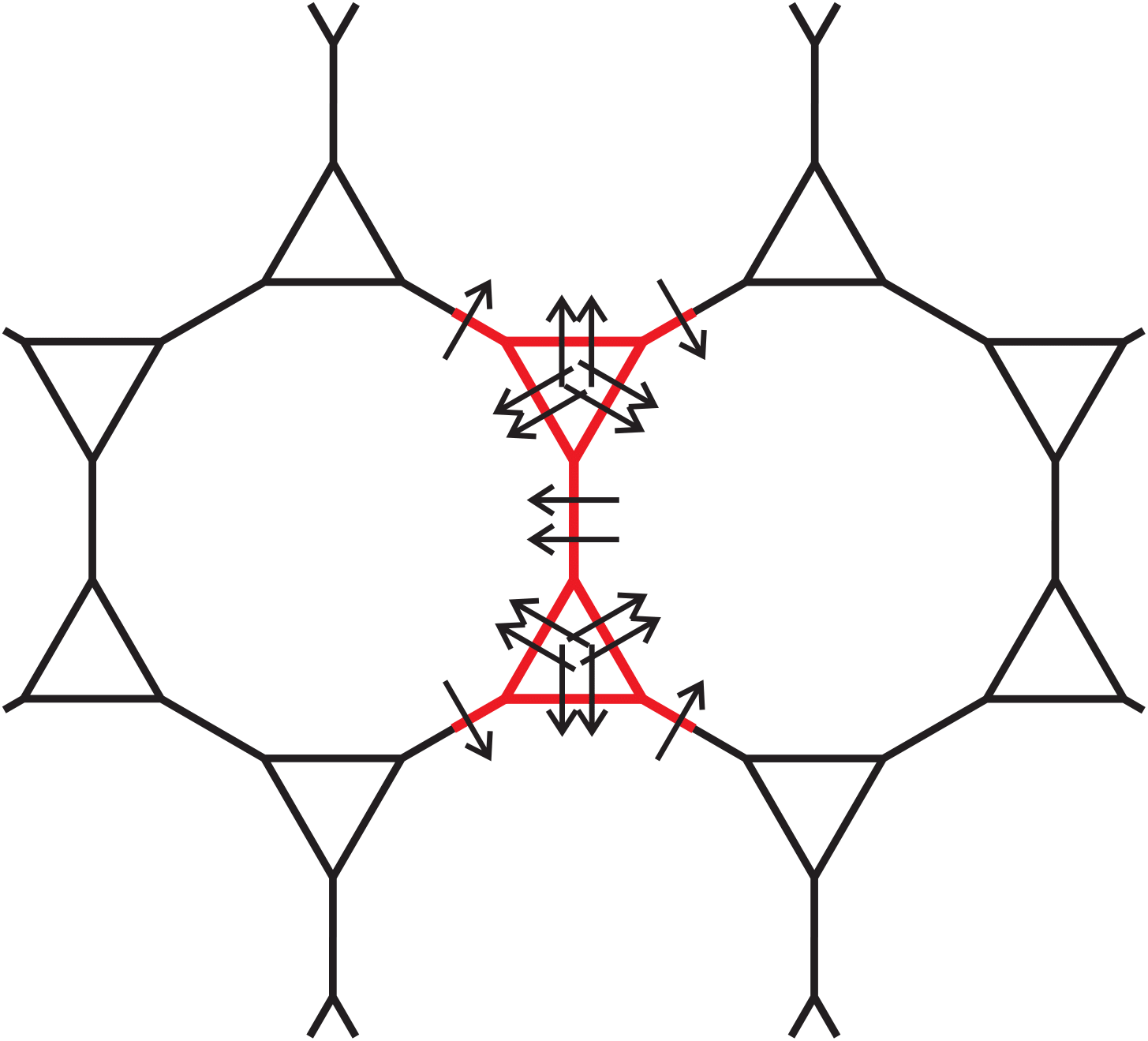}
		%\label{fig:plaqhoneycomb}
	}
	\subfigure[]{
		\includegraphics[width=0.22\textwidth]{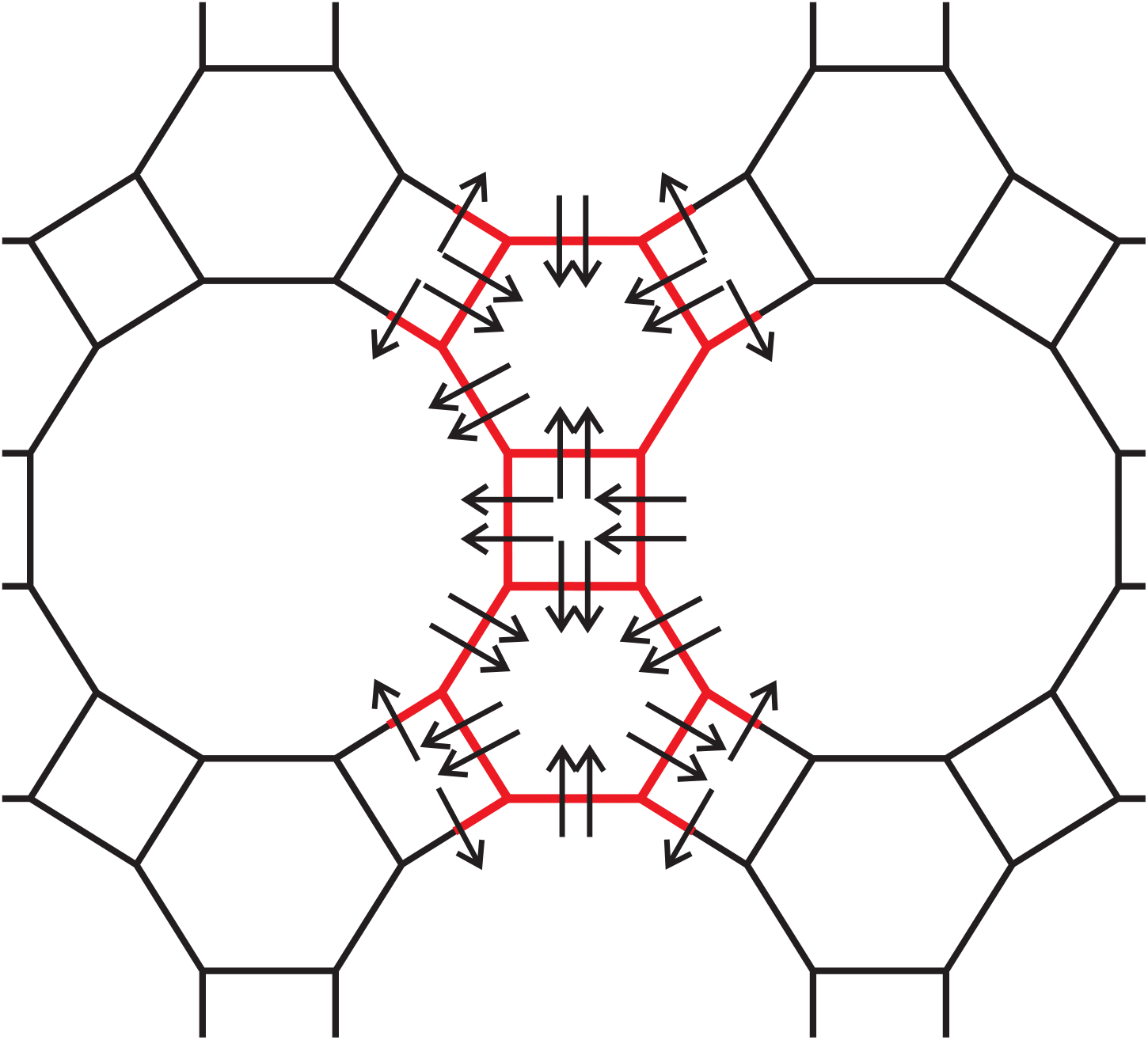}
		%\label{fig:plaqhoneycombbond}
	}
	\subfigure[]{
		\includegraphics[width=0.22\textwidth]{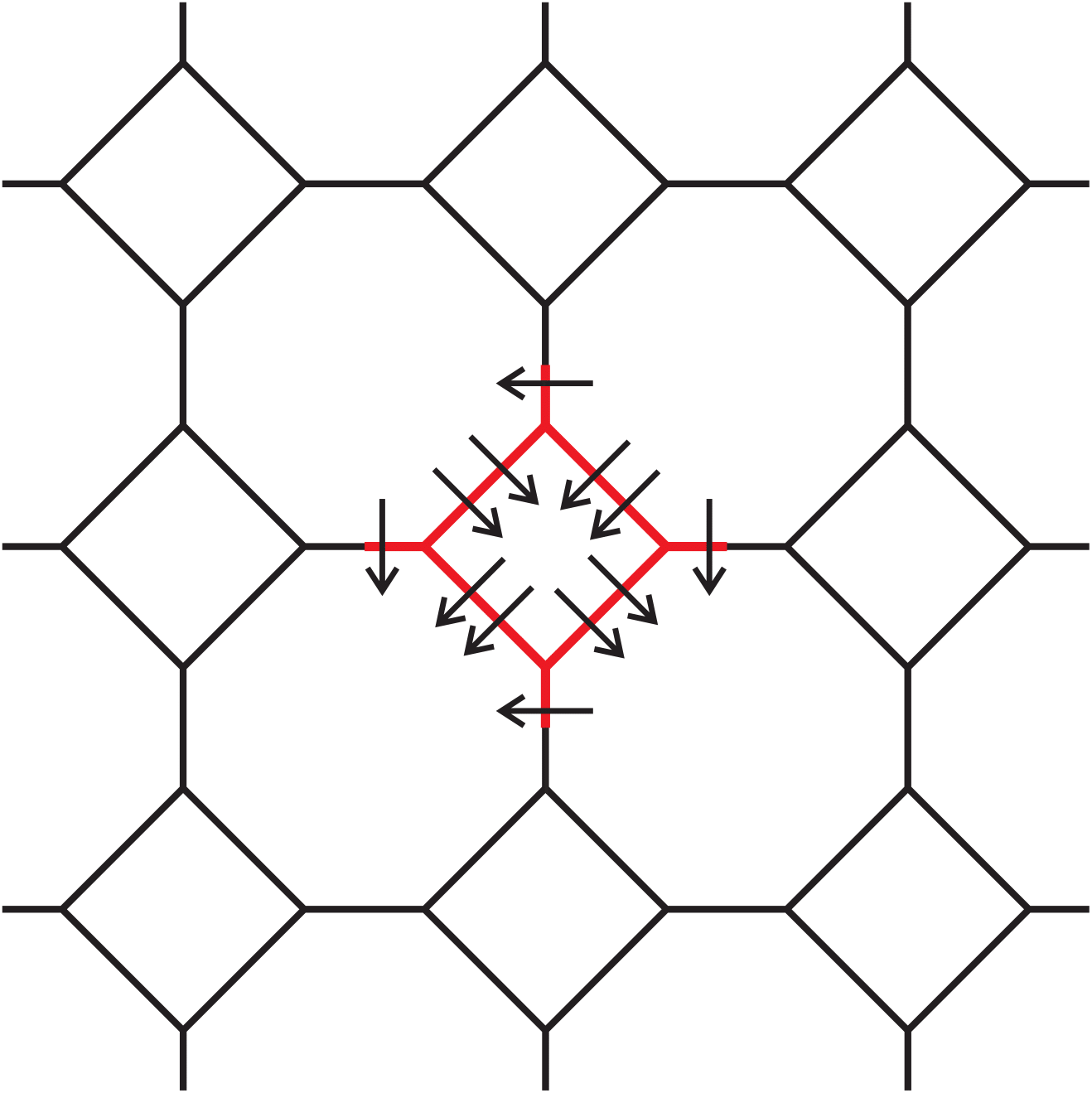}
		%\label{fig:plaqkagome}
	}
	\subfigure[]{
		\includegraphics[width=0.22\textwidth]{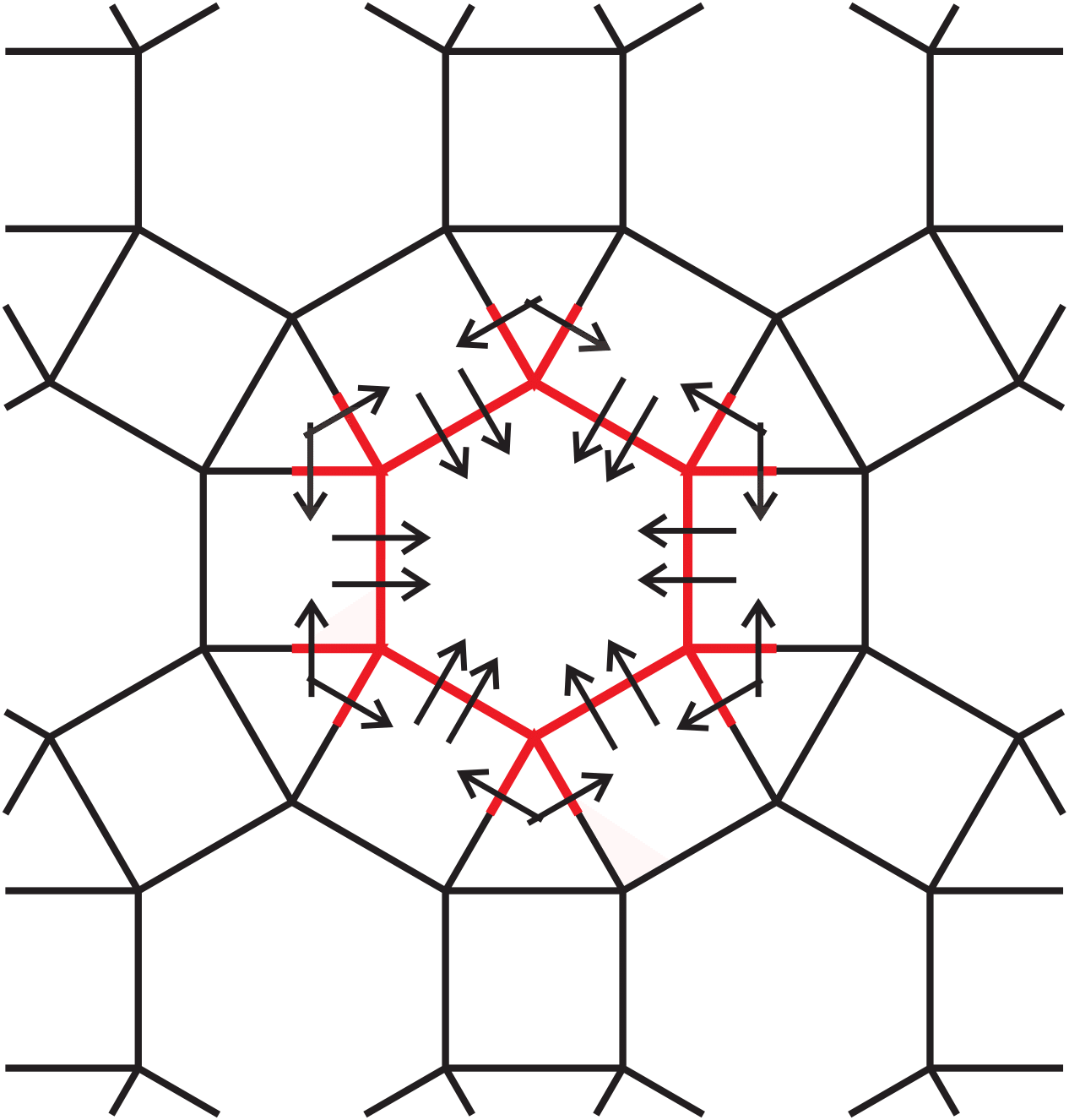}
		%\label{fig:plaqkagomebond}
	}
	\subfigure[]{
		\includegraphics[width=0.2\textwidth]{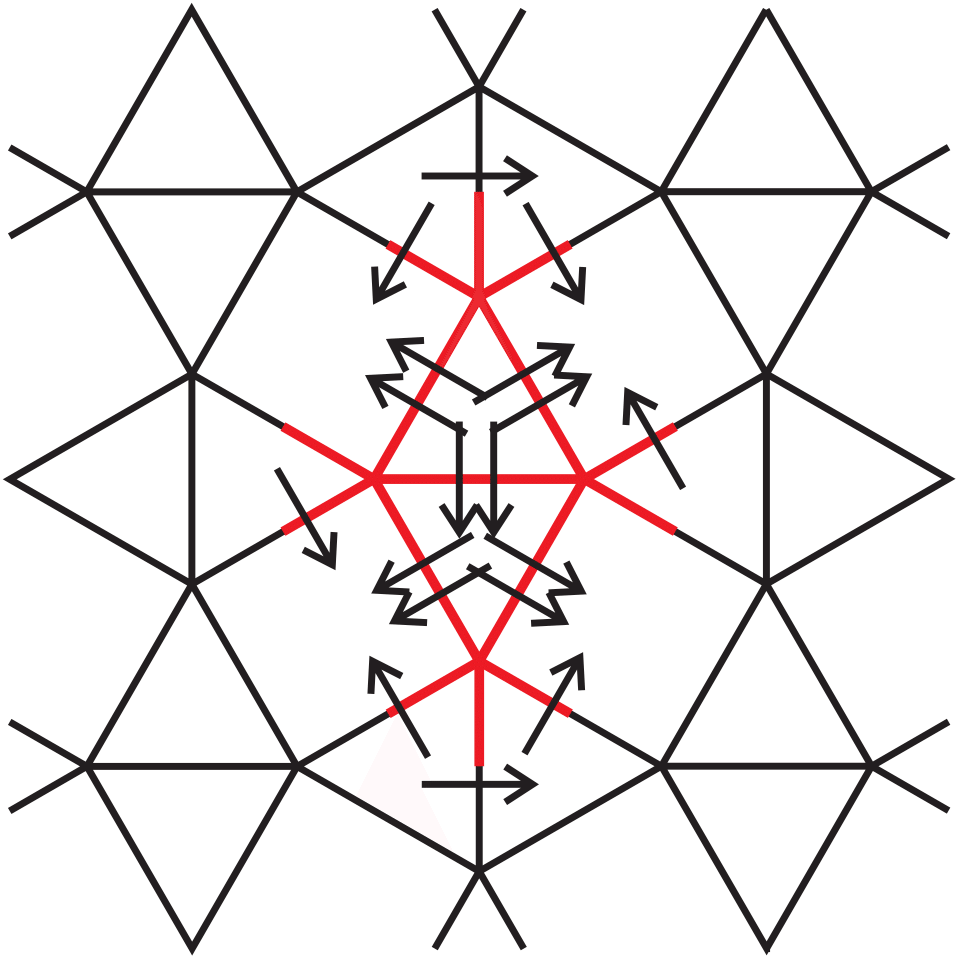}
		%\label{fig:plaqkagomebond}
	}\subfigure[]{
	\includegraphics[width=0.2\textwidth]{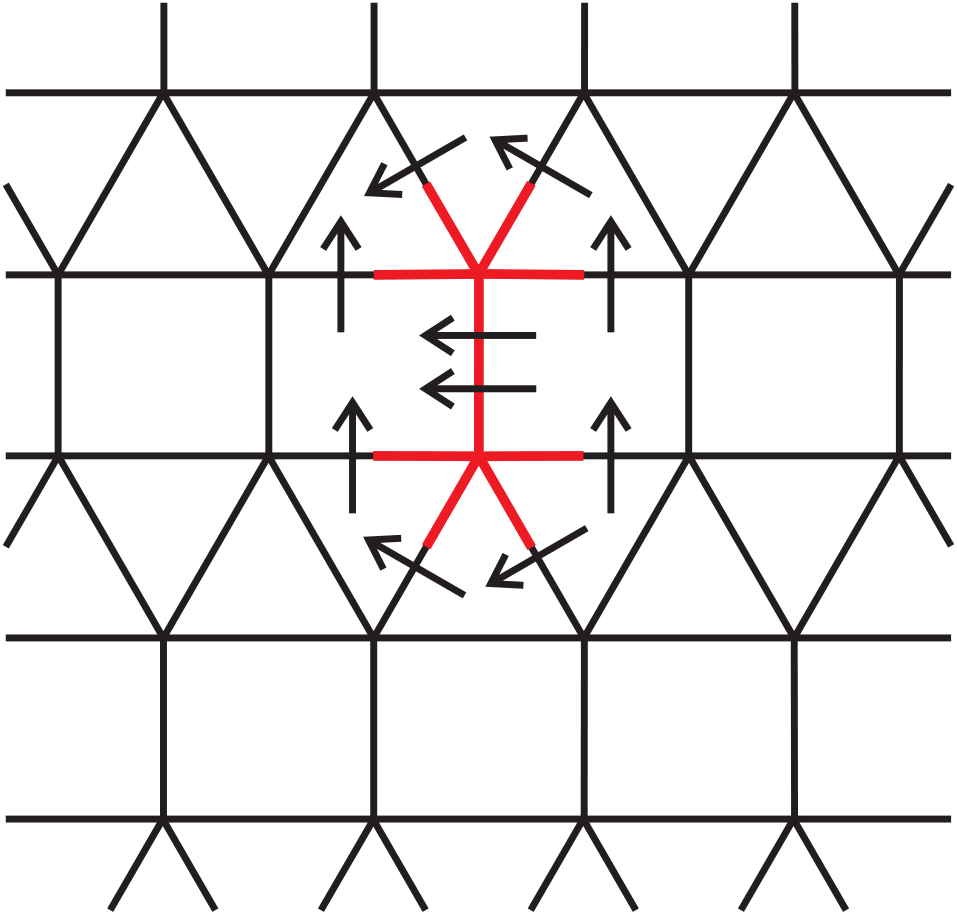}
	%\label{fig:plaqkagomebond}
}
\caption{(Color online) (a) The unit cell of the triangular lattice with plaquettes is indicated in red. Arrows represent graphically the on-site action of a symmetry representation which is a global symmetry of the corresponding plaquette state. (b) Simplified notation for (a). (c)-(h) Exemplary unit cells on various lattices together with symmetry representations that give rise to global symmetries on corresponding plaquette states. Each vertex within a unit cell defines a sublattice.}\label{fig:globalsym}
\end{figure}

In Fig.~\ref{fig:globalsym} we illustrate the branched structures on various lattices, based on the general proof above. There we simplify the drawing and do not show the individual partons, except in Fig.~\ref{fig:globalsym}a; the arrows represent the ordering.

%%%%%%%%%%%%%%%%%%%%%%%%%%%%%%%%%%%%%%%%%%%%%%%%%%%%%%%%%%%%%%%%%%%%%%%%%%%%%%%%%%%%%%%%%%%%%%%%%%%%%%%%%%%%%%%%%%%%%%%%%%%%%%%%%%%%%%%%%%%%%%%%%%%%%%%%%%%%%%%%%%%%%%%%%%%%%%%%%%%%%%%%%%%%%%%%%%%%%%%%%%%%%%%%%%%%%%%%%%%%%%%%%%%%%%%%%%%%%%%%%%%%%%%%%%%%%%%%%%%%%%%%%%%%%%%%%%%%%%%%%%%%%%%%%%%%%%%%%%%%%%%%%%%%%%

\subsection{Action of $U^I_i(g)$ on $\ket{\psi_{gs}}$ with boundary}\label{actionboundary}
\begin{figure}
	\centering
	\subfigure[]{
		\includegraphics[width=0.44\textwidth]{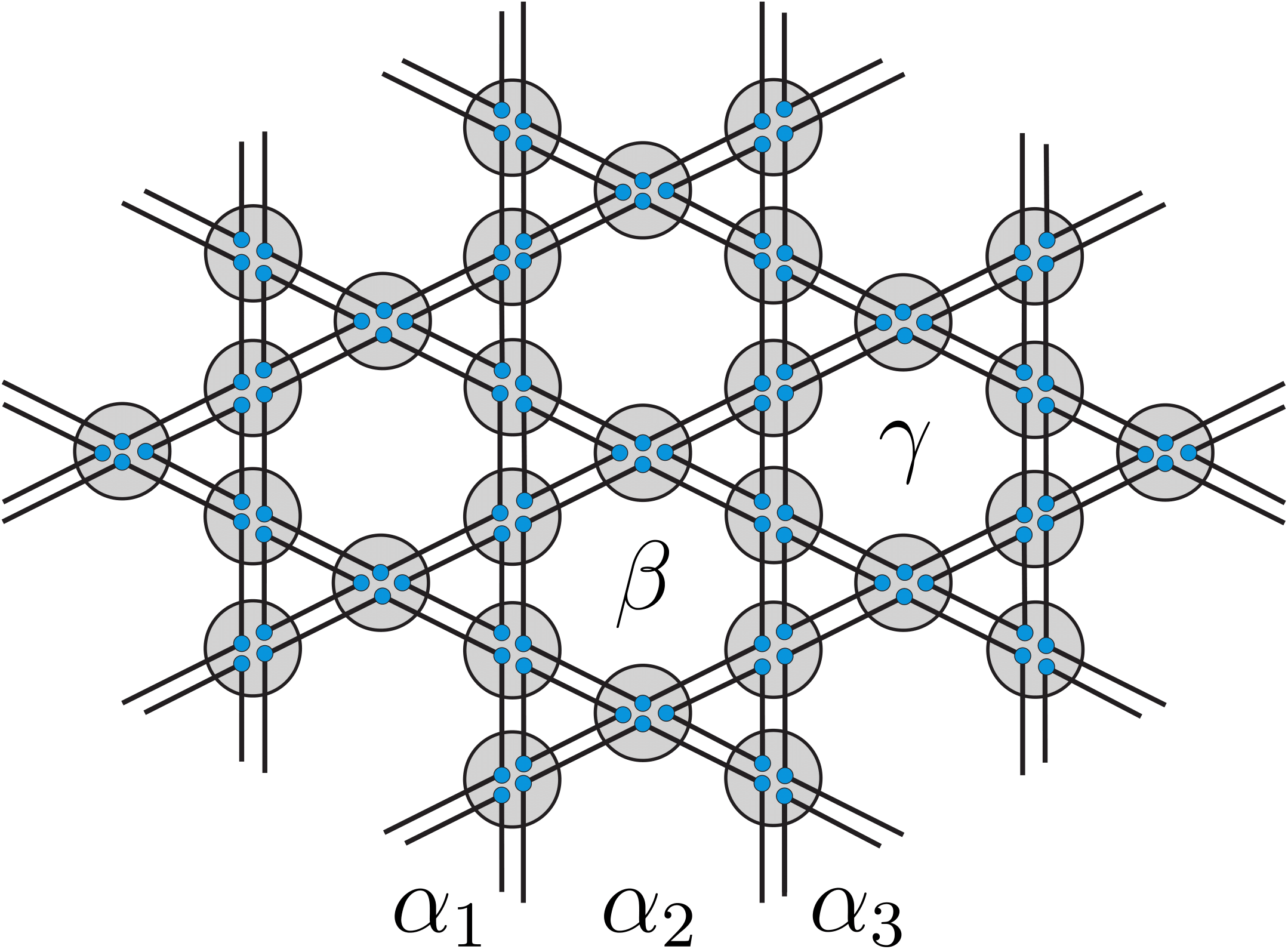}
	}	
	\subfigure[]{
		\includegraphics[width=0.44\textwidth]{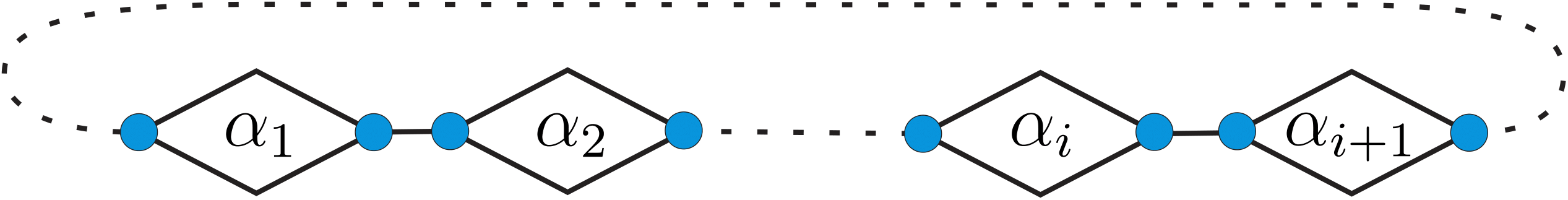}
	}
	\caption{(Color online) (a) The kagome lattice with exemplary boundary. Partons within one plaquette are in the same state. The ground state is $\sum_{\{\alpha_i\}}\ket{\{\alpha_i\}}_{gs}$. (b) The boundary represented by a 1D chain with periodic boundary conditions. Bonds represent the action of the symmetry on the effective low-energy degrees of freedom $\ket{\alpha_1,\alpha_2,...}$.}\label{boundarykagome}
\end{figure}

Now, let us consider the action of $U_i^I(g)$ on the system $\ket{\psi_{gs}}$ with boundary (see Fig. \ref{boundarykagome}). We choose $(I,\{i_a\},\{i_b\})$ such that the symmetry representation defines a global symmetry. Analogous to the previous analysis, cochains $\nu_3$ that depend solely on partons within full plaquettes or cochains that involve arguments from both full plaquettes and the boundary cancel. However, phase factors acting exclusively on the boundary persist. Let us relabel the state of plaquettes on the boundary by $\ket{\alpha_1,\alpha_2,...}$ counterclockwise such that $\ket{\alpha_i},\ket{\alpha_j}$ are adjacent plaquettes if $|i-j|=1$. Then, the ground state takes the form
\begin{align}
\ket{\bar{\psi}_{gs}}&=\sum_{\{\alpha_i\}}\ket{\{\alpha_i\}}_{gs}\nonumber\\
&\equiv\sum_{\{\alpha_i\}\in G}\sum_{\beta,\gamma,...\in G}\ket{\alpha_1,\alpha_2,...,\beta,\gamma,...}.
\end{align}

The effective action of $U_i^I(g)$ on the boundary depends on the direction of the bonds between the effective degrees of freedom $\{\alpha_i\}$ in Fig. \ref{boundarykagome}. The bonds are identified with phase factors arising between partons on the boundary. The direction reflects the branching of the orientated edges in the graphic representation of the $3$-cochains between such partons. Then, the global effective (periodic) action of $U^I_i(g)$ on the boundary is 
\begin{align}
&\underset{\text{i},\text{I}}{\bigotimes}U^\text{I}_\text{i}\ket{\{\alpha_i\}}_{gs}\nonumber\\&=\prod_{\langle ij\rangle}\nu_3^{s_{ij}}(\alpha_i, \alpha_j, g^{-1}\bar{g},\bar{g})\ket{\{g\alpha_i\}}_{gs}\label{boundary}
\end{align}
where the product runs over all nearest neighbors with respect to the direction of the bond and $s_{ij}=1$ if $i>j$ and $s_{ij}=-1$ if $i<j$. Analogous to the case without boundary, we can verify that $U^I_i(g)$ is a linear representation on the boundary.

Eq.~(\ref{boundary}) coincides with the results in Ref. \cite{cohomology}. That is, the action of $U^I_i(g)$ on the one-dimensional boundary can be represented by a matrix-product unitary operator (MPUO) $U(g)$,
\begin{align}
U(g)=\sum_{\{i_k\},\{i_k'\}}& Tr(T^{i_1,i_1'}(g)T^{i_2,i_2'}(g)\cdots T^{i_d,i_d'}(g))\nonumber\\&\ket{i_1'i_2'...i_d'}\bra{i_1i_2...i_d}.
\end{align}
Considering Eq. \ref{boundary}, the matrices $T^I_i(g)$ are found to be
\begin{align}
(T^\text{I}_\text{i})^{\alpha_i,g\alpha_i}(g)=\sum_{\alpha_{i+1}}\nu_3^{-1}(\alpha_i,\alpha_{i+1},g^{-1}\bar{g},\bar{g})\ket{\alpha_i}\bra{\alpha_{i+1}}
\end{align}
for all $\alpha_i$ if the bond goes from $\alpha_i$ to $\alpha_{i+1}$. Other terms are zero. If the bond direction is $\alpha_{i+1}\rightarrow\alpha_{i}$, then
\begin{align}
(T^\text{I}_\text{i})^{\alpha_i,g\alpha_i}(g)=\sum_{\alpha_{i+1}}\nu_3(\alpha_{i+1},\alpha_{i},g^{-1}\bar{g},\bar{g})\ket{\alpha_i}\bra{\alpha_{i+1}}
\end{align}
for all $\alpha_i$ and other terms are zero.

Consequently, plaquette states on arbitrary lattices exhibit the same nontrivial properties with respect to the symmetry representation $U^I_i(g)$ as the original model on a square lattice. That is, if $\nu_3$ is nontrivial, the model we constructed has a nontrivial boundary which cannot have a SRE, gapped and symmetric ground state. On the other hand, if $\nu_3$ is trivial, the boundary effective symmetry does allow SRE, gapped and symmetric states.

%%%%%%%%%%%%%%%%%%%%%%%%%%%%%%%%%%%%%%%%%%%%%%%%%%%%%%%%%%%%%%%%%%%%%%%%%%%%%%%%%%%%%%%%%%%%%%%%%%%%%%%%%%%%%%%%%%%%%%%%%%%%%%%%%%%%%%%%%%%%%%%%%%%%%%%%%%%%%%%%%%%%%%%%%%%%%%%%%%%%%%%%%%%%%%%%%%%%%%%%%%%%%%%%%%%%%%%%%%%%%%%%%%%%%%%%%%%%%%%%%%%%%%%%%%%%%%%%%%%%%%%%%%%%%%%%%%%%%%%%%%%%%%%%%%%%%%%%%%%%%%%%%%%%%%

%%%%%%%%%%%%%%%%%%%%%%%%%%%%%%%%%%%%%%%%%%%%%%%%%%%%%%%%%%%%%%%%%%%%%%%%%%%%%%%%%%%%%%%%%%%%%%%%%%%%%%%%%%%%%%%%%%%%%%%%%%%%%%%%%%%%%%%%%%%%%%%%%%%%%%%%%%%%%%%%%%%%%%%%%%%%%%%%%%%%%%%%%%%%%%%%%%%%%%%%%%%%%%%%%%%%%%%%%%%%%%%%%%%%%%%%%%%%%%%%%%%%%%%%%%%%%%%%%%%%%%%%%%%%%%%%%%%%%%%%%%%%%%%%%%%%%%%%%%%%%%%%%%%%%%
\section{Symmetry-protected topologically ordered states as universal resources for MBQC}\label{universality}
Now that we have shown  in the previous section that plaquette states on arbitrary lattices are nontrivial SPT states, we will now show that these states are also universal resources for MBQC. We first use the qubit plaquette state on the square lattice as an example to illustrate how it is a universal resource  and then go on to deal with plaquette states on arbitrary 2D lattices. Subsequently, we consider the qudit case, which turns out to be a simple generalization of the qubit case.  Along the way, we also discuss how one has to modify the CZX model~\cite{edgemodes} for arbitrary 2D lattices.
\subsection{$2$-level MBQC on the square-lattice plaquette state}\label{example}
\noindent {\bf Hamiltonian and symmetry}.
Before we proceed to the general case in two spatial dimensions let us consider an illustrative example, the CZX model. The CZX model was first described in terms of SPT order in Ref.~\cite{edgemodes}. This model has a ground state in the previously discussed canonical form and hence, has an entanglement structure given by plaquettes, see Eq.~(\ref{SREstate}), represented by squares in Fig.~\ref{fig:square}. It is defined on a square lattice with four virtual qubits on each physical site as shown in Fig.~\ref{fig:square} where qubits are identified with blue dots and sites with shaded circles.

Given the discussion in Sec.~\ref{construction}, we can find an on-site representation $U_i$ of the symmetry group $\mathbb{Z}_2$ such that this model exhibits nontrivial SPT order. In Ref.~\cite{edgemodes} it is shown that one such representation is
\begin{align}
U_{CZX}=U_XU_{CZ}
\end{align}
where
\begin{align}
U_X&= X_1\otimes X_2\otimes X_3\otimes X_4\\
U_{CZ}&= CZ_{12}CZ_{23}CZ_{34}CZ_{41}
\end{align}
with $CZ_{ct}=\ket{0}_c\bra{0}\otimes \mathbbm{1}_t+\ket{1}_c\bra{1}\otimes Z_t$ being the controlled-phase gate, acting on $c$(ontrol) and $t$(arget) qubit and $X,Z$ being Pauli $X$ and $Z$ operators. Instead, we could consider a symmetry representation constructed from the original model of Ref. \cite{cohomology}. Then, $U_{CZ}$ would be substituted by
\begin{align}
U_{sCZ}=sCZ_{12}\:sCZ_{23}\:sCZ_{43}\:sCZ_{14}
\end{align}
where we introduced a special controlled-phase gate $sCZ_{ct}=\ket{0}_c\bra{0}\otimes Z_t+\ket{1}_c\bra{1}\otimes \mathbbm{1}_t$. This result assumes the specific $3$-cocycle representation defined in Eq.~(\ref{cocyclerepr}) with $c=1$ and $\bar{g}$ being the nontrivial element. One can easily check that each matrix-product unitary operator (see Ref.~\cite{edgemodes, cohomology, MPS}), representing either one of the two symmetry representation on the boundary, corresponds to the same nontrivial $3$-cocycle.

The canonical plaquette state is the unique ground state of an exactly solvable Hamiltonian $H=\sum_{p_j}H_{p_j}$ defined on the $j$th full plaquette $p$ and the neighboring half plaquettes, i.e. local regions as depicted in Fig.~\ref{fig:square}. The local terms $H_{p_j}$ are given by
\begin{align}
\label{eqn:HCZX}
H_{p_j}=-X_4\otimes P_2^l\otimes P_2^r\otimes P_2^u\otimes P_2^d
\end{align}
where $X_4=\ket{0000}\bra{1111}+\ket{1111}\bra{0000}$ acts on a full plaquette, $P_2=\ket{00}\bra{00}+\ket{11}\bra{11}$ are projections acting on the four surrounding half plaquettes and the Hamiltonian acts as identity on the virtual qubits at the corners. Terms other than $X_4$ just ensure that $H_{p_i}$ is symmetric under the symmetry representations. On the one hand, this is because the projections $P$ can effectively be treated as $\mathbbm{1}$ with respect to controlled gates not acting on the full plaquette. On the other hand, controlled gates acting on $P$ and $X_4$ from the left or the right leave the local terms invariant for the same reason for that it is a global symmetry. Considering the prescription in this paper, this argumentation holds on any other lattice. The Hamiltonian is indeed just the projection on the SRE ground state
\begin{align}
\ket{\psi_{gs}}=\frac{1}{2^{n_p/2}}\bigotimes_{j}(\ket{0000}+\ket{1111})_{p_j}\label{examplegs}
\end{align}
where $n_p$ is the number of plaquettes. 

\begin{figure}
	\centering
	\subfigure[]{
		\includegraphics[width=0.22\textwidth]{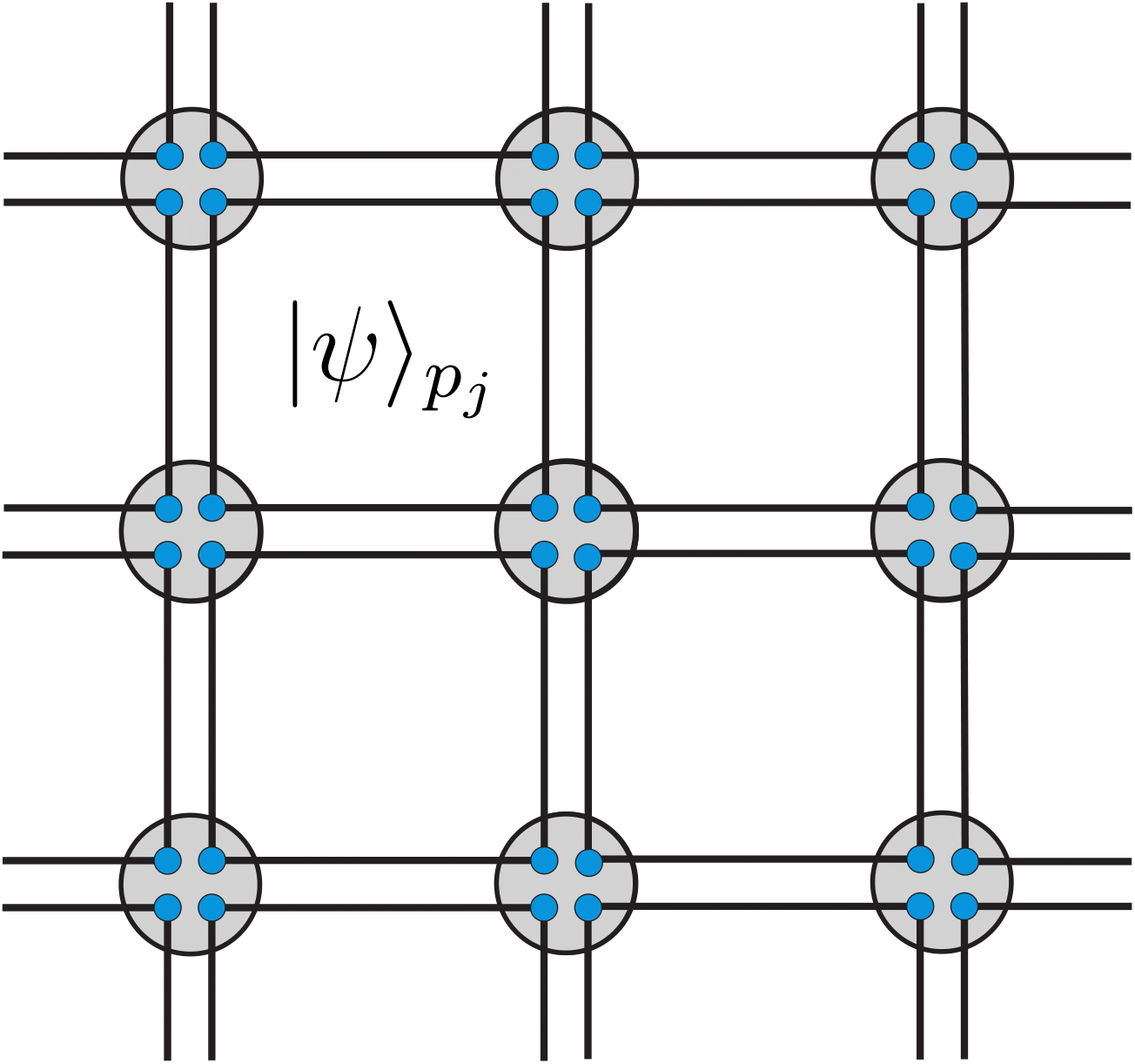}
		
		\label{fig:plaqsquare}
	}
	\subfigure[]{
		\includegraphics[width=0.22\textwidth]{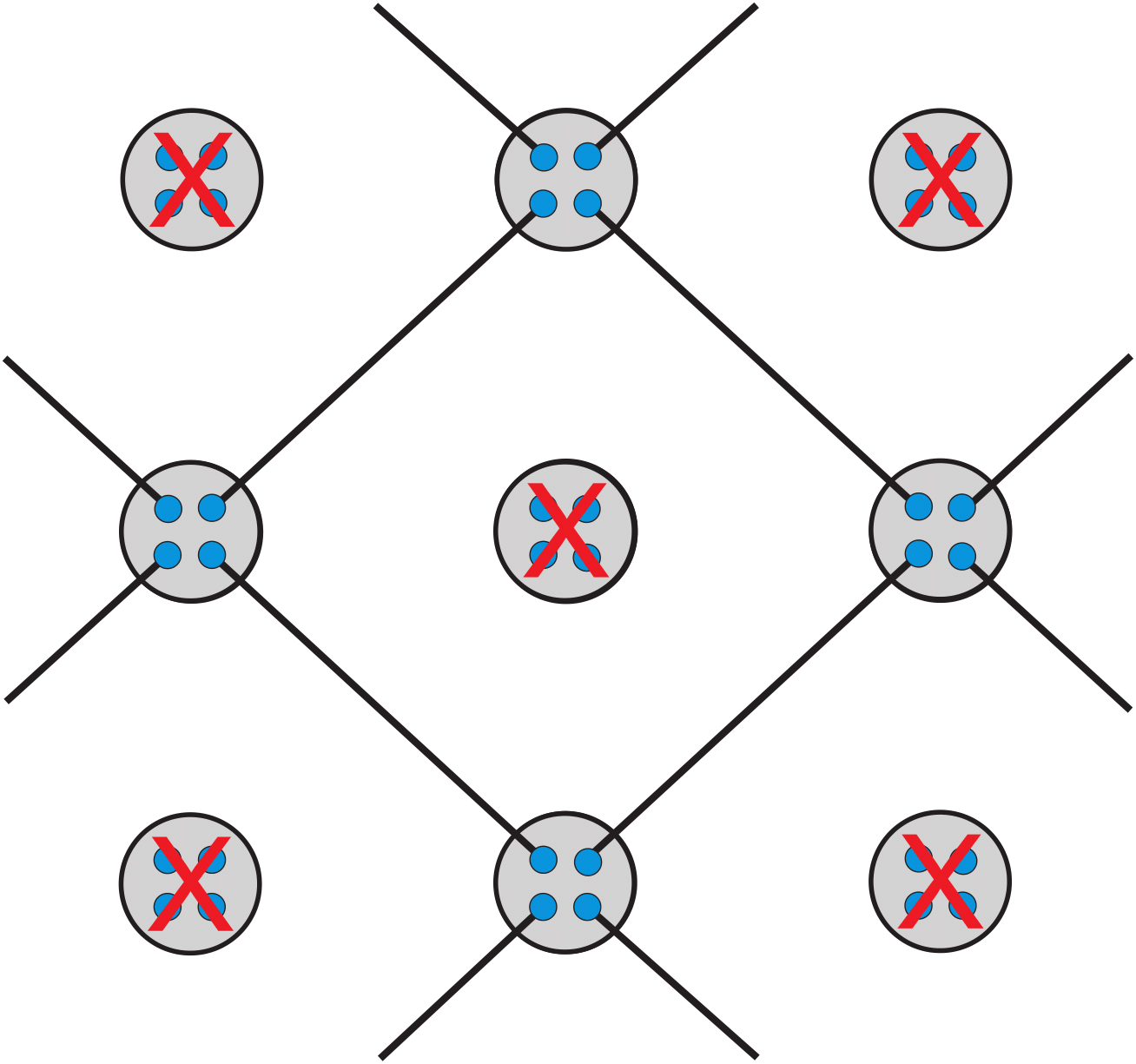}
		\label{fig:plaqsquarebond}}
	\caption{(Color online) (a) An example for two-dimensional plaquette states $\ket{\psi}_{p_j}$ defined in the context of the CZX model on a square lattice. (b) Measurement pattern that reduces the plaquette state to a bond state. Red crosses mark sites on which all virtual qubits are measured in the $\ket{\pm}$-basis.}\label{CZXbond}
\end{figure}

\medskip\noindent {\bf Quantum computational universality}.
So far, we have predominantly reviewed the construction in \cite{edgemodes}. Now, let us show that the ground state is indeed universal for MBQC. In order to process the information adequately, we define a local projective measurement pattern that reduces the plaquettes between four qubits on a square lattice to bonds between two qubits on another square lattice. As we will show, this configuration constitutes a universal resource for MBQC. In Fig.~\ref{CZXbond} we propose a measurement pattern that produces the desired result by measuring each virtual qubit on physical qubits along parallel diagonal lines. 

One such strategy to show that the plaquette state is a universal resource for MBQC is to construct a local measurement that reduces the state to a known universal resource state, such as a valence-bond state or a graph state. For the square lattice each physical site contains four qubit partons and we can express the measurement in terms of the degrees of freedom of partons. 

\begin{figure}
	\centering
	%\subfigure[]{
	%	\includegraphics[width=0.22\textwidth]{plaqtrianglebond1.pdf}
	%	\label{fig:plaqtriangle}
	%}
	\subfigure[]{
		\includegraphics[width=0.22\textwidth]{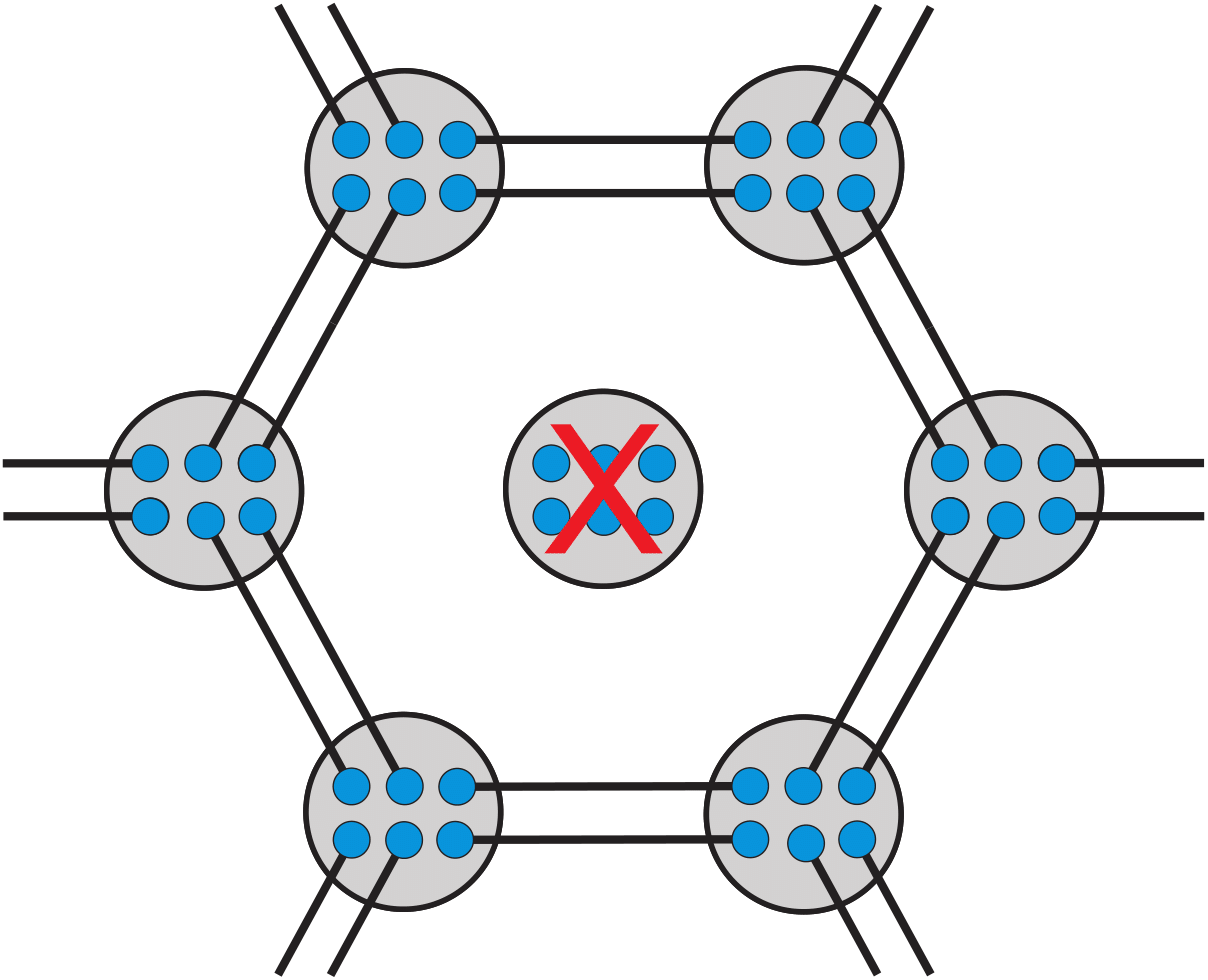}
		\label{fig:plaqtrianglebond}
	}
	%
	%\subfigure[]{
	%	\includegraphics[width=0.22\textwidth]{plaqhoneycombbond1.pdf}
	%	\label{fig:plaqhoneycomb}
	%}
	\subfigure[]{
		\includegraphics[width=0.22\textwidth]{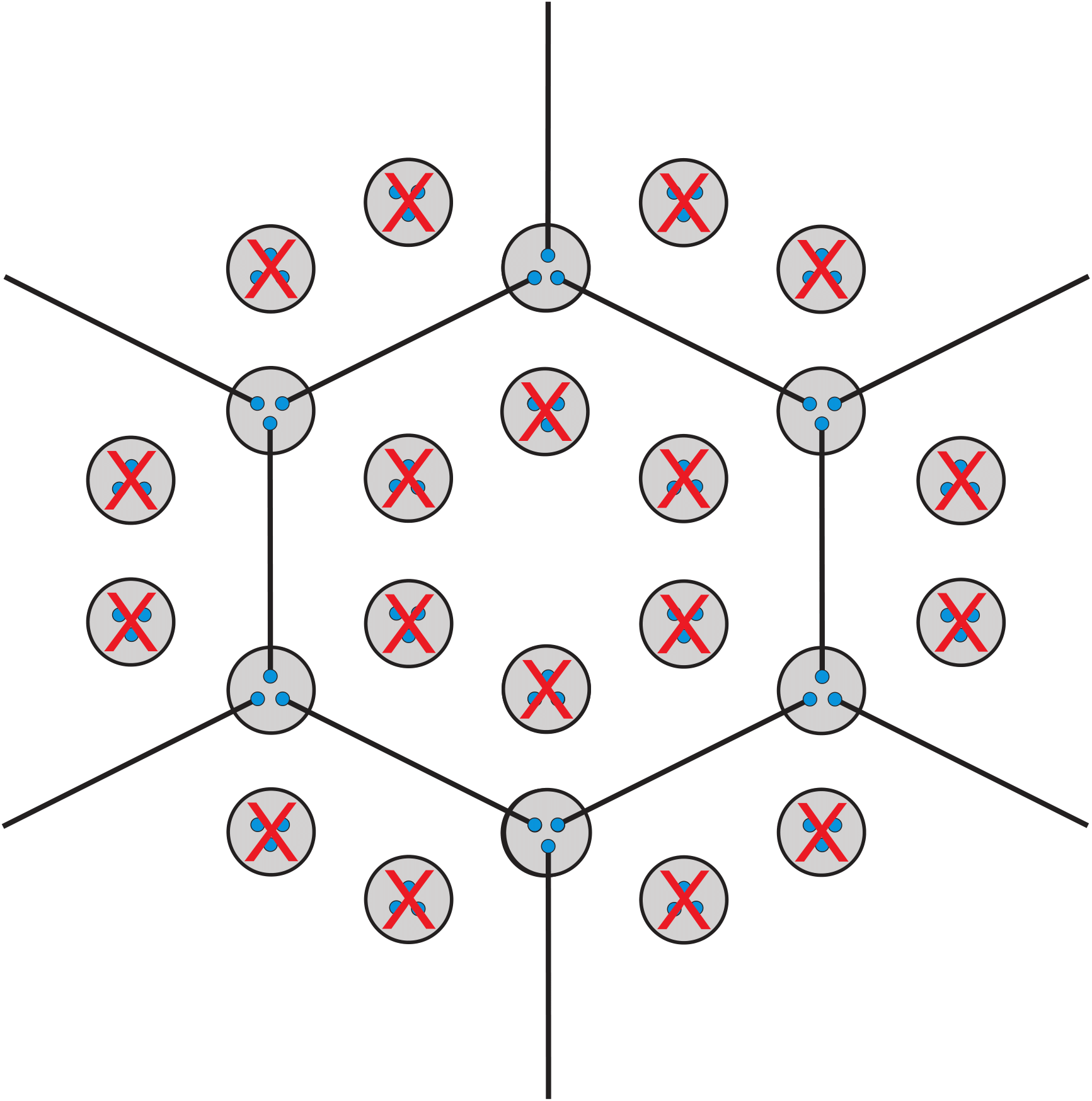}
		\label{fig:plaqhoneycombbond}
	}
	%
	%\subfigure[]{
	%	\includegraphics[width=0.38\textwidth]{plaqkagomebond1.pdf}
	%	\label{fig:plaqkagome}
	%}
	
	\subfigure[]{
		\includegraphics[width=0.38\textwidth]{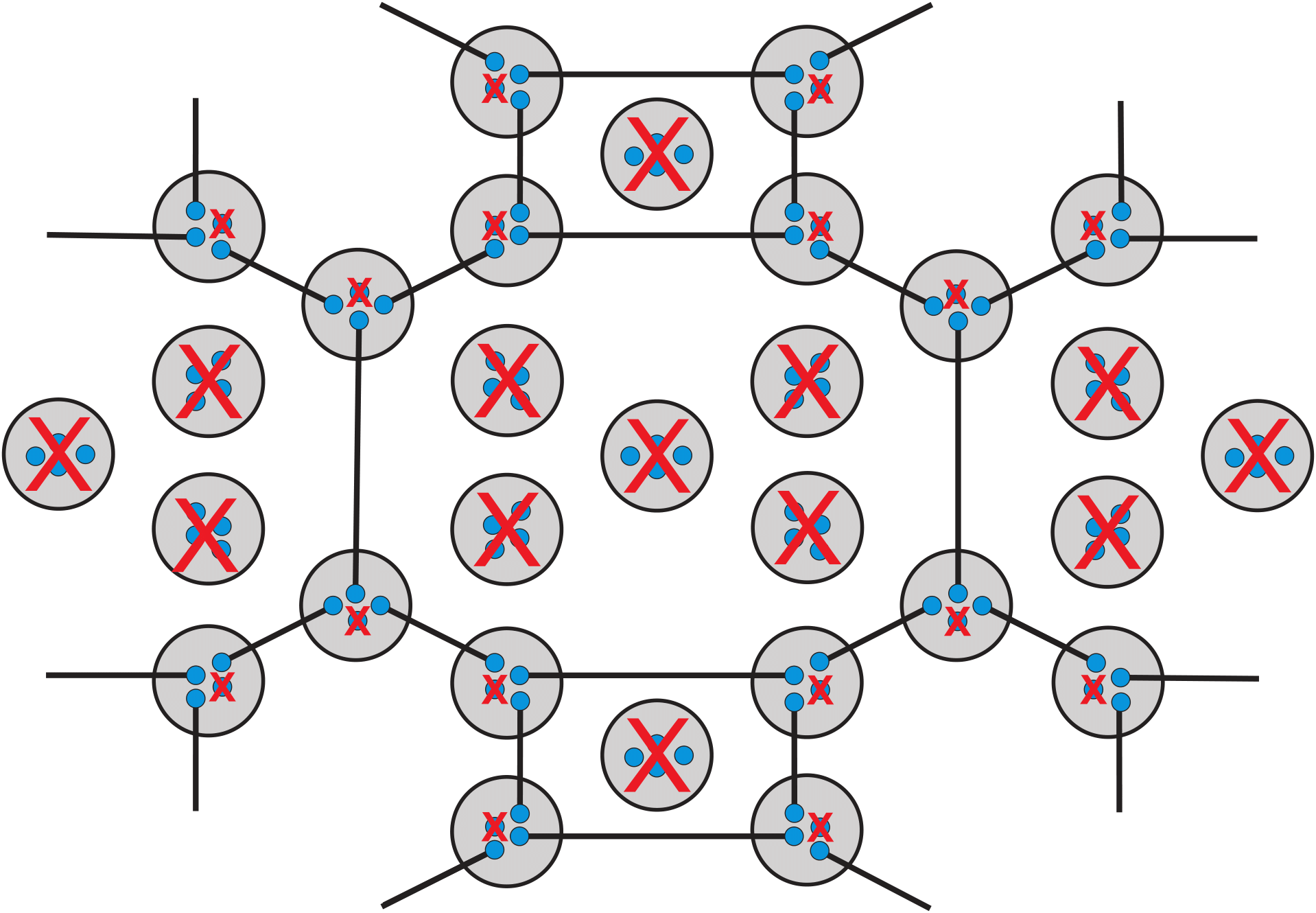}
		\label{fig:plaqkagomebond}
	}
	\caption{(Color online) Illustrated measurement patterns that reduce (a) triangular, (b) hexagonal and (c) kagome plaquette states respectively to bond states. Small red crosses mark virtual qudits that are measured individually in the $\ket{\pm}$- or $\tilde{Z}^m\ket{\tilde{+}}$-basis and big red crosses identify physical sites on which all virtual qudits are measured in said basis.}\label{bonds}
\end{figure}

The measurement $M_s$, used in our first step of reduction, is a local projective measurement on certain  sites $s$ (see Fig.~\ref{fig:plaqsquarebond})
\begin{align}
M_{s}(m_1,...,m_4)= \prod_{i=1}^4Z_i^{m_i}\ket{+}_i\bra{+}Z_i^{m_i}\label{measurementpattern}
\end{align}
with measurement outcomes $m_i\in\{0,1\}$ on virtual qubits $k=1,2,3,4$ and $\ket{\pm}\equiv 1/\sqrt{2}(\ket{0}\pm\ket{1})$ being the eigenstates of the Pauli $X$ operator. This constitutes a valid measurement as $\sum_{\{m_i\}} M^\dagger_{s}M_{s}=\mathbbm{1}_s$ is the idenity operator on site $s$. The rationale behind this pattern is that if a qubit entangled in a $n$-qubits GHZ state is measured in the $X$-basis, it becomes disentangled while leaving the remaining qubits still entangled in a $(n-1)$-qubits GHZ state up to an inconsequential local Pauli $Z$ byproduct operator in case of an outcome $m_k=1$. This can be understood with the following example. For the four-qubit GHZ state $|\psi_{\rm 4GHZ}\rangle=|0000\rangle+|1111\rangle$ (where we suppress the normalization), a measurement in the $X$ basis  on the first qubit can have two outcomes: $\{|\pm\rangle\}$. The $|+\rangle$ outcome projects the other three qubits to (suppressing normalization)
\begin{align}
{}_1\langle +|\cdot |\psi_{\rm 4GHZ}\rangle_{1234}=|000\rangle_{234}+|111\rangle_{234},
\end{align}
whereas the $|-\rangle$ outcome projects the other three qubits to
\begin{align}
{}_1\langle -|\cdot |\psi_{\rm 4GHZ}\rangle_{1234}=|000\rangle_{234}-|111\rangle_{234},
\end{align}
where the latter is the same as the former up to an Pauli $Z$ operator on any of the three remaining qubits.
By such a measurement on all but two qubits, the $n$-GHZ entanglement can be concentrated to any two-qubit standard Bell state $|00\rangle + |11\rangle$ (up to possible Pauli correction). 
As illustrated in Fig. ~\ref{CZXbond}, measuring sites in a checkerboard pattern in the $X$ basis reduces the entanglement structure from plaquettes to bonds between neighboring sites given by two-qubit Bell states $\ket{B}=\ket{00}+\ket{11}/\sqrt{2}$. The resultant state is a valence-bond state (with the bond state being $\ket{B}$) and is, via Eq.~(\ref{eqn:Hbond}), equivalent to the one defined in Ref.~\cite{Verstraete}. The latter was  already shown  to be a universal resource state for MBQC. The above reduction to a valence-bond state via local measurement can be applied more generally to plaquette states on other lattices; see, e.g., Fig.~\ref{bonds}. The significant difference to the measurement in Eq.~\ref{measurementpattern} is that one does not necessarily measure all virtual qubits within one physical site (see e.g. Fig. \ref{fig:plaqkagomebond}).

To show the universality we further convert the valence-bond state shown in Fig.~\ref{fig:plaqsquarebond} to a cluster state by local measurement,
\begin{eqnarray}
\label{eqn:Mc}
&& M_{c}(m_1,m_2,m_3)=H_1H_2\times\\
&&\,\left\{\prod_{i=1}^3X_i^{m_i}\big(\ket{0000}\bra{0000}+\ket{1111}\bra{1111}\big)\prod_{i=1}^3X_i^{m_i}\right\} H_1 H_2,\nonumber
\end{eqnarray}
where the Hadamard gate is $H\equiv(\ket{+}\bra{0}+\ket{-}\bra{1})/\sqrt{2}$, and $H_1$ and $H_2$ are used to flip the valence-bond state to the equivalent one used in Ref.~\cite{Verstraete}. In fact, the bonds in the two models are related by a Hadamard transformation:
\begin{align}
\label{eqn:Hbond}
H\otimes\mathbbm{1}\ket{B}=\mathbbm{1}\otimes H\ket{B}
=(\ket{00}+\ket{01}+\ket{10}-\ket{11})/2.
\end{align}
The choice of where the Hadamard gates appear is not unique, as long as each bond contains one such gate before the projector. This measurement projects into one of the eight two-dimensional subspaces. Regardless of the outcome, the resultant state is a cluster state (up to local Pauli $Z$ operators), to be proven below. Note that there are only three Pauli $X$ operators in Eq.~(\ref{eqn:Mc}) precisely due to dividing a 16-dimensional space to 8 two-dimensional subspaces. The Pauli $X$ operators can act on any three of the four partons. The parameters $m_1,m_2,m_3\in\{0,1\}$ represent the measurement outcomes. We can define the logical zero and one by the mapping (omitting the Hadamard gates)
\begin{align}
(|0\rangle \langle 0000|+|1\rangle\langle 1111|)X_1^{m_1}X_2^{m_2}X_3^{m_3}.
\end{align}
When $m_1=m_2=m_3=0$ on all sites, the effective state in terms of the logical $0$ and $1$ is exactly the cluster state~\cite{Verstraete}. We now consider the effect of some $m_i=1$.
Due to the commutation relation between $X_a$ and $CZ_{ab}$: 
\begin{align}
X_a CZ_{ab}= Z_b\, CZ_{ab} X_a,
\end{align}
an outcome $m_i=1$ generates an additional $Z$ byproduct operator acting on the qubit at the other end of the bond, opposite to parton $i$:
\begin{align}
X_a CZ_{ab}|++\rangle= Z_b\, CZ_{ab}|++\rangle.
\end{align}
  But on any site, if there is an additional $Z$ to the right of the projection $|0\rangle\langle 0000|+|1\rangle\langle 1111|$, then
\begin{align}
(|0\rangle\langle 0000|+|1\rangle\langle 1111|)Z_i=Z(|0\rangle\langle 0000|+|1\rangle\langle 1111|),
\end{align}
i.e., the effect on the logical qubit can be described by a Pauli $Z$ gate. Hence, we can convert the effect of outcomes $m_i=1$ to $Z$ gates acting on qubits at the other end of the bonds. In other words, the effect is given by logical $Z$ gates on the corresponding logical sites. If there is an even number of $Z$ gates on a site, the effect is an identity gate $\mathbbm{1}=Z^2$. On the other hand, if there is an odd number of $Z$ gates on a site, the effect is a $Z$ gate. Hence, we have proven that, regardless of outcomes $m_i$, the post-measurement state following a measurement~(\ref{eqn:Mc}) on all sites is a cluster state up to Pauli $Z$ gates. We remark that in the above argument there is an irrelevant overall minus sign if there are Pauli $X$ operators acting on both ends of the bond. This is due to the commutation of $X_a X_b$ and $CZ_{ab}$: 
\begin{align}
X_a X_b CZ_{ab}= - Z_a Z_b CZ_{ab} X_a X_b.
\end{align}

\medskip\noindent {\bf Alternative proof}. Another approach to prove universality after the measurement~(\ref{measurementpattern}) is to construct a universal set of gates using the valence-bond picture as in Ref.~\cite{Verstraete}. For completeness, we also demonstrate this here. Since $\ket{B}$ obviously resembles the bonds in the cluster state, the proof of universality is related to the argument for universality in the cluster state (see e.g. Ref.~\cite{Verstraete} or Ref.~\cite{qudits}). That is, we can effectively treat our bond state as a cluster state by adjusting the measurement bases. As remarked earlier,  the bonds in the two models are related by a Hadamard transformation shown in Eq.~(\ref{eqn:Hbond}). 
Henceforth, we will omit site labels if it is clear from context which site is discussed.

\begin{figure}
	\centering
	\subfigure[]{
		\includegraphics[width=0.22\textwidth]{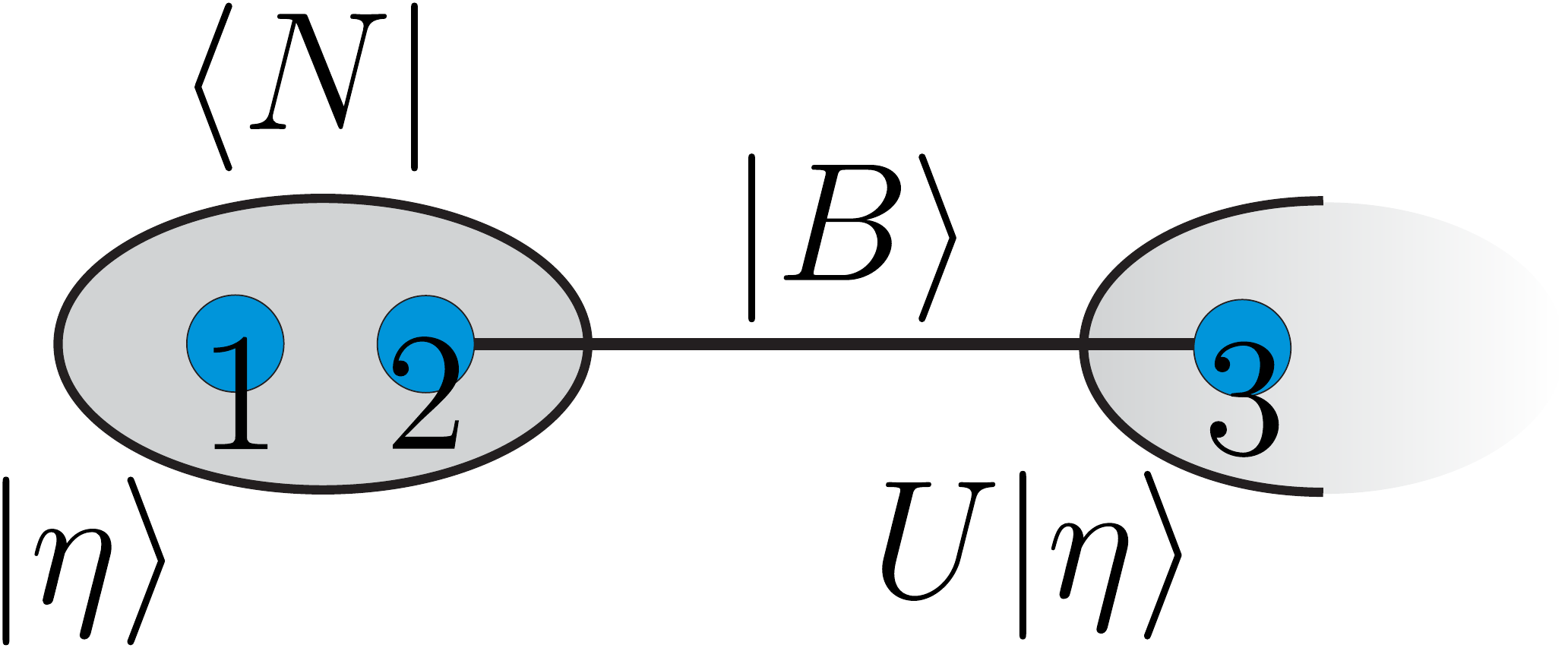}
		\label{onequbit}}
	\subfigure[]{
		\includegraphics[width=0.22\textwidth]{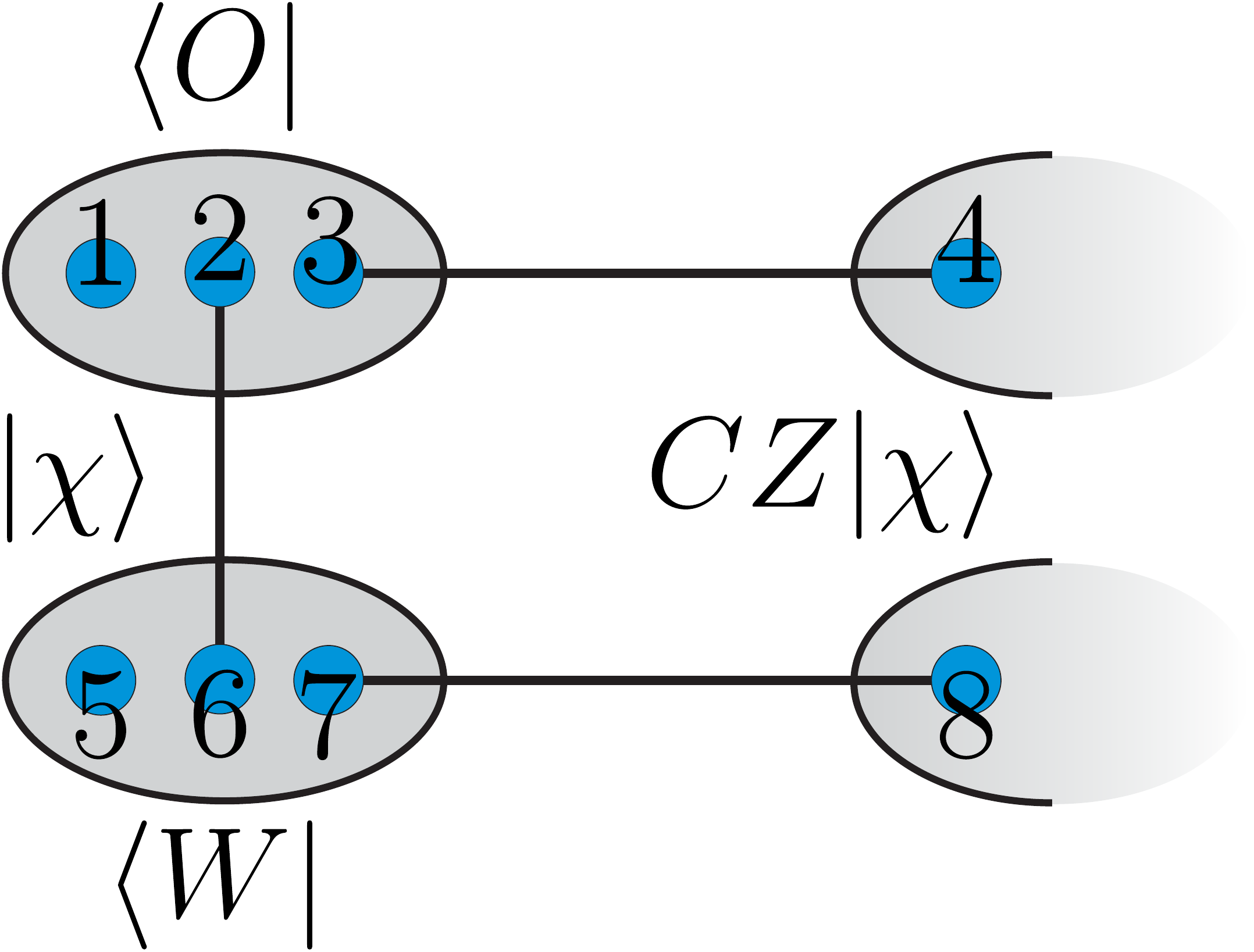}
		\label{twoqubit}
	}
	\caption{(Color online) (a) Implementation of any one-qubit unitary $U$ via joint measurement in basis $\ket{N}$ on qubits $1,2$. Qubits $2,3$ are entangled in $\ket{B}$. (b) Implementation of the controlled-phase gate $CZ$ via joint measurement in basis $\ket{O}$ on qubits $1,2,3$ and $\ket{W}$ on qubits $5,6,7$. Connected qubits are entangled pairwise in $\ket{B}$.}\label{universal}
\end{figure}

We already encountered the measurement in the $\ket{\pm}$-basis that can be implemented to prepare a suitable graph from the bond state. Similarly, initialization and readout can be performed in the computational basis. Application of the theorem in Ref. \cite{teleportation} gives that  any one-qubit operation $U$ can be achieved by measuring qubits $1$ and $2$ of a prepared state $\ket{\eta}_1\otimes(\ket{00}+\ket{11})_{23}/2$ in a twisted Bell basis (see Fig. \ref{onequbit})

\begin{align}
\ket{N(r,s)}=(U^\dagger X^rZ^s\otimes\mathbbm{1})\ket{B}.\label{Nqubit}
\end{align}
with measurement outcomes $r,s\in\{0,1\}$. This implements the unitary operation $U\ket{\eta}_3$ on an arbitrary input state $\ket{\eta}=a\ket{0}+b\ket{1}$ up to inconsequential Pauli byproduct operators $Z$ or $X$ in case of outcomes $r=1$ and $s=1$ respectively.

It is well known that we can perform any two-qubit transformation by combining one-qubit operations and the controlled-phase gate. The latter one can be implemented up to Pauli corrections by performing two distinct measurements on two adjacent sites in the following twisted GHZ-bases (see Fig. \ref{twoqubit})
\begin{align}
&\ket{O(r,s,t)}\nonumber\\
&\equiv(\mathbbm{1}\otimes H^\dagger\otimes \mathbbm{1})(X^r\otimes Z^s\otimes X^t)(\ket{000}+\ket{111})/\sqrt{2}.
\end{align}
on qubits $1,2,3$ with measurement outcomes $r,s,t\in \{0,1\}$ and
\begin{align}
\ket{W(u,v,w)}\equiv(X^u\otimes Z^v\otimes X^w)(\ket{000}+\ket{111})/\sqrt{2}.
\end{align}
on qubits $5,6,7$ with measurement outcomes $u,v,w\in \{0,1\}$. It is easy to verify that this does indeed teleport and entangle an arbitrary two-qubit input state $\ket{\chi}_{1,2}$ to $CZ\ket{\chi}_{4,8}$ up to Pauli operators. This completes the alternative proof of universality.

Note that the logical structure of this proof implies that we can perform the quantum computation on the plaquette state first and subsequently apply the measurement pattern that reduces the plaquette state to a valence-bond state. 
This procedure works equally well since the reductive measurement pattern is independent of other measurement outcomes.
 Hence, such measurements drop out of the temporal order in this model.

\subsection{2-level plaquette states on arbitrary 2D lattices}\label{arbitrarylattices}
Here we show that, in addition to the square lattice, plaquette states on arbitrary 2D lattices are indeed universal resource states for MBQC. By lattices we mean arbitrary 2D graphs that are polygon triangulation of two-tori with a macroscopic number of vertices and edges (as well as faces). (Note polygons here need not be convex, as long as they are topologically equivalent.) This definition also implies that the graphs we consider for universality are in the supercritical phase of 2D percolation. That is, all the regular 2D lattices, such as the eleven Archimedean lattices and their dual lattices, host plaquette states that both exhibit nontrivial SPT order and serve as universal resource for quantum computation. 

\medskip\noindent {\bf Hamiltonian and symmetry}.
We can also generalize the CZX model to other lattices. Hamiltonians similar to the one introduced in Eq.~(\ref{eqn:HCZX}) can be defined on arbitrary lattices so that the unique ground states are plaquette states on the same lattices. However, if the vertex degree is odd (such as in the honeycomb lattice) then the symmetry generated by $U_{CZX}=U_{CZ} U_{X}$ is no longer a linear representation of ${Z}_2$, as for example,
\begin{eqnarray}
&& U_X=X_1X_2X_3, \ U_{CZ}=CZ_{12} CZ_{23} CZ_{31}, \\
&& U_{CZX}^2=-\mathbbm{1},
\end{eqnarray}
due to $U_{CZ} U_X=-U_X U_{CZ}$ when there is an odd number of partons.  But as remarked in a footnote of Ref.~\cite{edgemodes} this can be remedied by replacing $X$ by $i X$, i.e., using $U_{iX}\equiv (iX_1)(iX_2)(iX_3)$ and the iCZX group $\{I, U_{iCZX}\equiv U_{CZ}U_{iX}\}$ gives a $Z_2$ symmetry group of the Hamiltonian. Alternatively, the  special CZX unitary can defined consistently and is a symmetry of the Hamiltonian. This is because it is guaranteed that
\begin{align}
U_{sCZ}U_X=U_XU_{sCZ}
\end{align}
since $X_cX_t\,sCZ_{ct}X_c X_t=Z_c Z_t \,sCZ_{ct}$ instead of $X_c X_t CZ_{ct}X_cX_t=-Z_c Z_t CZ_{ct}$. 

\begin{figure}
	\includegraphics[width=0.48\textwidth]{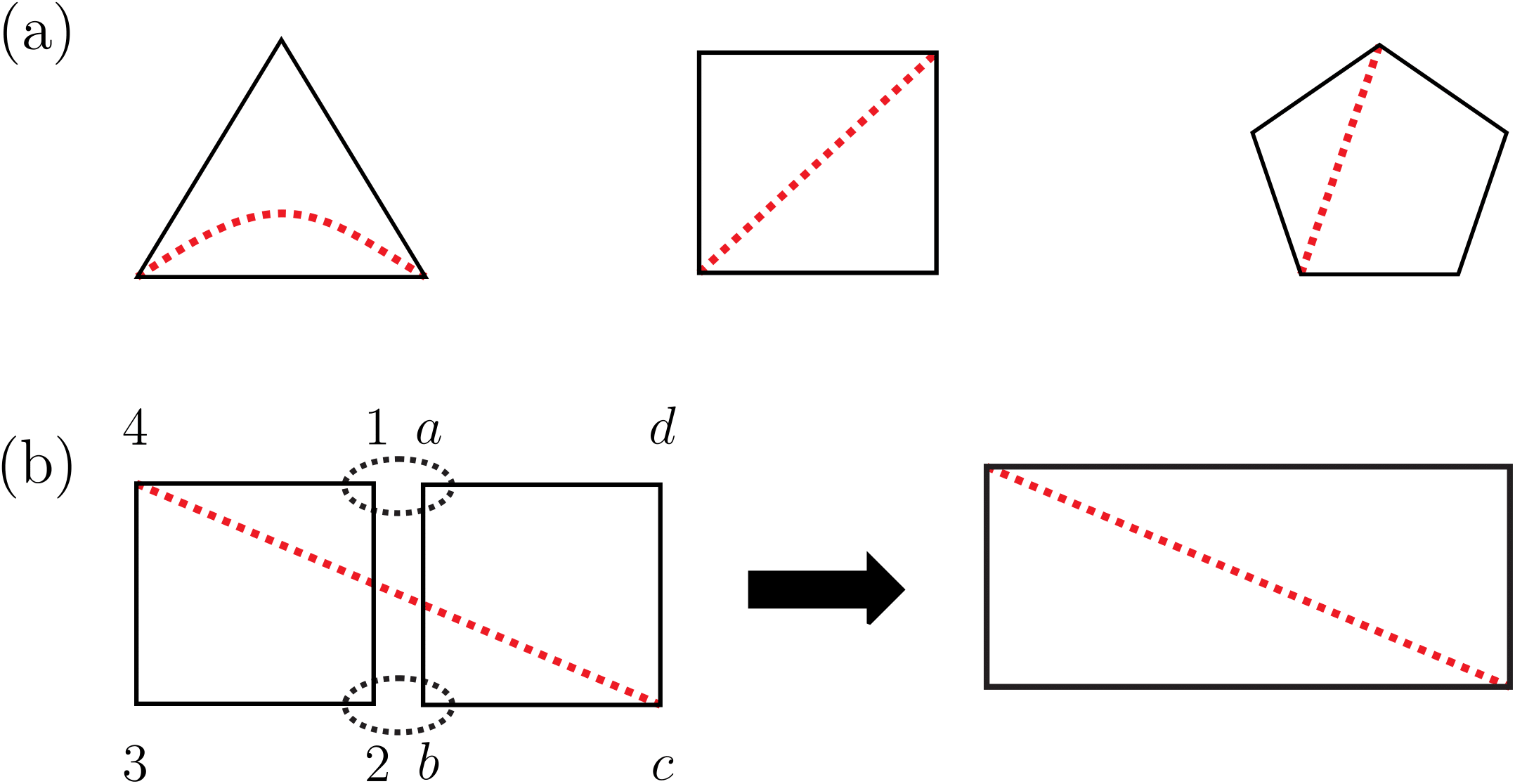}
	\caption{(Color online) Illustration of operations (a) O1 and (b) O2 as described in the text.  For the O1 operation, entanglement can be concentrated to any two partons on the plaquette. For the O2 operation, two plaquettes can be merged into one single plaquette and entanglement can be concentrated to any two partons on these two plaquettes.
		\label{fig:operations} }
\end{figure}

\medskip\noindent {\bf Quantum computational universality}.
In the previous section we discuss the concentration of GHZ entanglement in a plaquette to a Bell state between any two partons on this plaquette. Schematically, we can use a dashed line connecting these two partons (or just the two sites containing these two partons) to represent the said concentration. This represents an operation (which we shall refer to as O1) that can be implemented by local measurement; see Fig.~\ref{fig:operations}(a) for illustration.

Moreover, suppose we have two pairs of 4-GHZ states, $|\psi_{\rm 4GHZ}\rangle_{1,2,3,4}$ and $|\psi_{\rm 4GHZ}\rangle_{a,b,c,d}$  and the qubits $1$ and $a$ belong to partons on the same site. Then, it is worth mentioning here that a measurement of $1$ and $a$ in the Bell-state basis $\{|00\rangle\pm|11\rangle,|01\rangle\pm|10\rangle\}$ (which is still local on the site containing $1$ and $a$) can merge the two GHZ states into one shared among qubits $2,3,4,b,c,d$, for example,
\begin{align}
&&{}_{1a}(\langle 00|\pm\langle 11|)\cdot |\psi_{\rm 4GHZ}\rangle_{1234}\psi_{\rm 4GHZ}\rangle_{abcd} \nonumber\\
&&=
|000000\rangle_{234bcd}\pm|111111\rangle_{234bcd}.
\end{align}
Other outcomes $|01\rangle\pm|10\rangle$ give equivalent merging up to Pauli $Z$ and $X$ operators.

\begin{figure}
	\includegraphics[width=0.48\textwidth]{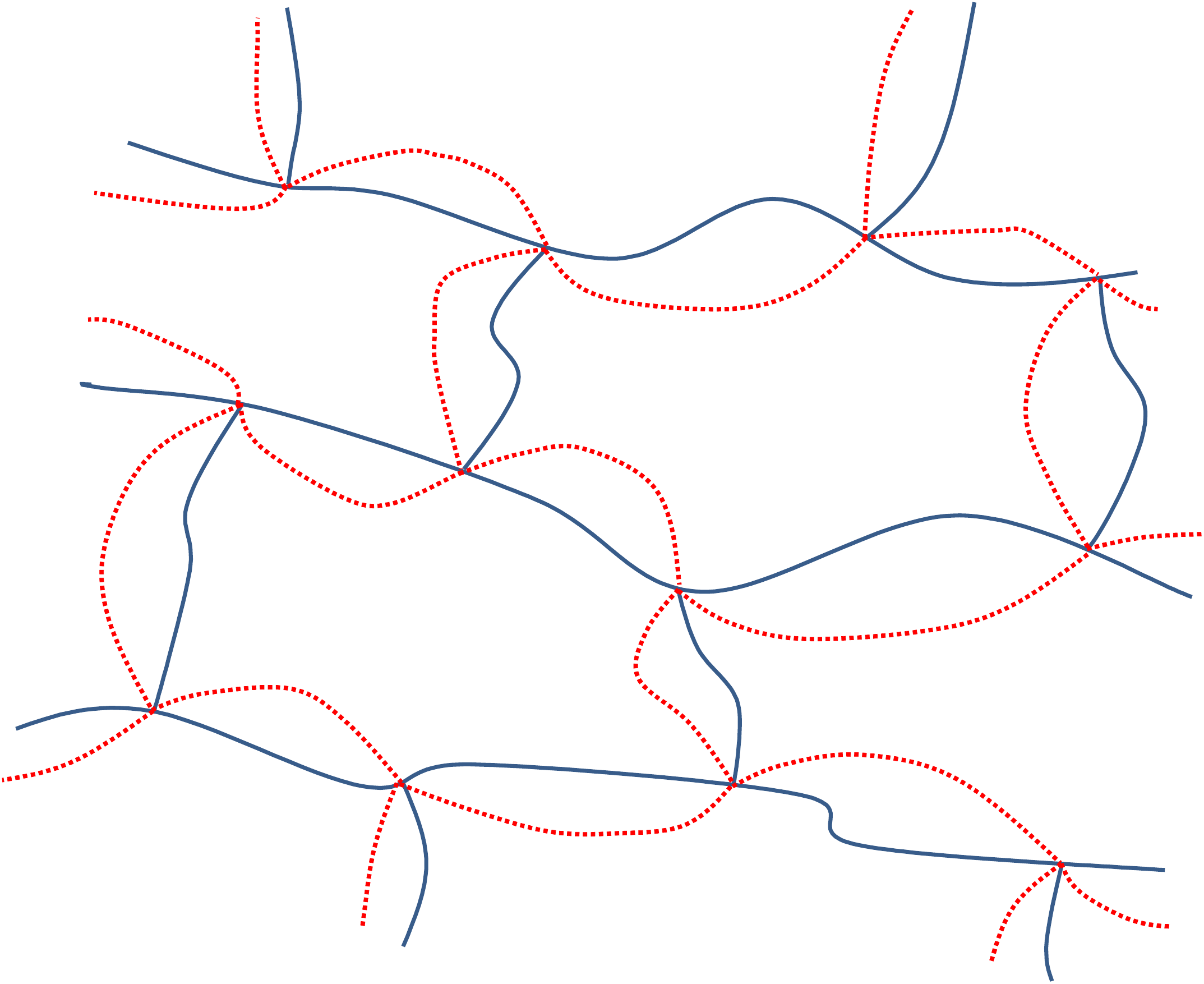}
	\caption{(Color online) Schematic illustration for the proof that plaquette states on arbitrary lattices are universal resources. Solid dark blue curves represent paths traversing edges on the lattice and dashed red curves represent concentration of entanglement via faces or polygons (which we did not draw explicitly). If the vertices where blue curves cross are sufficiently far apart, it is guaranteed that no two red dashed lines are to cross the same face (or polygon). Hence, a valence-bond state on a honeycomb lattice can be generated via local measurement. 
		\label{fig:general} }
\end{figure}

\begin{figure}
	\centering
	\subfigure[]{
		\includegraphics[width=0.22\textwidth]{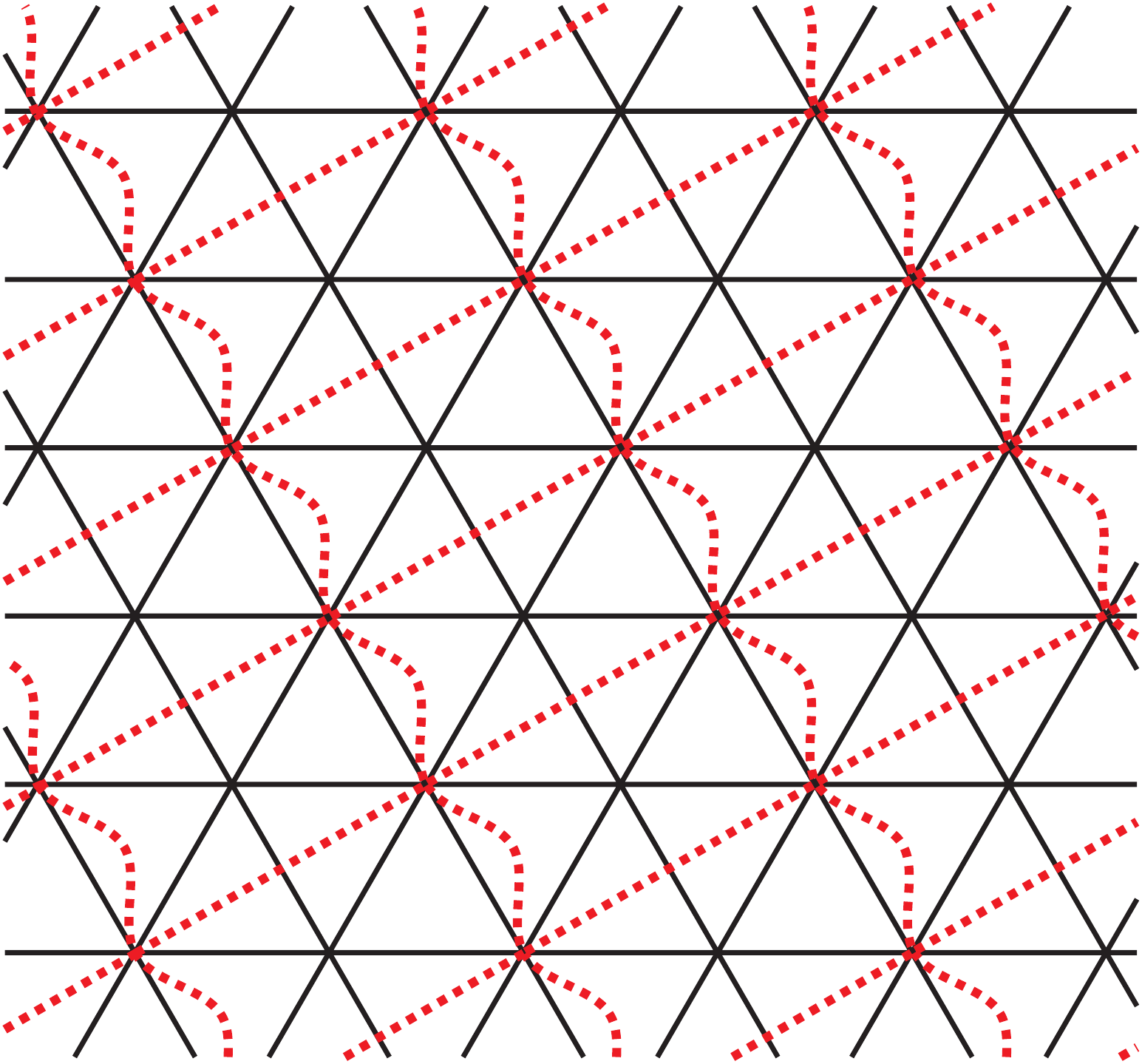}
		
	}
	\subfigure[]{
		\includegraphics[width=0.22\textwidth]{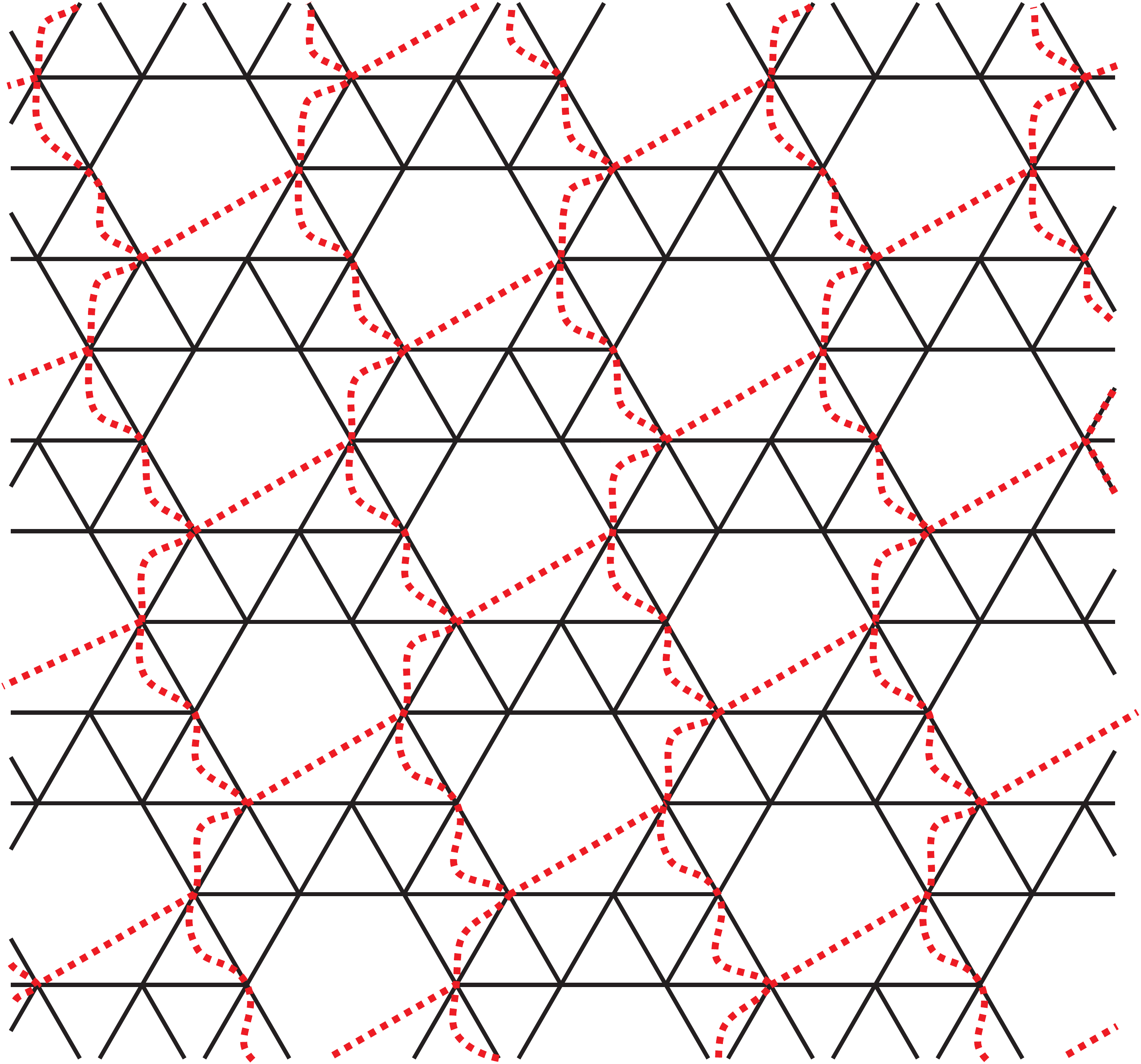}
		%\label{fig:plaqtrianglebond}
	}
	
	\subfigure[]{
		\includegraphics[width=0.22\textwidth]{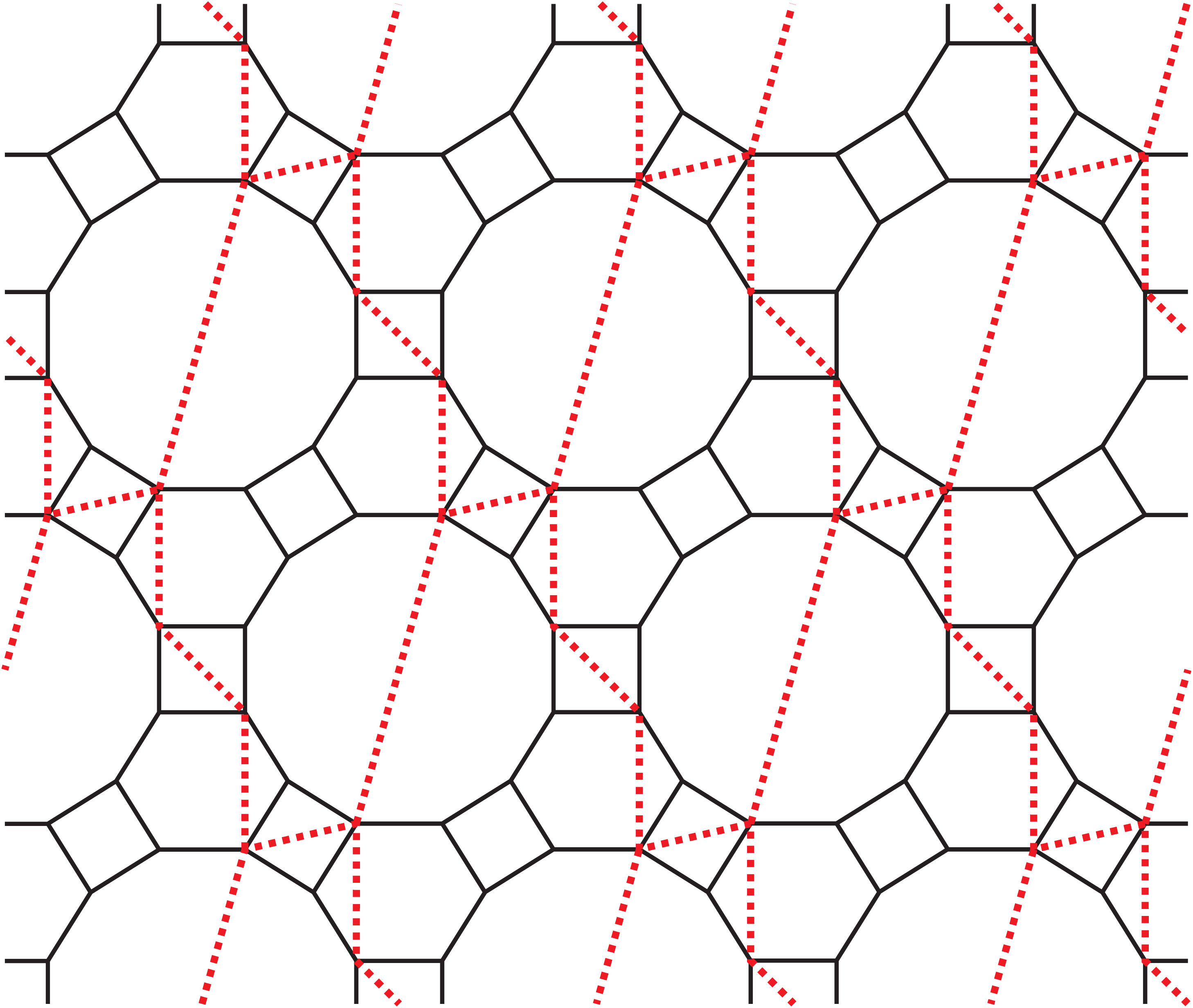}
		%\label{fig:plaqhoneycomb}
	}
	\subfigure[]{
		\includegraphics[width=0.22\textwidth]{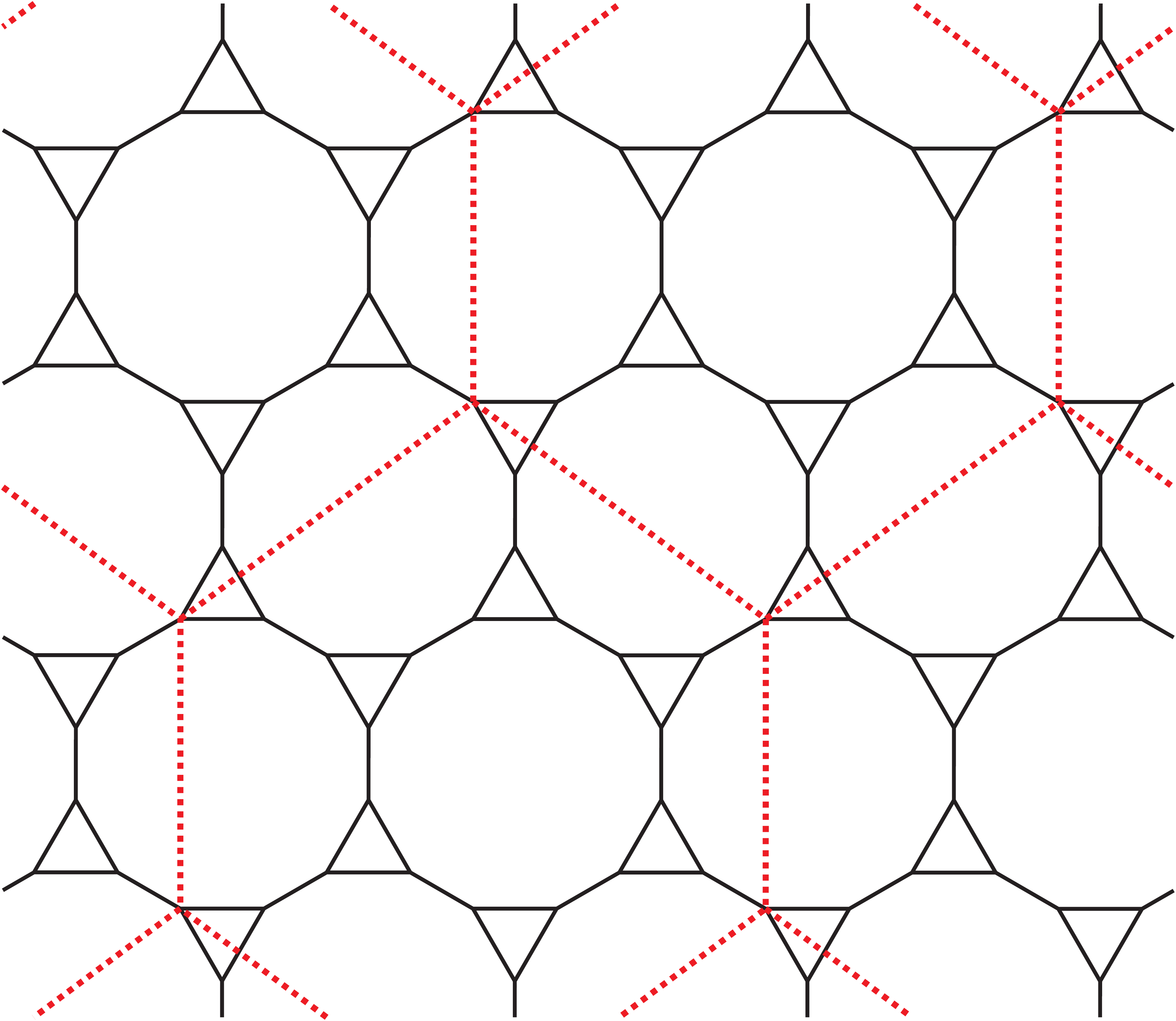}
		%\label{fig:plaqhoneycombbond}
	}
	\subfigure[]{
		\includegraphics[width=0.22\textwidth]{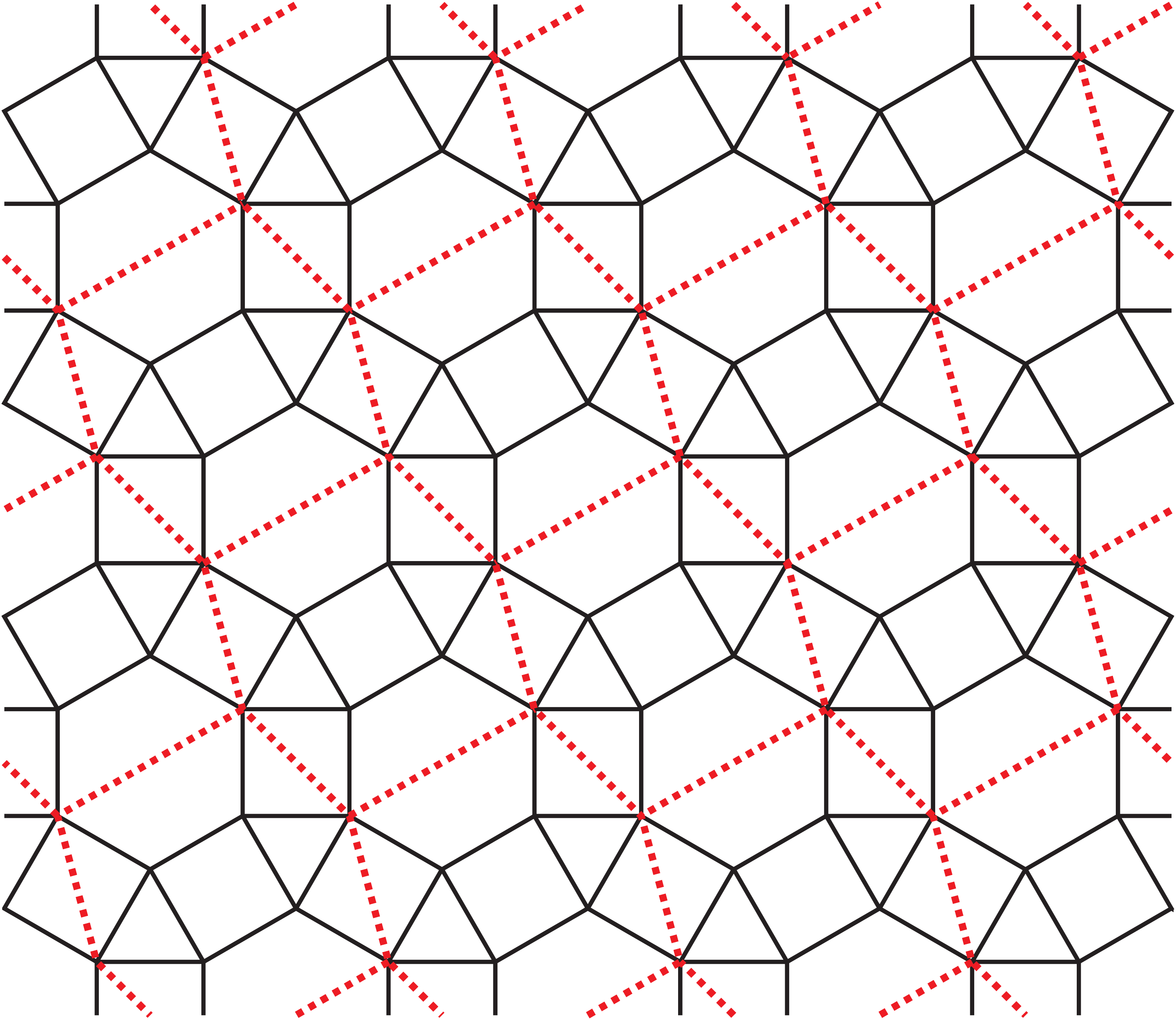}
		%\label{fig:plaqkagome}
	}
	\subfigure[]{
		\includegraphics[width=0.22\textwidth]{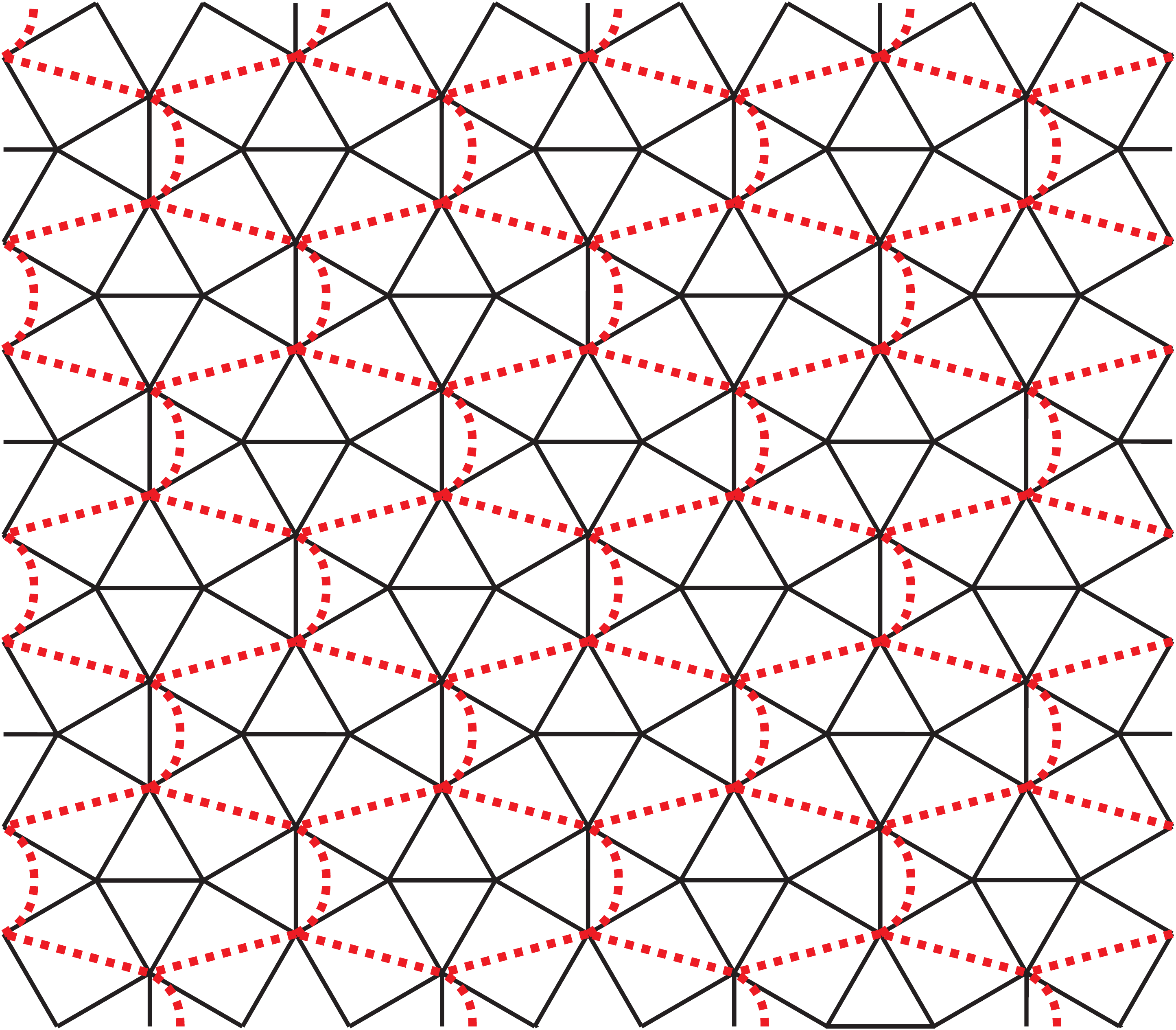}
		%\label{fig:plaqkagomebond}
	}
	\subfigure[]{
		\includegraphics[width=0.22\textwidth]{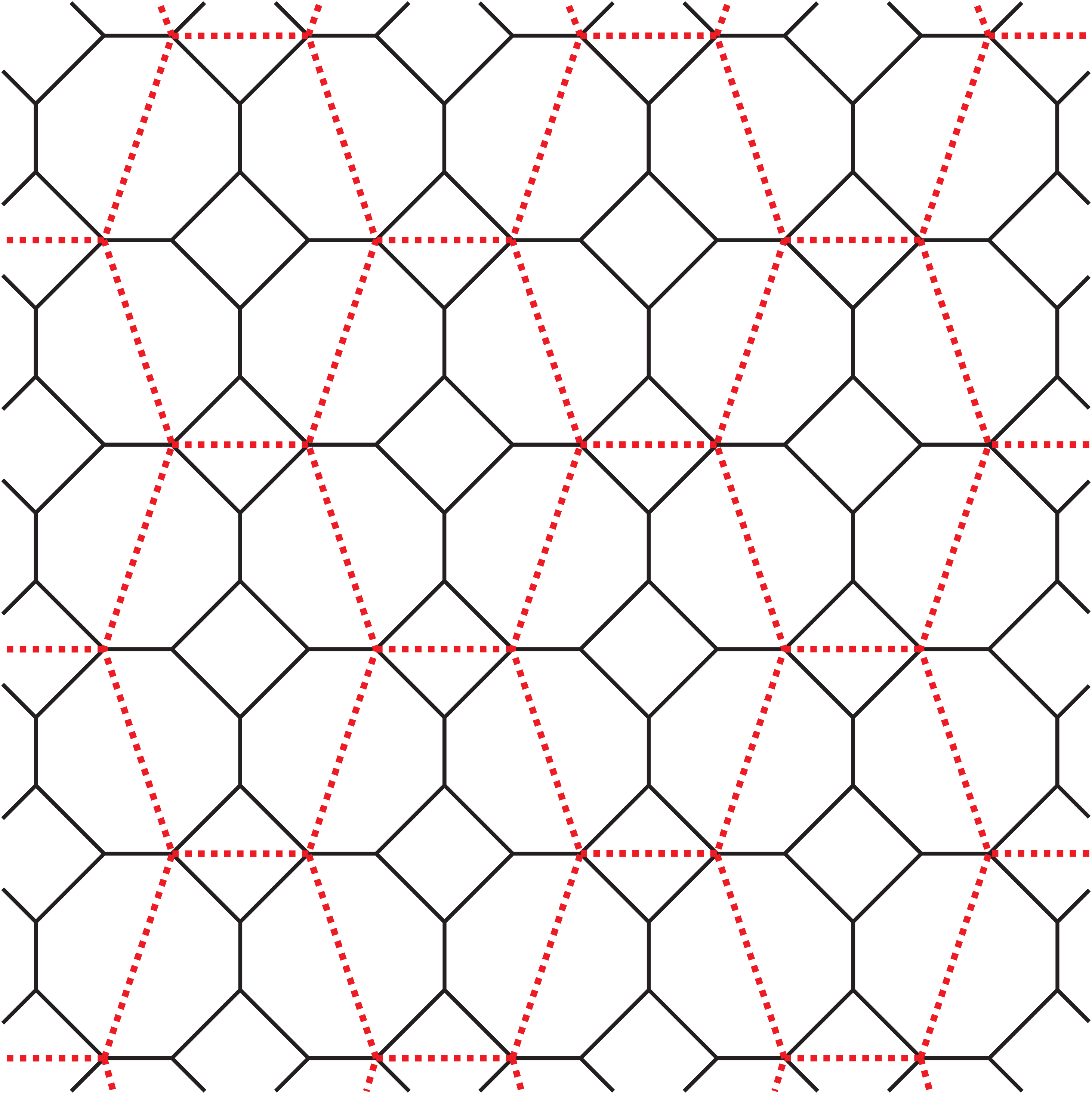}
		%\label{fig:plaqkagomebond}
	}\subfigure[]{
	\includegraphics[width=0.22\textwidth]{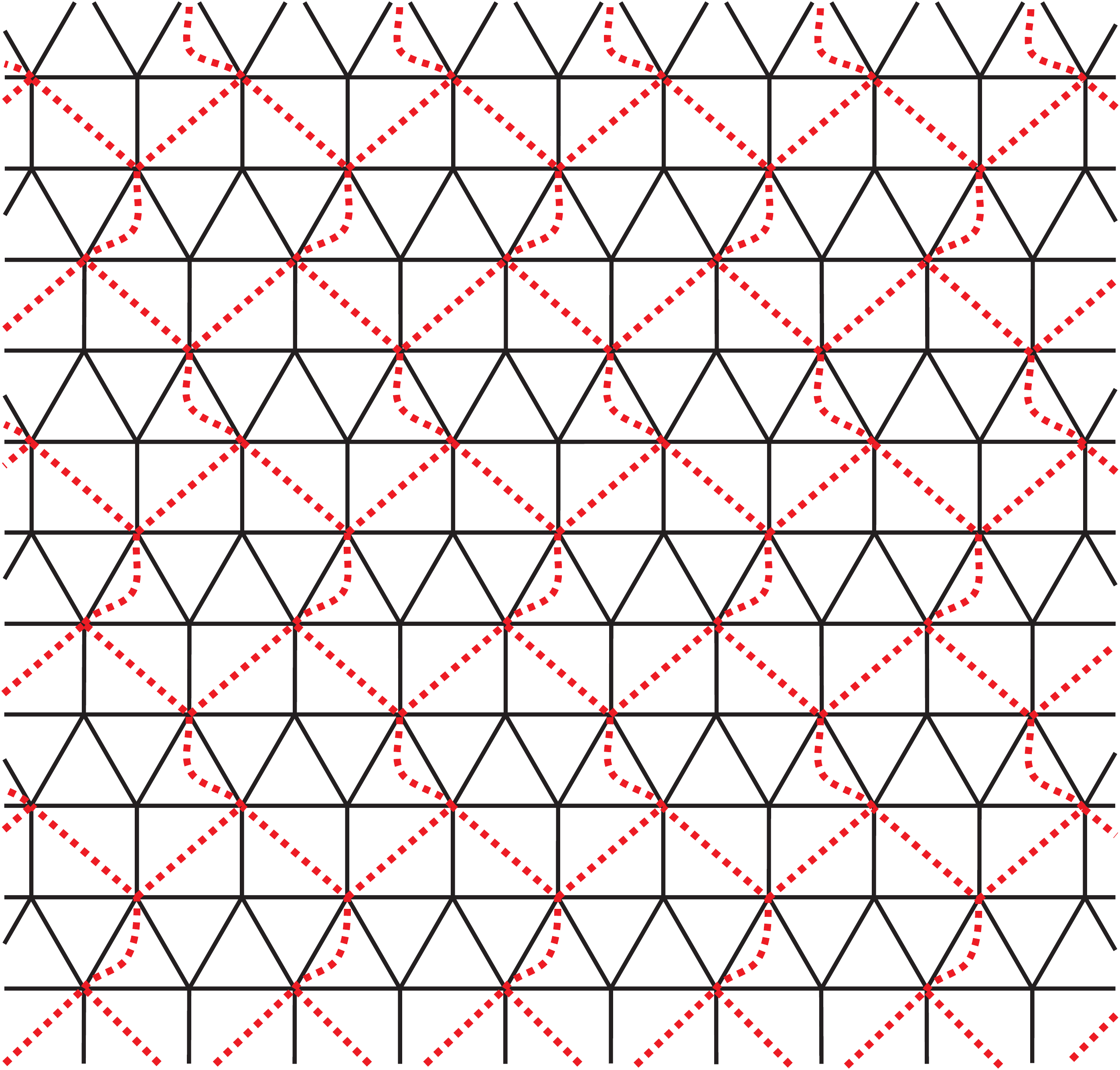}
	%\label{fig:plaqkagomebond}
}
\caption{(Color online) Various lattices for the reduction of plaquette states to valence-bond states. Utilizing the method described in the text and illustrated in Fig.~\ref{fig:operations}, plaquette states on black lattices are reduced to valence-bond states on lattices depicted by dashed red lines.}\label{fig:lattices}
\end{figure}

Suppose further that qubits 2 and $b$ also belong to one other site that shares two plaquette edges with the site containing qubits 1 and $a$. Then, as illustrated schematically in Fig.~\ref{fig:operations}(b), the two plaquettes are merged into one. (For completeness, note that one can also perform a measurement on qubits 2 and $b$  to remove both qubits from the plaquette or just on one of them to keep the other parton.) Together with the previous concentration operation, the entanglement of two plaquettes (or more) 
can be concentrated to one Bell state between any two parton qubits on any two separate locations of the two (or more) plaquettes. One can schematically draw a dashed line connecting these two partons or, for simplicity, the two physical sites containing the two partons, as shown in Fig.~\ref{fig:operations}(b) and later in Figs.~\ref{fig:general} and~\ref{fig:lattices}. This represents another operation, which we shall refer to as O2, that can be implemented by local measurement.

The above schematic operations O1 and O2 are sufficient to show that plaquette states on arbitrary 2D lattices are universal resources for MBQC. (In fact O1 is already sufficient, but the inclusion of O2 will be useful.) If the entanglement of a plaquette is concentrated to two partons, no other partons can be entangled. This implies that there cannot be a second dashed line drawn inside the same plaquette or polygon (i.e., there can be at most one),  nor can there be more than one dashed line across any adjacent plaquettes. The key point of our proof is that among arbitrary 2D graphs in the supercritical phase, one can always find a minor graph that is topologically equivalent to a honeycomb lattice~\cite{Browne}, with vertices separated from one another by a few graph distances and their number being only a fraction of the original number of vertices. If they are far enough apart, one can always connect these vertices with dashed lines cutting through polygons such that there is at most one line inside each polygon. This is illustrated in Fig.~\ref{fig:general}. Thereby, we obtain a valence-bond state on a honeycomb lattice, which can be further converted to a cluster state on the same lattice by local measurement, simply extending the proof described earlier in Sec.~\ref{example} for the square-lattice valence-bond state to the honeycomb case. Thus, we have proven that plaquette states on arbitrary 2D lattices (those whose graphs are in the supercritical phase of percolation) are universal resources for MBQC.

As a remark, partons within plaquettes that do not participate in the entanglement concentration (i.e., in those plaquettes where no dashed lines were drawn), can be measured in any basis so as to remove them from the resource state and computation. 
We illustrate the reduction procedure from plaquette states to valence-bond states on various Archimedean lattices in Fig.~\ref{fig:lattices}. As one can see, the operation O1 is sufficient. But for convenience, dashed lines corresponding to an operation O2, i.e. those that cross two polygons, were used in Fig.~\ref{fig:lattices}(d). They could be replaced by combinations of O1 dashed lines. The depicted prescriptions are not necessarily optimal but are used to demonstrate the proof described above.  

%%%%%%%%%%%%%%%%%%%%%%%%%%%%%%%%%%%%%%%%%%%%%%%%%%%%%%%%%%%%%%%%%%%%%%%%%%%%%%%%%%%%%%%%%%%%%%%%%%%%%%%%%%%%%%%%%%%%%%%%%%%%%%%%%%%%%%%%%%%%%%%%%%%%%%%%%%%%%%%%%%%%%%%%%%%%%%%%%%%%%%%%%%%%%%%%%%%%%%%%%%%%%%%%%%%%%%%%%%%%%%%%%%%%%%%%%%%%%%%%%%%%%%%%%%%%%%%%%%%%%%%%%%%%%%%%%%%%%%%%%%%%%%%%%%%%%%%%%%%%%%%%%%%%%%

\subsection{$d$-level MBQC on plaquette states}\label{general}
Originally, cluster states were defined for 2-level systems~\cite{Cluster} but they have been generalized to $d$-level systems for MBQC as well~\cite{dlevel,qudits}. 
Here, we generalize the results of the previous sections for qubit plaquette states to $d$-level systems with ground states $\ket{\psi_{gs}}$ given by Eq.~(\ref{SREstate}), i.e. with virtual  qu\textit{dits} replacing qu\textit{bits}:
\begin{align}
\ket{\psi_{gs}}=\frac{1}{d^{n_p/2}}\bigotimes_{\text{j}}\left(\sum_{g=0}^{d-1}\ket{\alpha_1=g,\beta_2=g,...,\zeta_k=g}_{p_j}\right),\label{generalgs}
\end{align}
which is constructed from the prescription in Sec.~\ref{construction} for a finite $d$-dimensional symmetry group on any plaquette, i.e. arbitrary $k$.

Let us first consider the square lattice, that is, $k=k^*=4$. The measurement $\tilde{M}$ applied in this case is just the extension of Eq.~(\ref{measurementpattern}) from qubit to qudit systems,
\begin{align}
\tilde{M}_s(m_1,...,m_{k^*})=\prod\limits_{i=1}^{k^*}\tilde{Z}^{m_i}_i\ket{\tilde{+}}_i\bra{\tilde{+}}\tilde{Z}^{m_i}_i\label{generalmeasure}
\end{align}
where $\tilde{Z}_i$'s are diagonal matrices (generalizing the Pauli $Z$ operator) acting on the $i$th virtual qudit within site $s$
\begin{align}
\tilde{Z}_i=\text{diag}(1,\Omega,\Omega^2,...,\Omega^{d-1})
\end{align}
with $\Omega=\exp(\frac{2\pi i}{d})$ such that $\tilde{Z}^d=\mathbbm{1}$. 
\begin{align}
\ket{\tilde{+}}_i\equiv\frac{1}{\sqrt{d}}\sum_{j=0}^{d-1}\ket{j}_i
\end{align}
is an eigenvector of the generalized Pauli $X$ operator $\tilde{X}$ that acts as
\begin{align}
\tilde{X}_i\ket{j}_i\equiv\ket{j+1\!\mod{d}}_i.
\end{align}
$\tilde{X}$ and $\tilde{Z}$ generate the generalized Pauli group or the Heisenberg-Weyl group on a single qudit system. To show that the generalized plaquette states are also universal resources, we will construct a similar measurement that reduces them to qudit valence-bond states. The valence-bond picture of MBQC~\cite{Verstraete} was generalized to qudits in Ref.~\cite{qudits}. The MBQC with qudit cluster states was also discussed in \cite{dlevel}. As for the qubit case universality can be proven in two different approaches.

It is easily verified that Eq.~(\ref{generalmeasure}) is a complete measurement that disentangles the ground state in Eq.~(\ref{generalgs}) analogous to the measurement in the preceding section. The resulting state is a valence-bond state with the two-qudit bonds being $\ket{\tilde{B}}=\sum_{j=0}^{d-1}\ket{jj}/\sqrt{d}$ up to inconsequential Pauli operatos.

Analogous to the discussion in Sec. \ref{arbitrarylattices}, we can extend the measurement (\ref{generalmeasure}) to arbitrary 2D lattices. For some simple lattices, we can find a similar measurement pattern that reduces systems defined on plaquettes to valence-bond states. In doing so, we can reduce, for example, the plaquette state on a triangular lattice to a bond state on a hexagonal lattice or the plaquette state on a honeycomb lattice to a bond state with the same hexagonal periodicity (see Fig.~\ref{bonds}).

Let us point out that such a measurement pattern is chosen due to its simplicity but one can think of patterns that reduce the square lattice to more complicated structures such as e.g. a decorated square lattice (which includes additional sites on edges of the square lattice). In that case, we do not necessarily measure each virtual qubit within one physical site and the post-measurement state retains some randomness on such sites. That is, the measurement $\tilde{M}_s$ on site $s$ acts on virtual qubits $i_1,...,i_l$ for some $l\leq k^*$ according to
\begin{align}
\tilde{M}_{s;i_1,...i_l}(m_{i_1},...,m_{i_l})=\prod_{k=1}^{l}\tilde{Z}^{m_{i_k}}\ket{\tilde{+}}_{i_k}\bra{\tilde{+}}\tilde{Z}^{m_{i_k}}\label{patterngeneral}
\end{align}
with outcomes $m_{i_1},...,m_{i_l}\in \{0,...,d-1\}$ and acts as unity everywhere else. Such a measurement pattern is necessary in order to reduce most lattices to valence-bond states (see e.g. Fig. \ref{fig:plaqkagomebond}). It has already been employed with $d=2$ in the context of operations O1 and O2 in the previous section. In fact, the schematic diagrams for this reduction are the same, regardless whether partons are qubits or qudits; see Fig.~\ref{fig:operations}.
Considering Eq.~(\ref{patterngeneral}), it is intuitively clear that we can reduce any generalized plaquette state to a generalized valence-bond state by local measurement (see Fig.~\ref{fig:plaqbond}). The schematic diagram for this proof is the same as for the qubit case; see Fig.~\ref{fig:general}. Hence, generalized plaquette states defined on arbitrary lattices can be reduced to generalized valence-bond states.

Analogous to the qubit case, we can employ local measurement to convert generalized valence-bond states to cluster states. For example, the generalized valence-bond state on the square lattice can be reduced to a cluster state via local measurement, similar to Eq.~\ref{eqn:Mc}, 
\begin{eqnarray}
\label{eqn:Md}
&& \tilde{M}_{d}(m_1,m_2,m_3)=F_1^\dagger F_2^\dagger\times\\
&&\,\left\{\prod_{i=1}^3\big(\tilde{X}_i^\dagger\big)^{m_i}\big(\sum_{j=0}^{d-1}\ket{jjjj}\bra{jjjj}\big)\prod_{i=1}^3\big(\tilde{X}_i\big)^{m_i}\right\} F_1 F_2,
\nonumber
\end{eqnarray}
where the Hadamard gate of the qubit case is substituted by a quantum Fourier transformation $F$:
\begin{align}
F|j\rangle\equiv\frac{1}{\sqrt{d}}\sum_{k=0}^{d-1} \Omega^{jk} |k\rangle.
\end{align}

We note that, similar to the qubit case, the choice of where the $F$ gates appear is not unique, as long as each bond contains one such gate before the projector; see also Eq.~(\ref{relation}) below. This measurement projects into one of the  $d^3$ $d$-dimensional subspaces. Regardless of the outcome, the resultant state is a $d$-level cluster state (up to local $Z$ operators), to be proven below. The parameters $m_1,m_2,m_3\in\{0,1,\cdots,d-1\}$ represent the measurement outcomes. Analogous to the qubit case, we can define the logical state by a mapping (omitting the $F$ gates)
\begin{align}
(\sum_{j=0}^{d-1}|j\rangle \langle jjjj|)\tilde{X}_1^{m_1}\tilde{X}_2^{m_2}\tilde{X}_3^{m_3}.
\end{align}
When $m_1=m_2=m_3=0$ on all sites, the effective state in terms of the logical $j$ is exactly the $d$-level cluster state~\cite{dlevel}. We now consider the effect of some $m_i\ne 0$.
The commutation relation between the $d$-level $\tilde{X}_a$ and $\tilde{CZ}_{ab}$ is
\begin{align}
\tilde{X}_a \,\tilde{CZ}_{ab}= \tilde{Z}_b^{-1}\, \tilde{CZ}_{ab} \tilde{X}_a,
\end{align}
where the $d$-dimensional $CZ$-gate acting on $c$(ontrol) and $t$(arget) qudit is given by 
\begin{align}\label{eqn:dCZ}
\tilde{CZ}\ket{jk}_{c,t}=\Omega^{jk}\ket{jk}_{c,t}.
\end{align}
From the above we see that
an outcome $m_i\ne 0$ generates an additional $\tilde{Z}^{-m_i}$ byproduct operator acting on the qudit  at the other end of the bond, opposite to parton $i$:
\begin{align}
\tilde{X}^{m_a}_a \,\tilde{CZ}_{ab}|\tilde{+}\tilde{+}\rangle= \tilde{Z}_b^{-m_a}\, \tilde{CZ}_{ab} |\tilde{+}\tilde{+}\rangle.
\end{align}

  But on any site, if there is an additional $\tilde{Z}^{-m_i}$ to the right of the projection $\sum_j|j\rangle\langle jjjj|$, then
\begin{align}
(\sum_{j=0}^{d-1}|j\rangle\langle jjjj|)\tilde{Z}_i^{-m_i}=\tilde{Z}^{-m_i}(\sum_{j=0}^{d-1}|j\rangle\langle jjjj|),
\end{align}
i.e., the effect on the logical qudit can be described by a $\tilde{Z}$ gate. Hence, we can convert the effect of outcomes $m_i\ne 0$ to $\tilde{Z}_i$ gates acting on qudits at the other end of the bonds. In other words, the effect is given by logical $\tilde{Z}$ gates on the corresponding logical sites.  Hence, we have proven that, regardless of the outcomes $m_i$, the post-measurement state following a measurement~(\ref{eqn:Mc}) on all sites is a $d$-level cluster state up to $\tilde{Z}$ gates (to some power depending on measurement outcomes).

Alternatively, we can also show that the post-measurement state subsequent to a measurement~(\ref{patterngeneral}) is universal for MBQC by explicitly constructing a universal set of quantum gates by local measurements. In Ref.~\cite{qudits} it has been shown that a $d$-level cluster state is universal for MBQC. Similar to the qubit case, the bonds $\ket{\tilde{B}}$ and the bonds in the cluster state are related by 
\begin{align}
%\label{eqn:Fbond}
(F\otimes\mathbbm{1})\ket{\tilde{B}}=(\mathbbm{1}\otimes F)\ket{\tilde{B}}=\frac{1}{\sqrt{d}}\sum_{k,j=0}^{d-1}\Omega^{kj}\ket{kj}.\label{relation}
\end{align}
Note that we ignored and will continue to ignore site labels. Due to the relation in Eq.~(\ref{relation}), we can again twist the measurement bases and projections by $F$ in order to show that the valence-bond state is universal for MBQC.

The measurement in the $\tilde{Z}^m\ket{\tilde{+}}$-basis can be deployed for preparation. Once again, measurements in the computational basis serve for initialization and readout. Since the proof in Ref. \cite{teleportation} is valid for any dimensionality $d<\infty$ of a single parton's Hilbert space, we can easily extend the measurement basis in Eq. \ref{Nqubit} to qudits:
\begin{align}
\ket{\tilde{N}(r,s)}=(U^\dagger \tilde{X}^r\tilde{Z}^s\otimes\mathbbm{1})(\sum_{j=0}^{d-1}\ket{jj})/\sqrt{d}\label{onequbitgeneral}
\end{align}
with measurement outcomes $r,s\in \{0,1,...,d-1\}$.
Universality is ensured if we can combine any one-qudit operation with the $\tilde{CZ}$-gate (see Ref. \cite{qudits}). The $\tilde{CZ}$-gate is implemented up to Pauli byproduct operators via two distinct measurements on two adjacent sites in the bases
\begin{align}
&\ket{\tilde{O}(r,s,t)}\nonumber\\
&=(\mathbbm{1}\otimes F^\dagger\otimes\mathbbm{1})(\tilde{X}^r\otimes \tilde{Z}^s\otimes \tilde{X}^t)(\sum_{j=0}^{d-1}\ket{jjj})/\sqrt{d}\label{CZ1}
\end{align}
on qudits $1,2,3$ with measurement results $r,s,t\in\{0,1,...,d-1\}$ and 
\begin{align}
\ket{\tilde{W}(u,v,w)}=(\tilde{X}^u\otimes \tilde{Z}^v\otimes \tilde{X}^w)(\sum_{j=0}^{d-1}\ket{jjj})/\sqrt{d}\label{CZ2}
\end{align}
on qudits $5,6,7$ with measurement results $u,v,w\in\{0,1,...,d-1\}$.

This completes the proof of universality for the general $d$-dimensional case. Since the argument is independent of the underlying lattice, we can utilize any generalized plaquette state for quantum computation.

Note that the logical structure of our proof implies again that we can perform the quantum computation on the plaquette state first and subsequently apply the discussed measurement pattern $\tilde{M}$ accordingly. This procedure works equally well since the measurement $\tilde{M}$ is independent of other measurement outcomes. Hence, such measurements drop out of the temporal order in this model. 

%%%%%%%%%%%%%%%%%%%%%%%%%%%%%%%%%%%%%%%%%%%%%%%%%%%%%%%%%%%%%%%%%%%%%%%%%%%%%%%%%%%%%%%%%%%%%%%%%%%%%%%%%%%%%%%%%%%%%%%%%%%%%%%%%%%%%%%%%%%%%%%%%%%%%%%%%%%%%%%%%%%%%%%%%%%%%%%%%%%%%%%%%%%%%%%%%%%%%%%%%%%%%%%%%%%%%%%%%%%%%%%%%%%%%%%%%%%%%%%%%%%%%%%%%%%%%%%%%%%%%%%%%%%%%%%%%%%%%%%%%%%%%%%%%%%%%%%%%%%%%%%%%%%%%%

%%%%%%%%%%%%%%%%%%%%%%%%%%%%%%%%%%%%%%%%%%%%%%%%%%%%%%%%%%%%%%%%%%%%%%%%%%%%%%%%%%%%%%%%%%%%%%%%%%%%%%%%%%%%%%%%%%%%%%%%%%%%%%%%%%%%%%%%%%%%%%%%%%%%%%%%%%%%%%%%%%%%%%%%%%%%%%%%%%%%%%%%%%%%%%%%%%%%%%%%%%%%%%%%%%%%%%%%%%%%%%%%%%%%%%%%%%%%%%%%%%%%%%%%%%%%%%%%%%%%%%%%%%%%%%%%%%%%%%%%%%%%%%%%%%%%%%%%%%%%%%%%%%%%%%

\section{Conclusion}\label{summary}

In this paper we extend the symmetry representation of the group cohomology approach for SPTO by Chen et al.~\cite{cohomology} to arbitrary lattices in order to show that the plaquette states exhibit nontrivial SPTO not only on the square lattice but also on arbitrary 2D lattices. In connection to quantum computation, we prove that plaquette states and their generalizations serve as universal resource states for measurement-based quantum computation as long as the underlying graphs reside in the supercritical phase of percolation. It is interesting to note that the deciding property for universality in graph states is exactly the same~\cite{universality,Browne,WeiAffleckRaussendorf12}. As a contrast, the AKLT states, which also exhibit nontrivial SPTO only if translation invariance is preserved~\cite{cohomology,Zengbook}, are known to be universal only for some regular lattices~\cite{WeiAffleckRaussendorf11,Miyake11,Wei13,WeiEtAl,WeiRaussendorf15}.

Our initial motivation was to see if one could reach similar conclusions as in the 1D case by Else et al. ~\cite{symmetryMBQC}, i.e. that certain quantum gates (or even better, a set of universal gates) may be protected within the entire 2D SPT phase (at least for a certain symmetry). If this was the case, then SPT phases would provide naturally protected resource states and could potentially yield better fault tolerance in comparison to other, topologically trivial universal resources. Measurements would then only needed to be performed in certain (symmetric) bases, giving rise to the protected gates and possibly reducing the types of error. But we have not succeeded in doing so. As for the plaquette states, the measurement that reduces them to cluster states can be done at one step, so the fault tolerance threshold the same as that of 2D cluster states. Unfortunately, this is of order $10^{-6}$ to $10^{-7}$. But for 3D cluster states, the threshold is as high as $10^{-2}$, since the measurement patterns can simulate braiding of anyons and the topological protection can be exploited~\cite{RHG}. Similarly, one can use a three-dimensional version of plaquette states and reduce them to 3D cluster states. This means that one can achieve the same threshold as for the 3D cluster state. However, we do not know whether quantum computation performed directly on plaquette states without reduction to the cluster state could result in a better threshold.

Similar to 2D graph states, the plaquette states can be defined on any 2D graph. However, the local Hilbert-space dimension can be large, depending on the number of partons on each site. Thus, it would be desirable to find projections from partons to a smaller physical subspace while at the same time preserving the nontrivial SPTO and quantum computational universality. Moreover, we have solely demonstrated the universality of plaquette states, which are nothing but representative states in SPT phases. One immediate question is whether all ground states in any or some nontrivial SPT phase are still universal. These are open-ended questions for future investigation.

In this paper, we solely discuss bosonic systems. However, it has been shown in Ref.~\cite{edgemodes} that SPT order exists for fermionic systems in 2D as well. In fact, plaquette states play a similar role if fermions are under consideration. In Ref.~\cite{supercohomology} the authors argue that SPT order in some fermionic systems can be classified by a special group supercohomology theory. Furthermore, it has already been shown in Ref.~\cite{fermionMBQC} that fermionic MBQC is possible provided that we consider additional information processing channels, namely parity modes, besides the conventional information mode. The additional modes carry the parity property of fermions. We suspect that a fermionic system defined on e.g. triangular plaquettes, Fig.~\ref{fig:plaqtrianglebond}, exhibits nontrivial SPT order and that it might also serve as a fermionic resource state.

As we were finishing up this work, we became aware of a related work by Miller and Miyake in Ref. \cite{universalSPT2D} where the authors construct a specific symmetry-protected qubit state and demonstrate that it is universal for MBQC. 

\begin{acknowledgments}
	The authors acknowledge useful discussions with Lukasz Fidkowski, Ching-Yu Huang, Abhishodh Prakash and Sebastian Dick.  This work was supported by the
	National Science Foundation under  Grant No. PHY
	1333903.
	Part of the results presented here were based on the unpublished Master Thesis by H.P.N., submitted to the Graduate School of the Stony Brook University on August 14th, 2015. 
	H.P.N. hereby acknowledges support from a Fulbright Program grant while at Stony Brook University sponsored by the Bureau of Educational and Cultural Affairs of the United States Department of State and administered by the Institute of International Education.

	%Grant No. PHY 1314748
\end{acknowledgments}

%\newpage

\end{document}